%% file: 20180601_jlt_dte2ecm.tex
\documentclass[journal]{IEEEtran}
\usepackage{cite}
\usepackage{rotating}
\usepackage{graphicx}
\usepackage{amsmath}
\usepackage{amssymb}
\usepackage{mathtools}
\usepackage{psfrag}
\usepackage{lipsum}
\usepackage{color}
\usepackage{xcolor}
\usepackage[nolist]{acronym}
\usepackage[caption=false,font=footnotesize]{subfig}
\usepackage{stfloats}
\usepackage{tikz,pgfplots}
\usepackage[ruled,lined,linesnumbered]{algorithm2e}
\usetikzlibrary{tikzmark}
\graphicspath{{figures/},
    {figures/end_to_end_pert_model/}
}

\newcommand{\figpath}{figures}
\newcommand{\figwidth}{\textwidth}
\let\oldnl\nl
\newcommand{\nonl}{\renewcommand{\nl}{\let\nl\oldnl}}

\newlength\fheight
\newlength\fwidth
\DeclareSymbolFont{AMSb}{U}{msb}{m}{n}
\DeclareSymbolFontAlphabet{\mathbb}{AMSb}
\DeclareSymbolFont{EULE}{U}{eur}{m}{n}
\DeclareSymbolFont{EUSE}{U}{eus}{m}{n}
\DeclareSymbolFont{EURE}{U}{eur}{m}{n}
\DeclareMathSymbol{\pdf}{\mathalpha}{EULE}{"66}
\DeclareMathSymbol{\pmf}{\mathalpha}{EULE}{"70}
\DeclareMathSymbol{\cdf}{\mathalpha}{EULE}{"46}
\DeclareMathSymbol{\qfac}{\mathalpha}{EUSE}{"51}
\DeclareMathSymbol{\gain}{\mathalpha}{EURE}{"67}
\DeclareMathSymbol{\accDD}{\mathalpha}{EUSE}{"42}
\DeclareMathSymbol{\logG}{\mathalpha}{EUSE}{"47}
\DeclareMathSymbol{\PP}{\mathalpha}{EUSE}{"50}
\DeclareMathSymbol{\XX}{\mathalpha}{EUSE}{"58}
\DeclareMathSymbol{\PWDD}{\mathalpha}{EUSE}{"49}
\DeclareMathSymbol{\probe}{\mathalpha}{EURE}{"1A}
\DeclareMathSymbol{\lu}{\mathalpha}{EURE}{"75}
\DeclareMathSymbol{\lU}{\mathalpha}{EURE}{"55}
\DeclareMathSymbol{\smap}{\mathalpha}{EUSE}{"53}
\DeclareMathSymbol{\zvar}{\mathalpha}{EURE}{"7A}

\def\ve#1{{\mathchoice{\mbox{\boldmath$\displaystyle #1$}}%
              {\mbox{\boldmath$\textstyle #1$}}%
              {\mbox{\boldmath$\scriptstyle #1$}}%
              {\mbox{\boldmath$\scriptscriptstyle #1$}}}}
\def\fr#1{\uppercase{#1}}%
\def\ul#1{\underline{\smash{\hbox{#1}}}}

\renewcommand{\equiv}{\triangleq}

\newcommand{\defeq}{%
  \mathrel{\vbox{\offinterlineskip\ialign{%
    \hfil##\hfil\cr
    {\fontsize{4}{4.8}\selectfont def}\cr
    \noalign{\kern.2ex}
    $=$\cr
}}}}

\newcommand{\norm}[1]{\left\| #1\right\|}

\newcommand{\fourier}[1]{\mathcal{F} \{\,#1\,\}}

\newcommand{\invfourier}[1]{\mathcal{F}^{-1} \{\,#1\,\}}

\newcommand{\dtft}[1]{\hat{\mathcal{F}} \{\,#1\,\}}
\newcommand{\idtft}[1]{\hat{\mathcal{F}}^{-1} \{\,#1\,\}}
\newcommand{\set}[1]{\{\,#1\,\} }
\newcommand{\expval}[1]{\mathrm{E}\{\,#1\,\}}
\newcommand{\sequence}[1]{\langle\,#1 \,\rangle}
\newcommand{\pert}[1]{\Delta#1}

\def\conv{*}

\newcommand{\alias}[1]{\text{\scshape Alias}\{\, #1 \,\}}
\newcommand{\fold}[1]{\text{\scshape Fold}\{\, #1 \,\}}
\newcommand{\dft}[1]{\text{\scshape DFT}\{\, #1 \,\}}
\newcommand{\idft}[1]{\text{\scshape DFT}^{-1}\{\, #1 \,\}}

\newcommand{\trf}{\mbox{\setlength{\unitlength}{0.1em}%
        \begin{picture}(19,10)%
            \put(4,3){\circle{4}}%
            \put(6,3){\line(1,0){8}}%
            \put(14,3){\circle*{4}}%
        \end{picture}%
    }%
}%
\newcommand{\itrf}{\mbox{\setlength{\unitlength}{0.1em}%
        \begin{picture}(19,10)%
            \put(4,3){\circle*{4}}%
            \put(6,3){\line(1,0){6}}%
            \put(14,3){\circle{4}}%
        \end{picture}%
    }%
}%
\newcommand*{\QEDB}{\hfill\ensuremath{\square}}%

\def\trfjoin@{\joinrel}

\def\PERT{\rm \scriptscriptstyle PERT}
\def\PERTs{\rm \scriptstyle PERT}
\def\REGs{\rm \scriptstyle REG}
\def\REGLOGs{\rm \scriptstyle REGLOG}
\def\FDs{\rm \scriptstyle FD}
\def\TDs{\rm \scriptstyle TD}
\def\SSFM{\rm \scriptscriptstyle SSFM}

\def\jj{\mathrm{j}}
\def\dd{\mathrm{d}}
\def\ee{\mathrm{e}}

\def\qq{\ve{\mathfrak{q}}}
\def\aa{\ve{a}}
\def\ts{\ve{s}}
\def\uu{\ve{u}}

\def\met{\mathrm{m}}
\def\watt{\mathrm{W}}
\def\mwatt{\mathrm{mW}}

\def\dB{\mathrm{dB}}
\def\dBm{\mathrm{dBm}}
\def\km{\mathrm{km}}
\def\ps{\mathrm{ps}}

\def\GHz{\mathrm{GHz}}
\def\GBd{\mathrm{GBd}}

\def\NN{\mathbb{N}}
\def\Z{\mathbb{Z}}
\def\R{\mathbb{R}}
\def\C{\mathbb{C}}

\def\F{\mathbb{F}}
\def\TT{\mathbb{T}}
\def\T{\mathsf{T}}
\def\H{\mathsf{H}}

\def\X{\mathsf{x}}
\def\Y{\mathsf{y}}
\def\STp{\smap_{\Tsym, \probe}}
\def\STnu{\smap_{\Tsym, \nu}}
\def\Nfft{N_{\rm \scriptscriptstyle DFT}}
\def\Nsym{N_{\rm \scriptscriptstyle SYM}}
\def\Tsym{T}
\def\Rs{R_\mathrm{s}}

\def\wNyq{\omega_{\rm {Nyq}}}

\def\rolloff{\rho}
\def\Eg{E_\mathrm{T}}
\def\Es{E_\mathrm{s}}

\def\Pow{P}
\def\Rm{{R_{\mathrm{m}}}}
\def\N0{N_\mathsf{0}}

\def\kap{\textsc{k}}

\def\w0{\omega_0}

\def\f0{f_0}
\def\lam0{\lambda_0}
\def\P0{P_\mathrm{o}}
\def\vg{v_\mathrm{g}}
\def\Leff{{L_{\mathrm{eff}}}}
\def\Leffa{{L_{\mathrm{eff,a}}}}
\def\Lsp{{L_\sp}}

\def\Ld{{L_\mathrm{D}}}
\def\Nsp{N_\sp}
\def\Nch{N_\mathrm{ch}}
\def\sp{\mathrm{sp}}

\def\Tx{\mathrm{T}}
\def\Rx{\mathrm{R}}
\def\Ch{\mathrm{C}}
\def\NL{\mathrm{NL}}
\def\phiT{\Phi}
\def\aphi{\bar{\phi}}
\def\aPhi{\bar{\Phi}}

\def\phinlnu{\phi_{\NL, \nu}}
\def\phinl{\phi_{\NL}}
\def\phinlp{\phi_{\NL, \probe}}

\def\accD{\accDD}

\def\dOm{\xi}
\def\luu{\ve{u}_{\rm \scriptscriptstyle LIN}}
\def\lUU{\ve{U}_{\rm \scriptscriptstyle\! LIN}}
\def\UU{\ve{U}}
\def\Hnl{H_{\NL}}
\def\bs{\ve{s}}
\def\bS{\ve{S}}
\def\superSCI{{\rm \scriptscriptstyle SCI}}
\def\superXCI{{\rm \scriptscriptstyle XCI}}
\def\ejwt{\ee^{\jj \omega \Tsym}}
\def\ejwwt{\ee^{\jj \ve{\omega} \Tsym}}
\def\setk{\mathcal{K}}
\begin{document}

\definecolor{myblue}{rgb}{0,0.4,0.8}
\definecolor{myred}{rgb}{0.8,0,0}
\definecolor{mygray}{rgb}{0.5,0.5,0.5}
\definecolor{LightGray}{rgb}{0.2,0.2,0.2}
\pgfplotsset{
grid style={solid,very thin},
tick style={color=gray},
label style={font=\footnotesize},
tick label style={font=\footnotesize,color=LightGray},
legend style={font=\small},
}
\begin{acronym}
\acro{2D}{\ul{two}-\ul{d}imensional}
\acro{3D}{\ul{three}-\ul{d}imensional}
\acro{4D}{\ul{four}-\ul{d}imensional}
\acro{A/D}{\ul{a}nalog-to-\ul{d}igital}
\acro{ACF}{\ul{a}uto\ul{c}orrelation \ul{f}unction}
\acro{ADC}{\ul{a}nalog-to-\ul{d}igital \ul{c}onverter}
\acro{ASE}{\ul{a}mplified \ul{s}pontaneous \ul{e}mission}
\acro{ASK}{\ul{a}mplitude-\ul{s}hift \ul{k}eying}
\acro{AWGN}{\ul{a}dditive \ul{w}hite \ul{G}aussian \ul{n}oise}
\acro{BER}{\ul{b}it \ul{e}rror \ul{r}atio}
\acro{BICM}{\ul{b}it-\ul{i}nterleaved \ul{c}oded \ul{m}odulation}
\acro{BRGC}{\ul{b}inary-\ul{r}eflected \ul{G}ray \ul{c}ode}
\acro{BRGL}{\ul{b}inary-\ul{r}eflected \ul{G}ray \ul{l}abeling}
\acro{CDF}{\ul{c}umulative \ul{d}istribution \ul{f}unction}
\acro{CD}{\ul{c}hromatic \ul{d}ispersion}
\acro{CM}{\ul{c}oded \ul{m}odulation}
\acro{CW}{\ul{c}ontinuous \ul{w}ave}
\acro{D/A}{\ul{d}igital-to-\ul{a}nalog}
\acro{DAC}{\ul{d}igital-to-\ul{a}nalog \ul{c}onverter}
\acro{DEC}{\ul{dec}oder}
\acro{DFT}{\ul{d}iscrete \ul{F}ourier \ul{t}ransform}
\acro{DGD}{\ul{d}ifferential \ul{g}roup \ul{d}elay}
\acro{DP}{\ul{d}ual-\ul{p}olarization}
\acro{DSP}{\ul{d}igital \ul{s}ignal \ul{p}rocessing}
\acro{DTFT}{\ul{d}iscrete-\ul{t}ime \ul{F}ourier \ul{t}ransform}
\acro{DU}{\ul{d}ispersion \ul{u}ncompensated}
\acro{E/O}{\ul{e}lectrical-to-\ul{o}ptical}
\acro{ECB}{\ul{e}quivalent \ul{c}omplex \ul{b}aseband}
\acro{EDFA}{\ul{e}rbium-\ul{d}oped \ul{f}iber \ul{a}mplifier}
\acro{ENC}{\ul{enc}oder}
\acro{FEC}{\ul{f}orward \ul{e}rror \ul{c}or-rection}
\acro{FFT}{\ul{f}ast \ul{F}ourier \ul{t}ransform}
\acro{FIR}{\ul{f}inite \ul{i}mpulse \ul{r}esponse}
\acro{FWM}{\ul{f}our-\ul{w}ave \ul{m}ixing}
\acro{GN}{\ul{G}aussian \ul{n}oise}
\acro{EGN}{\ul{e}xtended \ul{G}aussian \ul{n}oise}
\acro{GVD}{\ul{g}roup \ul{v}elocity \ul{d}ispersion}
\acro{IDRA}{\ul{i}deal \ul{d}istributed \ul{R}aman \ul{a}mplification}
\acro{IIR}{\ul{i}nfinite \ul{i}mpulse \ul{r}esponse}
\acro{IQ}{\ul{i}nphase-\ul{q}uadrature}
\acro{ISI}{\ul{i}nter-\ul{s}ymbol \ul{i}nterference}
\acro{LDPC}{\ul{l}ow-\ul{d}ensity \ul{p}arity-\ul{c}heck}
\acro{LLR}{\ul{l}og \ul{l}ikelihood \ul{r}atio}
\acro{LMS}{\ul{l}east \ul{m}ean \ul{s}quare}
\acro{LO}{\ul{l}ocal-\ul{o}scillator}
\acro{LP}{\ul{l}ogarithmic \ul{p}erturbation}
\acro{MIMO}{\ul{m}ultiple-\ul{i}nput/\ul{m}ultiple-\ul{o}utput}
\acro{ML}{\ul{m}aximum \ul{l}iklihood}
\acro{NLPN}{\ul{n}on\ul{l}inear \ul{p}hase \ul{n}oise}
\acro{MSE}{\ul{m}ean-\ul{s}quared \ul{e}rror}
\acro{NLSE}{\ul{n}on\ul{l}inear \ul{S}chr{\"o}dinger \ul{e}quation}
\acro{NLIN}{\ul{n}on\ul{l}inear \ul{i}nterference \ul{n}oise}
\acro{NLI}{\ul{n}on\ul{l}inear \ul{i}nterference}
\acro{O/E}{\ul{o}ptical-to-\ul{e}lectrical}
\acro{OA}{\ul{o}ptical \ul{a}mplification}
\acro{ODE}{\ul{o}rdinary \ul{d}ifferential \ul{e}quation}
\acro{OSNR}{\ul{o}ptical \ul{s}ignal-to-\ul{n}oise \ul{r}atio}
\acro{PAM}{\ul{p}ulse-\ul{a}mplitude \ul{m}odulation}
\acro{PDE}{\ul{p}artial \ul{d}ifferential \ul{e}quation}
\acro{PDF}{\ul{p}robability \ul{d}ensity \ul{f}unction}
\acro{PDM}{\ul{p}olarization-\ul{d}ivision \ul{m}ultiplex}
\acro{PMD}{\ul{p}olarization \ul{m}ode \ul{d}ispersion}
\acro{PMF}{\ul{p}robability \ul{m}ass \ul{f}unction}
\acro{PN}{\ul{p}hase \ul{n}oise}
\acro{PRBS}{\ul{p}seudo \ul{r}andom \ul{b}it \ul{s}equence}
\acro{PSD}{\ul{p}ower \ul{s}pectral \ul{d}ensity}
\acro{PWDD}{\ul{p}ower-\ul{w}eighted \ul{d}ispersion \ul{d}istribution}
\acro{QAM}{\ul{q}uadrature \ul{a}mplitude \ul{m}odulation}
\acro{QPSK}{\ul{q}arternary \ul{p}hase-\ul{s}hift \ul{k}eying}
\acro{RBW}{\ul{r}esolution \ul{b}and\ul{w}idth}
\acro{RC}{\ul{r}aised \ul{c}osine}
\acro{RP}{\ul{r}egular \ul{p}erturbation}
\acro{eRP}{\ul{e}nhanced \ul{r}egular \ul{p}erturbation}
\acro{RRC}{\ul{r}oot-\ul{r}aised \ul{c}osine}
\acro{RV}{\ul{r}andom \ul{v}ariable}
\acro{SCI}{\ul{s}elf-\ul{c}hannel \ul{i}nterference}
\acro{MCI}{\ul{m}ulti-\ul{c}hannel \ul{i}nterference}
\acro{SER}{\ul{s}ymbol \ul{e}rror \ul{r}atio}
\acro{SNR}{\ul{s}ignal-to-\ul{n}oise \ul{r}atio}
\acro{SPM}{\ul{s}elf-\ul{p}hase \ul{m}odulation}
\acro{SP}{\ul{s}et \ul{p}artitioning}
\acro{SOP}{\ul{s}tate \ul{o}f \ul{p}olarization}
\acro{SSFM}{\ul{s}plit-\ul{s}tep \ul{F}ourier \ul{m}ethod}
\acro{SSMF}{\ul{s}tandard \ul{s}ingle-\ul{m}ode \ul{f}iber}
\acro{VSTF}{\ul{V}olterra \ul{s}eries \ul{t}ransfer \ul{f}unction}
\acro{WDM}{\ul{w}avelength-\ul{d}ivision \ul{m}ultiplexing}
\acro{XCI}{\ul{cross}-\ul{c}hannel \ul{i}nterference}
\acro{XPM}{\ul{cross}-\ul{p}hase \ul{m}odulation}
\acro{XPolM}{\ul{cross}-\ul{p}olarization \ul{m}odulation}
\acro{iXPM}{\ul{i}ntra-channel \ul{cross}-\ul{p}hase \ul{m}odulation}
\end{acronym}

\title{On Discrete-Time/Frequency-Periodic End-to-End Fiber-Optical Channel Models}
%
%
%

\author{Felix~Frey,
        Johannes~K.~Fischer,~\IEEEmembership{Senior Member,~IEEE,}
        and~Robert~F.H.~Fischer,~\IEEEmembership{Senior Member,~IEEE}
\thanks{F.~Frey and R.\,F.H.~Fischer are with the Institute of Communications Engineering, Ulm University, 89081 Ulm,
    Germany
    (e-mail: \{felix.frey, robert.fischer\} @uni-ulm.de) }
\thanks{J.K.~Fischer is with the Photonics Networks and Systems department at Fraunhofer Heinrich-Hertz-Institute, Einsteinufer 37, 10587 Berlin, Germany.}
\thanks{Funded by the Deutsche Forschungsgemeinschaft (DFG, German Research Foundation)---grant FI~982/14-1.}
}


\maketitle

\begin{abstract}
    A discrete-time end-to-end fiber-optical channel model is derived based on the first-order perturbation approach.
    The model relates the discrete-time input symbol sequences of co-propagating wavelength channels to the received symbol sequence after matched filtering and $\ve{\Tsym}$-spaced sampling.
    To this end, the interference from both self- and cross-channel nonlinear interactions of the continuous-time \textsl{optical} signal is represented by a single discrete-time perturbative term.
    Two equivalent discrete-time models can be formulated---one in the time-domain, the other in the $\ve{1/\Tsym}$-periodic continuous-frequency domain.
    The time-domain formulation coincides with the well-known \textsl{pulse-collision picture}.
    The novel frequency-domain picture incorporates the sampling operation via an aliased and periodic kernel description.
    This gives rise to an alternative perspective on the end-to-end input/output relation between the spectrum of the discrete-time transmit symbol sequence and the spectrum of the receive symbol sequence.
    Both views can be extended from a regular, i.e., solely additive model, to a combined regular-logarithmic model to take the multiplicative nature of certain distortions into consideration.
    An alternative formulation of the \textsl{Gaussian Noise} model is provided to take the aliasing of frequency components correctly into account.
    A novel algorithmic implementation of the discrete and periodic frequency-domain model is presented.
    The derived end-to-end model requires only a single computational step and shows good agreement in the mean-squared error sense compared to the oversampled and inherently sequential split-step Fourier method.
\end{abstract}


\IEEEpeerreviewmaketitle

\section{Introduction}
\IEEEPARstart{I}{n} communication theory, discrete-time end-to-end channel models play a fundamental role in developing advanced transmission and equalization schemes.
Most notable the discrete-time \textsl{linear}, dispersive channel with \ac{AWGN} is often used to model \textsl{point-to-point} transmission scenarios.
In the last decades, a large number of transmission methods for such linear channels have emerged and are now applied in many digital transmission standards.
With the advent of high-speed CMOS technology, those schemes have also been adopted in applications for \textsl{fiber-optical} transmission with digital-coherent reception \cite{Savory2010}.

However, many of the applied techniques (e.g., coded modulation, signal shaping, and equalization) are designed for linear channels whereas the fiber-optical channel is inherently \textsl{nonlinear}.
An exact model to obtain the output sequence from a given input sequence by an \textsl{explicit} input/output relation is highly desirable to make further advances in developing strategies optimized for fiber-optical transmission.

Indeed, in the past two decades, considerable effort was spent developing channel models for fiber-optical transmission with good trade-offs between computational complexity and numerical accuracy.
Most of the prior work is, however, concerned with \textsl{continuous-time} models. The corresponding formulation of the \textsl{discrete-time} equivalent is incomplete---as detailed below.

Due to physical properties, the optical and analog electrical signals are continuous-time signals and analog processing is adequate.
Since all communication signals are bandlimited, when obeying the sampling theorem, the analog signals can be processed in discrete-time domain without loss of information.
Since the bandwidth of a single communication signal is typically larger than the symbol rate $\Rs \defeq 1/T$ (where $T$ is the duration of the modulation interval), the sampling frequency has to be larger than $1/T$, so-called oversampling.
However, in any digital receiver for $T$-spaced pulse-amplitude modulation, $T$-spaced sampling and further $T$-spaced discrete-time signal processing are performed.
Thus, for communication signals the sampling theorem is in general \textsl{not} fulfilled.
In fact, aliasing of frequency components is an essential part in recovering the data.
Hence, all discrete-time end-to-end channel models have to incorporate this sampling step in order to fully capture the effects.

In fiber-optics, starting from the \ac{NLSE}, approximate solutions of the \textsl{optical} end-to-end channel (i.e., continuous-time) can be obtained following either a \textsl{perturbative approach} (cf.~\cite[P.~610]{Zwillinger1998}) or the equivalent method of \ac{VSTF} (cf.~\cite{Peddanarappagari1997, Vannucci2002a}).
These (continuous-time) channel models can approximate the nonlinear distortion---there commonly termed \ac{NLI}---up to the order of the series expansion of the \ac{NLSE}.
A comprehensive summary of recent developments on channel models can be found in \cite[Sec.~I]{Ghazisaeidi2017a}.

One particular class of channel models---based on a first-order time-domain perturbative approach---has been published in the early 2000s in a series of contributions by A.~Mecozzi et al.~\cite{Mecozzi2000, Mecozzi2000a, Mecozzi2001}.
The results, however, were limited to transmission schemes that were practical at that time (e.g., dispersion-managed transmission, intensity-modulation, and direct-detection) and did not include the receiver-side matched filter and sampling step. The details of the theory and its derivation were published recently in \cite{Mecozzi2011}.

A follow-up seminal paper by A.~Mecozzi and R.-J.~Essiambre \cite{Mecozzi2012a} extends the former work by including the \textsl{matched filter} and $T$-spaced sampling after ideal coherent detection%
.
This work constitutes the first \textsl{discrete-time} end-to-end formulation using a time-domain approach.
One central result is the integral formulation of the (Volterra/perturbation) \textsl{kernel coefficients} in time-domain providing a first-order approximation of the \textsl{per-modulation-interval} equivalent end-to-end input/output relation.
Based on this paper, R.~Dar et al.~\cite{Dar2013, Dar2014b, Dar2015b, Dar2015c} derived the so-called \textsl{pulse-collision picture} of the nonlinear fiber-optical channel.
Here, the properties of cross-channel \ac{NLI} were properly associated with certain types of pulse collisions in \textsl{time}-domain. 
In particular, the importance of separating additive and multiplicative distortions were discussed.

In this contribution, we complement the view on $T$-spaced end-to-end channel models for optical transmission systems by an equivalent \textsl{frequency}-domain description.
We provide an integral solution which relates the periodic spectrum of the transmit \textsl{symbol} sequence to the periodic spectrum of the receive \textsl{symbol} sequence, i.e., after (linear) channel matched filtering and aliasing to frequencies within the Nyquist interval.
The time discretization with symbol spacing $T$ translates to a $1/T$-periodic representation in frequency.
This is fundamentally different from prior well-known frequency-domain approaches, e.g.~\cite{Peddanarappagari1997, Vannucci2002a, Poggiolini2012}, that use a continuous-time or \textsl{oversampled} representation of both the signal \textsl{and} the kernel coefficients%
\footnote{In prior work (e.g.~nonlinear compensators based on Volterra models, or the \ul{G}aussian \ul{N}oise (GN)-model and extended models derived thereof) signal and system descriptions are usually based on oversampled representations at rates equal or larger than the \textsl{bandwidth} $B \ge \Rs$ of the continuous-time signal, or respectively, at rates equal or larger than the bandwidth of the whole WDM signal. The relation of the proposed model to the GN/EGN-model is discussed in Section \ref{sec:gn}.}%
.
Remarkably, the \textsl{frequency matching} which is imposed by the general \ac{FWM} process in the optical domain is still maintained in the $1/T$-periodic frequency-domain.

We believe that both the existing time-domain end-to-end channel model according to the pulse collision picture, and the novel $1/T$-periodic frequency-domain end-to-end model have potential application in a variety of fields.
Among those is the application as a \textsl{forward} channel model for the optimization of detection schemes that operate on a per-symbol basis, e.g., recovery of phase distortions or determination of symbol likelihood values.
Similarly, in a \textsl{backward-propagation-sense}, both models can find application in fiber nonlinearity compensation which requires \textsl{real-time} processing using fixed-point arithmetic, i.e., implementation and computational complexity is of particular interest.
Those methods for fiber nonlinearity compensation based on variants of the time-domain perturbation theory have already been proposed and demonstrated by various groups in various different flavours, cf.~\cite{Tao2011,Oyama2014,Gao2014, Zhuge2014a, Frey2018}.
In both forward and reverse applications, the kernel coefficients can be pre-calculated or pre-trained, whereas for the latter case, i.e., for fiber nonlinearity compensation, additionally adaptation of the kernel coefficients is required.


The paper is organized as follows.
In Section \ref{sec:II}, the notation is briefly introduced and the system model of coherent fiber-optical transmission is presented.
In Section \ref{sec:III}, starting from the \textsl{continuous-time} end-to-end relation of the \textsl{optical} channel---an intermediate result following the perturbation approach---the \textsl{discrete-time} end-to-end relation is derived.
We particularly highlight the relation between the time and frequency representation and point out the connection to other well-known channel models.
The relevant system parameters, i.e., \textsl{memory} and \textsl{strength}, of the nonlinear response are identified which lead to design rules for potential applications.
For such applications, a novel algorithmic implementation in $1/T$-periodic frequency-domain is introduced.
In analogy to linear systems, we argue that---depending on the particular system scenario---a frequency-domain implementation of the $T$-spaced channel model can potentially be advantageous in terms of computational complexity compared to a time-domain implementation.
This is expected for systems where the nonlinear memory is \textsl{large}---similar as for linear systems
\footnote{In this contribution, we will not provide a complexity analysis, as the scope of this work is the analytical derivation and the validity of the proposed algorithms. A thorough complexity analysis is part of future investigations.}%
.

Similar to the pulse-collision picture, certain mixing products in frequency-domain can be attributed to a pure phase and polarization rotation.
This in turn motivates the extension of the original \textsl{regular} perturbation model in frequency-domain to a combined \textsl{regular-logarithmic} model taking the multiplicative nature of certain distortions properly into account.
In Section \ref{sec:IIII}, the theoretical considerations are complemented by numerical simulations which are in accordance with results obtained by the \ac{SSFM}.
Here, the relevant metric to assess the match between both models is the \ac{MSE} between the two $T$-spaced output sequences for a given input sequence.
Section \ref{sec:IIIII} presents some conclusions and an outlook.

\section{Notation and System Model}\label{sec:II}
This section briefly introduces the notation and the overall system model to make this contribution as self-contained as possible.
\subsection{Notation and Basic Definitions}
Sets are denoted with calligraphic letters, e.g., $\mathcal{A}$ is the set of data symbols, i.e., the symbol \textsl{alphabet} or \textsl{signal constellation}. A \textsl{set of numbers} or \textsl{finite fields} are typeset in blackboard bold typeface, e.g., the set of real numbers is $\R$, and the set of non-negative real numbers is $\R_{\ge 0}$.
Bold letters, such as $\ve{x}$, indicate vectors.
If not stated otherwise, a vector $\ve{x} = [x_1, x_2, \dots , x_{n}]^\T$ of dimension $n$ is a column vector\footnote{$(\cdot)^{\T}$ denotes transposition and $(\cdot)^\H$ is the Hermitian transposition.}, and the set of indices to the elements of the vector is $\mathcal{I} \defeq \{1, \ldots, n\}$.
Non-bold italic letters, like $x$, are scalar variables, whereas non-bold Roman letters refer to constants, e.g., the imaginary number is $\jj$ with $\jj^2 = -1$.

A \textsl{real} (bandpass) signal is typically described using the \ac{ECB} representation, i.e., we consider the complex envelope $x(t) \in \C$ with \textsl{inphase} (real) and \textsl{quadrature} (imaginary) component.

The $n$-dimensional \textsl{Fourier transform} of a continuous-time signal $x(\ve{t}) = x(t_1,t_2, \dots, t_n)$ depending on the $n$-dimensional time vector $\ve{t} = [t_1, t_2, \dots, t_n]^\T \in \R^n$ (in \textsl{seconds}) is denoted by $\fr{x}(\ve{\omega}) = \fourier{x(\ve{t})}$, and defined as \cite[Ch.~4]{Oppenheim1983}
\begin{align}
    \fr{x}(\ve{\omega}) &= \fourier{x(\ve{t})} \defeq \int\limits_{\R^n} x(\ve{t})\,\ee^{-\jj \ve{\omega} \cdot \ve{t}} \;\dd^n \ve{t} \\
    x(\ve{t}) &= \invfourier{\fr{x}(\ve{\omega})} = \frac{1}{(2 \pi)^n}\int\limits_{\R^n} \fr{x}(\ve{\omega})\,\ee^{\jj \ve{\omega} \cdot \ve{t}} \;\dd^n \ve{\omega} .
\end{align}
Here, $\fr{x}(\ve{\omega})$ is a continuous function of 
\textsl{angular frequencies} $\ve{\omega} = [\omega_1, \omega_2, \dots, \omega_n]^\T \in \R^n$ with $\omega = 2 \pi f$ and frequency $f\in \R$ (in \textsl{Hertz}).
In the exponential we made use of the dot product of vectors in $\R^n$ given by $\ve{\omega}\,\cdot\,\ve{t} \defeq \omega_1 t_1 + \omega_2 t_2 + \dots + \omega_n t_n$.
The integral is an $n$-fold multiple integral over $\R^n$ and the integration boundaries are at $-\infty$ and $\infty$ in each dimension. We use the expression $\dd^n \ve{t}$ as shorthand for $\dd t_1 \dd t_2 \dots \dd t_n$.
For the one-dimensional case with $n=1$ the variable subscript is dropped.
We may also write the correspondence as $x(t) \trf \fr{x}(\omega)$ for short.

The $n$-dimensional \ac{DTFT} of a discrete-time sequence\footnote{If \textsl{whole} (finite-length) sequence is treated, this is indicated by the square bracket notation, i.e., $\sequence{x[k]}$.} $\sequence{x[\ve{k}]}$ with $\ve{k} = [k_1, k_2, \dots, k_n]^\T \in \Z^n$ with spacing $\Tsym$ between symbols is periodic with $1/\Tsym$ in frequency-domain and denoted as $X(\ee^{\jj \ve{\omega} \Tsym}) = \dtft{x[\ve{k}]}$, and defined as\footnote{The notation $\sum_{\ve{k}\in \Z^n}$ is short for $\sum_{k_1=-\infty}^{\infty}\sum_{k_2=-\infty}^{\infty} \dots \sum_{k_n=-\infty}^{\infty}$ and $\mathbb{Z}$ is the set of integers.}
\begin{equation}
    \fr{x}(\ee^{\jj \ve{\omega} T}) = \dtft{x[\ve{k}]} \defeq \sum\limits_{\ve{k}\in \Z^n} x[\ve{k}]\,\ee^{-\jj \ve{\omega}\cdot \ve{k} T}
\end{equation}
\begin{equation}
    x[\ve{k}] = \idtft{X(\ee^{\jj \ve{\omega} T})}
    = \left(\frac{T}{2 \pi}\right)^n \int_{\TT^n} X(\ee^{\jj \ve{\omega} T})\,\ee^{\jj \ve{\omega} \cdot \ve{k} T} \;\dd^n \ve{\omega}.
\end{equation}
The set of frequencies in the Nyquist interval is $\TT \defeq \set{\omega \in \R \mid -\wNyq \le \omega < \omega_{\rm Nyq}}$ with the Nyquist (angular) frequency $\omega_{\rm Nyq} \defeq 2 \pi /(2 \Tsym) $.

In the present work the so-called \textsl{engineering} notation of the Fourier transform with a negative sign in the complex exponential (in the forward, i.e., time-to-frequency, direction) is used. 
This has immediate consequences for the solution of the electro-magnetic wave equation (cf.~\textsl{Helmholtz equation}), and therefore also for the \ac{NLSE}. 
In the optical community, there exists no fixed convention with respect to the sign notation, e.g., some of the texts are written with the physicists' (e.g., \cite[Eq.~(2.2.8)]{Agrawal2010} or \cite{Mecozzi2012a}) and others with the engineering (e.g., \cite{Kaminow2013},\cite[Eq.~(A.4)]{Engelbrecht2015} ) notation in mind. Consequently, the derivations shown here may differ marginally from some of the original sources.

Continuous-time signals are associated with meaningful physical units, e.g., the electrical field has typically units of \textsl{volts per meter} (V/m).
The \ac{NLSE} and the \textsl{Manakov equation} derived thereof are carried out in \textsl{Jones space} over the quantity $\ve{u}(t) = [u_\X(t), u_\Y(t)]^\T \in \C^2$ called the \textsl{optical field envelope}.
The optical field envelope has the same orientation as the associated electrical field but is normalized s.t.~$\ve{u}^\H \ve{u}$ equals the instantaneous power given in \textsl{watts} (W).
In this work, signals are instead generally treated as dimensionless entities as this considerably simplifies the notation when we move between the various signal domains (see, e.g., discussion in \cite[P.~11]{Fischer2002} or \cite[P.~230]{Kammeyer2008}).
To this end, $\ve{u}^\H \ve{u}$ is re-normalized to be dimensionless. 
Similarly, the \textsl{nonlinearity coefficient} $\gamma$ (commonly given in $\mathrm{W}^{-1}\mathrm{m}^{-1}$) is also re-normalized to have units of $\mathrm{m}^{-1}$.

To distinguish a two-dimensional complex vector $\ve{u} = [u_\X, u_\Y]^\T \in \C^2$ in Jones space from its associated three-dimensional real-valued vector in \textsl{Stokes space}, we use decorated bold letters $\vec{\ve{u}} = [{u}_{1}, {u}_{2}, {u}_{3}]^\T\in \R^3$.
The (permuted) set of Pauli matrices is given by \cite{Gordon2000}
\begin{equation}
    \ve{\sigma}_1 \defeq \begin{bmatrix} 1 & 0 \\ 0 & -1\end{bmatrix}
    \quad
    \ve{\sigma}_2 \defeq \begin{bmatrix} 0 & 1 \\ 1 & 0\end{bmatrix}
    \quad
    \ve{\sigma}_3 \defeq \begin{bmatrix} 0 & -\jj \\ \jj & 0\end{bmatrix},
    \label{eqn:pauli}
\end{equation}
and the Pauli vector is $\vec{\ve{\sigma}} \defeq [\ve{\sigma}_1, \ve{\sigma}_2, \ve{\sigma}_3]^\T$ where each vector component is a $2\times2$ Pauli matrix.
The relation between Jones and Stokes space can then be established by the concise (symbolic) expression $\vec{\ve{u}} = \ve{u}^\H \vec{\ve{\sigma}} \ve{u}$ to denote the element-wise operation $u_{i} = \ve{u}^\H \ve{\sigma}_i \ve{u}$ for all Stokes vector components $i = 1,2,3$. 
The Stokes vector $\vec{\ve{u}}$ can also be expanded using the dot product with the Pauli vector to obtain the complex-valued $2\times2$ matrix with
\begin{equation}
    \vec{\ve{u}} \cdot \vec{\ve{\sigma}} = 
    u_{1} \ve{\sigma}_1 + u_{2} \ve{\sigma}_2 + u_{3} \ve{\sigma}_3 = 
    \begin{bmatrix}
        u_\X u_\X^* \!- \!u_\Y u_\Y^* &2 u_\X u_\Y^*\\
        2u^*_\X u_\Y  &u_\Y u_\Y^*\! -\! u_\X u_\X^*
    \end{bmatrix},
    \label{eqn:pauliexpand}
\end{equation}
which will later be used to describe the \textsl{instantaneous} polarization rotation around the Stokes vector $\vec{\ve{u}}$ using the Jones formalism. 
We may also use the equality \cite[Eq.~(3.9)]{Gordon2000}
\begin{equation}
    \ve{u} \ve{u}^\H = \frac{1}{2} \left( \ve{u}^\H \ve{u} \, \mathbf{I} + \vec{\ve{u}} \cdot \vec{\ve{\sigma}} \right)
    \label{eqn:pauliproject}
\end{equation}
with the identity matrix $\mathbf{I}$ and $\norm{\ve{u}}^2 = \ve{u}^\H \ve{u} = u_\X u_\X^* + u_\Y u_\Y^*$.


\subsection{System Model}
In this work we consider \textsl{point-to-point} coherent optical transmission over two planes of polarization in a single-mode fiber.
This results in a complex-valued $2\times2$ \ac{MIMO} transmission which is typically used for multiplexing.
One of the major constraints of today's fiber-optical transmission systems is the bandwidth of electronic devices which is orders of magnitude smaller than the available bandwidth of optical fibers.
It is hence routine to use \ac{WDM}, where a number of so-called \textsl{wavelength channels} are transmitted simultaneously through the same fiber. 
Each wavelength signal is modulated on an individual laser operated at a certain wavelength (or respectively at a certain frequency) such that neighboring signals do not share the same frequency band when transmitted jointly over the same fiber medium.

\begin{figure*}[t]
    \centering
    \vspace*{-1mm}
    \psfrag{qb}[c][c]{\footnotesize $\F_2$}
    \psfrag{ab}[c][c]{\footnotesize $\mathcal{A} \subset \C^2$}
    \psfrag{ob}[c][c]{\footnotesize $\C^2$}
    \psfrag{sb}[c][c]{\footnotesize $\C^2$}
    \psfrag{ub}[c][c]{\footnotesize $\C^2$}
    \psfrag{var}[r][c]{\footnotesize $\Lsp~$}
    \psfrag{NF}[c][c]{\scriptsize $~$}
    \psfrag{DA}[c][c]{\small $\mathsf{D/A}$}
    \psfrag{EO}[c][c]{\small $\mathsf{E/O}$}
    \psfrag{OE}[c][c]{\small $\mathsf{O/E}$}
    \psfrag{AD}[c][c]{\small $\mathsf{A/D}$}
    \psfrag{OA}[c][c]{\small $\mathsf{OA}$}
    \psfrag{qk}[c][c]{\footnotesize $\mathfrak{q}[\kap]$}
    \psfrag{M}[c][c]{\footnotesize $\mathcal{M}$}
    \psfrag{ak}[c][c]{\footnotesize $\ve{a}[k]$}
    \psfrag{st}[c][c]{\footnotesize $\ve{s}_\probe(t)$}
    \psfrag{s1}[c][c]{\footnotesize $\ve{s}_\nu(t)$}
    \psfrag{a1}[c][c]{\footnotesize $\ve{b}_\nu[k]$}
    \psfrag{u0}[c][c]{\footnotesize $\ve{u}_\probe(0,t)$}
    \psfrag{um}[c][c]{\footnotesize $\ve{u}(0,t)$}
    \psfrag{u3}[c][c]{\footnotesize $\ve{u}_{\Nch-1}(0,t)$}
    \psfrag{u2}[c][c]{\footnotesize $\ve{u}_0(0,t)$}
    \psfrag{u1}[c][c]{\footnotesize $\ve{u}_\nu(0,t)$}
    \psfrag{uL}[c][c]{\footnotesize $\ve{u}(L,t)$}
    \psfrag{rt}[c][c]{\footnotesize $\ve{r}(t)$}
    \psfrag{yk}[c][c]{\footnotesize $\ve{y}[k]$}
    \psfrag{nt}[c][c]{\footnotesize $N_{\mathrm{sp}} \times$}
    \psfrag{eeu}[c][c]{\footnotesize \textbf{optical end-to-end channel}}
    \psfrag{ees}[c][c]{\footnotesize \textbf{electrical end-to-end channel}}
    \psfrag{eea}[c][c]{\footnotesize \textbf{discrete-time end-to-end channel}}
    \includegraphics[width=0.90\figwidth]{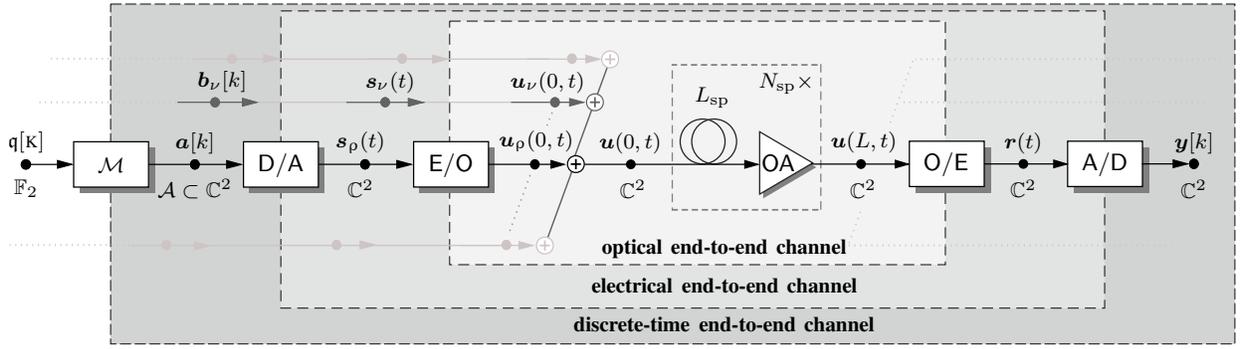}
    \caption{Generic fiber-optical transmission system model.}%
    \label{fig:generic_model} %
\end{figure*}

Fig.~\ref{fig:generic_model} shows the block diagram of a coherent optical transmission system exemplifying the digital, analog electrical, and optical domains of a single wavelength channel. 
Within the bandwidth of a wavelength channel, we can consider the \textsl{optical end-to-end} $2\times2$ \ac{MIMO} channel as frequency-flat if we neglect the effects of bandlimiting devices (e.g., switching elements in a routed network).
The nonlinear property of the fiber-optical transmission medium is the source of interference within and between different wavelength channels.
In the following, we will call the channel under consideration the \textsl{probe} channel, while a co-propagating wavelength channel is called \textsl{interferer}.
This allows us to discriminate between \ac{SCI} and \ac{XCI}.
In Fig.~\ref{fig:generic_model} the probe channel in the optical domain is denoted by a subscript $\probe$, whereas interferers are labeled by the channel index $\nu$ with $\nu \in \set{1,2,\ldots,\Nch \mid \nu \ne \probe}$. 
The various domains and its entities are discussed in the following. 

\subsubsection{Transmitter Frontend}
The transmission system is fed with equiprobable source bits of the probe (and interferer) channel.
The binary source generates uniform i.i.d.~\textsl{information bits} $\mathfrak{q}[\kap]\in \F_2$ at each discrete-time index $\kap\in \Z$.
$\F_2$ denotes the Galois field of size two and $\Z$ is the set of integers.
The binary sequence $\sequence{\mathfrak{q}[\kap]}$ is partitioned into binary tuples of length $\Rm$, s.t.~$\ve{\mathfrak{q}}[k] = [\mathfrak{q}_1[k], \ldots, \mathfrak{q}_{\Rm}[k]] \in \{0,1\}^\Rm$, where $k \in \Z$ is the discrete-time index of the \textsl{data symbols}.
Here, $\Rm$ is called the rate of the modulation and is equivalent to the number of bits per transmitted data symbol, if we neglect channel coding and assume that the size of the symbol set is a power of two.
Each $\Rm$-tuple is associated with one of the possible data symbols $\aa = [a_\X, a_\Y]^\T \in \mathcal{A} \subset \C^2$, i.e., with one of the \textsl{constellation} points.
We say that the binary $\Rm$-tuples are mapped to the data symbols $\aa \in \mathcal{A}$ by a bijective mapping rule $\mathcal{M}:\qq\mapsto\aa$.

The size of the data symbol set is $M = |\mathcal{A}| = 2^\Rm$ and we can write the alphabet as $\mathcal{A} \defeq \left\{ \ve{a}_1, \dots, \ve{a}_{M} \right\} \subset \C^2$.
The symbol set has zero mean if not stated otherwise, that is $\expval{{\aa}} = \ve{0}$, and we deliberately normalize the variance of the symbol set to $\sigma_a^2 \defeq \expval{\norm{\aa}^2} = 1$ (the expectation is denoted by $\expval{\cdot}$ and the Euclidean vector norm is $\norm{\cdot}$).
For reasons of readability we denote the data symbols of the interfering channels by $\ve{b}_\nu[k]$.

The discrete-time data symbols $\aa[k]$ are converted to the continuous-time transmit signal $\ts(t)$ by means of pulse-shaping constituting the \ac{D/A} transition, cf.~Fig.~\ref{fig:generic_model_submodel} (a).
We can express the transmit signal $\ts(t) = [s_1(t), s_2(t)]^\T \in \C^2$ as a function of the data symbols by
\begin{equation}
    \ts(t) = \Tsym \cdot \sum_{k\in \Z} \aa[k] h_\Tx(t-k\Tsym),
    \label{eqn:modulation}
\end{equation}
where $\ts(t)$ is a superposition of a time-shifted (with symbol period $\Tsym$) basic pulse $h_\Tx(t)$ weighted by the data symbols.
The pre-factor $\Tsym$ is required to preserve a dimensionless signal in the continuous-time domain (cf. \cite[P.~11]{Fischer2002}).
We assume that the transmit pulse has $\sqrt{\text{Nyquist}}$ property, i.e., $|H_\Tx(\omega)|^2$ has Nyquist property with the Fourier pair $h_\Tx(t) \trf H_\Tx(\omega)$.
To keep the following derivations tractable, all wavelength channels transmit at the same \textsl{symbol rate} $\Rs\defeq 1/\Tsym$ as the probe channel.
The pulse energy $\Eg$ of the probe channel is given by \cite[Eq.~(2.2.22)]{Fischer2002}
\begin{equation}
    \Eg = \int_{-\infty}^{\infty} |\Tsym \cdot h_\Tx(t)|^2 \dd t = \frac{1}{2\pi}\int_{-\infty}^{\infty} | \Tsym \cdot H_\Tx(\omega)|^2 \dd \omega.
    \label{eqn:pulseEnergy}
\end{equation}
The pulse energy $\Eg$ has the unit \textsl{seconds} due to the normalization of the signals.
Using the symbol energy $\Es \defeq \sigma_a^2 \Eg$, the average signal power $P$ calculates to \cite[Eq.~(4.1.1)]{Fischer2002}
\begin{equation}
    \Pow \defeq \frac{1}{\Tsym} \int_0^\Tsym \expval{\norm{\ts(t)}^2} \dd t = \frac{\sigma_a^2}{\Tsym} \Eg = \frac{\Es}{T}.
    \label{eqn:transmitPower}
\end{equation}
Since, see above, the variance of the data symbols $\sigma_a^2$ is fixed to $1$, the transmit power $P$ is directly adjusted via the pulse energy $\Eg$.
The corresponding quantities related to one of the interferers are indicated by the subscript $\nu$.

\subsubsection{Optical Channel}\label{sec:nlch}
\begin{figure}[tb]
    \centering
    \subfloat[Transmitter frontend of the \textsl{probe}]{
        \psfrag{qb}[c][c]{\footnotesize $\F_2$}
        \psfrag{ab}[c][c]{\footnotesize $\mathcal{A}$}
        \psfrag{sb}[c][c]{\footnotesize $\C^2$}
        \psfrag{ub}[c][c]{\footnotesize $\C^2$}
        \psfrag{axp}[c][c]{\footnotesize $\ee^{\jj \Delta \omega_\probe t} \equiv 1$}
        \psfrag{TX}[c][c]{\footnotesize $\Tsym \cdot H_\Tx(\omega)$}
        \psfrag{exp}[c][c]{\footnotesize $\sqrt{\Pow}$}
        \psfrag{ak}[c][c]{\footnotesize $\ve{a}[k]$}
        \psfrag{st}[c][c]{\footnotesize $\ve{s}_\probe(t)$}
        \psfrag{u0}[c][c]{\footnotesize $\ve{u}_\probe(0,t)$}
        \includegraphics[width=0.24\figwidth]{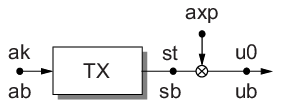}
    }
    \\
    \subfloat[Optical channel]{
        \psfrag{sb}[c][c]{\footnotesize $\C^2$}
        \psfrag{ub}[c][c]{\footnotesize $\C^2$}
        \psfrag{u0}[c][c]{\footnotesize $\ve{u}_1(0,t)$}
        \psfrag{uR}[c][c]{\footnotesize $\ve{u}_\probe(0,t)$}
        \psfrag{uS}[c][c]{\footnotesize $\ve{u}(0,t)$}
        \psfrag{uL}[c][c]{\footnotesize $\ve{u}(L,t)$}
        \psfrag{MA}[c][c]{\footnotesize \shortstack{Nonlinear\\ Optical Channel}}
        \psfrag{uN}[c][c]{\footnotesize $\ve{u}_{\Nch}(0,t)$}
        \includegraphics[width=0.34\figwidth]{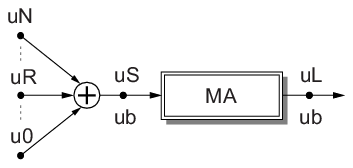}
    }
    \\
    \subfloat[Receiver frontend of the \textsl{probe} and variables (top row) associated with the regular perturbation model, cf.~Section III]{
        \psfrag{qb}[c][c]{\footnotesize $\F_2$}
        \psfrag{ab}[c][c]{\footnotesize $\C^2$}
        \psfrag{sb}[c][c]{\footnotesize $\C^2$}
        \psfrag{ub}[c][c]{\footnotesize $\C^2$}
        \psfrag{axp}[c][c]{\footnotesize $\ee^{-\jj \phi_{\rm SPM}}$}
        \psfrag{uL}[c][c]{\footnotesize $\ve{u}(L,t)$}
        \psfrag{rt}[c][c]{\footnotesize $\ve{r}(t)$}
        \psfrag{yk}[c][c]{\footnotesize $\ve{y}[k]$}
        \psfrag{dr}[c][c]{\footnotesize $\Delta\ve{s}(t)$}
        \psfrag{dy}[c][c]{\footnotesize $\Delta\ve{a}[k]$}
        \psfrag{du}[c][c]{\footnotesize $\Delta\ve{u}(L,t)$}
        \psfrag{CD}[c][c]{\footnotesize $H^*_{\Ch}(L, \omega)$}
        \psfrag{RX}[c][c]{\footnotesize $\frac{\Tsym}{\Eg}\cdot H^*_\Tx(\omega)$}
        \psfrag{RZ}[c][c]{\footnotesize $F(\zvar)$}
        \psfrag{RR}[c][c]{\footnotesize Receive Filter $H_\Rx(\omega)$}
        \psfrag{kT}[c][c]{\footnotesize $kT$}
        \includegraphics[width=0.46\figwidth]{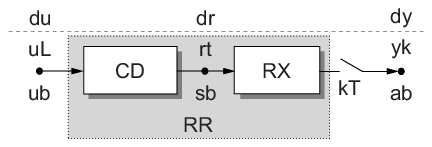}
    }
    \caption{Components of the generic fiber-optical transmission system model.}%
    \label{fig:generic_model_submodel} %
\end{figure}

The \ac{E/O} conversion is performed by an ideal \ac{DP} \ac{IQ} converter. 
The two elements of the transmit signal $\ts_\nu(t)$ correspond to the modulated \textsl{optical} signals in the $\X$- and $\Y$-polarization.
The \textsl{optical field envelope} $\uu_\nu(z,t)$ of each wavelength channel
\begin{equation}
    \uu_\nu(0,t) = \ve{s}_\nu(t) \exp(\jj \Delta \omega_\nu t),
    \label{eqn:opticalInterferer}
\end{equation}
is modulated at its angular carrier frequency $\omega_\nu = \omega_0 + \Delta \omega_\nu$ at the input of the optical transmission line $z=0$.
Here, $\omega_0 = 2\pi f_0$ is the center frequency of the signaling regime of interest.
For the probe channel, we require that the carrier frequency $\omega_\probe$ coincides with $\omega_0$ such that $\Delta \omega_\probe = 0$ and $\ve{u}_\probe(0,t) = \ve{s}_\probe(t)$. The transmitter frontend of the probe channel is shown in Fig.~\ref{fig:generic_model_submodel} (a).


The $\Nch$ wavelength signals $\uu_\nu(0,t)$ at $z=0$ are combined by an ideal optical multiplexer to a single \ac{WDM} signal, cf.~Fig.~\ref{fig:generic_model_submodel} (b). 
The optical field envelope before transmission is then
\begin{align}
    \uu(0,t) &= \sum_{\nu=1}^{\Nch} \uu_\nu(0,t) = \sum_{\nu=1}^{\Nch} \ts_\nu(t) \exp(\jj \Delta \omega_\nu t)
    \label{eqn:wdmsignal_time}\\
        &\reflectbox{\rotatebox[origin=c]{-90}{$\trf$}}\nonumber\\
    \ve{U}(0,\omega) &= \sum_{\nu=1}^{\Nch} \ve{U}_\nu(0, \omega) = \sum_{\nu=1}^{\Nch} \ve{S}_\nu(\omega-\Delta\omega_\nu), 
    \label{eqn:wdmsignal_freq}
\end{align}
with the Fourier pairs $\ts_\nu(t) \trf \ve{S}_\nu(\omega)$ and $\uu(0,t) \trf \UU(0, \omega)$.
Any initial phase and laser \ac{PN} are neglected to focus only on deterministic distortions.

The optical field envelope is the \ac{ECB} representation of the \textsl{\ul{o}ptical field} $\uu_{\rm o}(z,t)$ in the passband notation
\begin{equation}
    \ve{u}_\mathrm{o}(z,t) \defeq \uu(z,t) \cdot \exp(\jj \omega_0 t - \jj \beta_0(z) z),
    \label{eqn:optical-field}
\end{equation}
which is known as the \textsl{slowly varying amplitude approximation} \cite[Eq.~(2.4.5)]{Agrawal2010}.
For consistency of notation we treat the optical field envelope as a dimensionless entity (in accordance with the electrical signals).
The optical field propagates in $z$-direction (the dimension $z$ has units of \textsl{meter}) with the \textsl{local} propagation constant $\beta_0(z) = \beta(z,\omega_0)$, and $\beta(z,\omega) \in \R$ is the space and frequency-dependent propagation constant. 
A Taylor expansion of $\beta(z,\omega)$ is performed around $\omega_0$ with the derivatives of $\beta(z, \omega)$ represented by the coefficients \cite[Eq.~(2.4.4)]{Agrawal2010}
\begin{equation}
    \beta_n(z) \defeq  \left.\frac{\partial^n \beta(z,\omega)}{\partial \omega^n}\right|_{\omega = \omega_0}, \qquad n \in \mathbb{N} .
        \label{eqn:beta2_taylor}
\end{equation}
Here, we only consider coefficients up to second order, i.e., $n\in \set{0,1,2}$.
We also introduce the path-average\footnote{We discriminate between \textsl{local} (i.e., $\alpha(z)$, $\beta(z)$, $\gamma(z)$) and \textsl{path-average} (i.e., $\alpha$, $\beta$, $\gamma$) properties of the transmission link. The latter are implicitly indicated if the $z$-argument of the local property is omitted, e.g., ${\beta}_2 \equiv \frac{1}{L}\int_0^L \beta_2(\zeta) \dd \zeta$.} \textsl{dispersion length} 
\begin{equation}
    \Ld \defeq \frac{1}{2 \pi |{\beta}_2| \Rs^2},
    \label{eqn:dispersion_length}
\end{equation}
which denotes the distance after which two spectral components spaced $B=\Rs$ \textsl{Hertz} apart, experience a differential group delay of $T = 1/\Rs$ due to \ac{CD}.
We can equivalently define the walk-off length of the probe and one interfering wavelength channel as
\begin{equation}
    L_{\mathrm{wo}, \nu} \defeq \frac{1}{ |\Delta \omega_\nu \beta_2| \Rs},
    \label{eqn:walkoff_length}
\end{equation}
which quantifies the fiber length that must be propagated in order for the $\nu^{\rm th}$ wavelength channel to walk off by one symbol from the probe channel.

\paragraph{Signal Propagation}
In the absence of noise, the two dominating effects governing the propagation of the optical signal in the fiber are \textsl{dispersion}---expressed by the $z$-profile of the fiber dispersion coefficient $\beta_2(z)$---and \textsl{nonlinear signal-signal interactions}.
Generation of the so-termed local \ac{NLI} depends jointly on the local fiber nonlinearity coefficient $\gamma(z) \in \R_{\ge 0}$ and the $z$-profile of the optical signal power.
For ease of the derivation, we assume that all $z$-dependent variation in $\gamma(z)$ can be equivalently expressed in a variation of either a local gain $g(z) \in \R_{\ge 0}$ or the local fiber attenuation $\alpha(z) \in \R_{\ge 0}$.
We also neglect the time- (and frequency-) dependency of the attenuation, gain, and nonlinearity coefficient.

The interplay between the optical signal, dispersion, and nonlinear interaction is all combined in the noiseless Manakov equation.
It is a coupled set of partial differential equations in time-domain for the optical field envelope $\uu(z,t)$ in the \ac{ECB}, and the derivative is taken w.r.t.~propagation distance $z \in \R$ and to the \textsl{retarded time} $t\in \R$.
The retarded time is defined as $t \defeq t' - z/\vg$, where $t'$ is the \textsl{physical} time and $\vg$ is the (path-average) group velocity $\vg = 1/{\beta}_1$ of the probe channel \cite[Eq.~(2.4.8)]{Agrawal2010}.
It can be understood as a time frame that moves at the same average velocity as the probe to cancel out any group delay at the reference frequency $\omega_\probe = \omega_0$.
All other frequencies experience a residual group delay relative to the reference frequency
due to \ac{CD}.

The propagation of $\uu(z,t)$ in the signaling regime of interest is governed by \cite[Eq.~(6.26)]{Engelbrecht2015}
\begin{equation}
    \frac{\partial}{\partial z}\uu =\jj \frac{\beta_2(z)}{2} \frac{\partial^2}{\partial t^2} \uu + \frac{g(z) - \alpha(z)}{2} \uu - \jj \gamma(z) \frac{8}{9}~ \norm{\uu}^2 \uu .
    \label{eqn:nlse}
\end{equation}
The space- and time-dependency of $\uu(z,t)$ is omitted here for compact notation.
By allowing the local gain coefficient $g(z)$ to contain Dirac $\delta$-functions one can capture the $z$-dependence of an amplification scheme, i.e., based on lumped \ac{EDFA} or Raman amplification.
Polarization-dependent effects such as birefringence and \ac{PMD} are neglected limiting the following derivations to the practically relevant case of low-PMD fibers.
We also assume that all wavelength channels are co-polarized, i.e., modulated on polarization axes parallel to the ones of the probe channel.

\paragraph{Dispersion Profile}
The accumulated dispersion is a function that satisfies \cite[Eq.~(8)]{Johannisson2013}
\begin{equation}
    \frac{\dd \accD(z)}{\dd z} = \beta_2(z).
\end{equation}
Here, $\accD(z)$ can be used to express a $z$-dependency in the dispersion profile, i.e., lumped dispersion compensation by in-line dispersion compensation or simply a transmission link with distinct fiber properties across multiple spans. We obtain
\begin{equation}
	\accD(z) = \int_{0}^{z} \beta_2(\zeta) \dd \zeta + \accD_0,
    \label{eqn:acc_dispersion}
\end{equation}
where $\accD_0 \defeq \accD(0)$ is the amount of pre-dispersion (in units of squared \textsl{seconds}, typically given in $\ps^2$) at the beginning of the transmission line. 

\paragraph{Power Profile}
To describe the power evolution of $\ve{u}(z,t)$, we introduce the \textsl{normalized} power profile \( \PP(z) \) as a function that satisfies \cite[Eq.~(7)]{Johannisson2013}
\begin{equation}
    \frac{\dd \PP(z)}{\dd z} = \left(g(z) - \alpha(z)\right)\, \PP(z),
    \label{eqn:norm-power-profile}
\end{equation}
with boundary condition $\PP(0) = \PP(L) = 1$, i.e., the last optical amplifier resets the signal power to the transmit power. The $z$-dependence on $\alpha(z)$ allows for varying attenuation coefficients over different spans.
In writing (\ref{eqn:norm-power-profile}) we assumed that both the local gain coefficient and attenuation coefficient are frequency-independent.
We may also define the logarithmic gain/loss profile as
\begin{equation}
    \label{eqn:logG}
    \logG(z) \defeq \ln\left(\PP(z)\right)
    = \int_0^z \left(g(\zeta) - \alpha(\zeta)\right) \;\dd \zeta.
\end{equation}
The last expression in (\ref{eqn:logG}) is obtained by solving (\ref{eqn:norm-power-profile}) for $\PP(z) = \ee^{\logG(z)}$.
The boundray conditions on $\PP(z)$ immediately give the boundary condition $\logG(0) = \logG(L) = 0$.

The \textsl{effective length} of the \textsl{whole} transmission link is defined as
\begin{equation}
    \Leff \defeq \int_0^L \PP(\zeta) \,\dd \zeta = \int_0^L \exp(\logG(\zeta)) \,\dd \zeta,
    \label{eqn:efflength}
\end{equation}
which is the length of a fictitious lossless fiber with the same integrated power profile, i.e., with the same nonlinear impact as the whole link.

We can now define the impulse response and transfer function of the \textsl{linear} channel---that is, when the fiber nonlinearity coefficient is zero, i.e., $\gamma = 0$ in (\ref{eqn:nlse}).
To that end, we define the optical field envelope $\luu(z,t) \trf \lUU(z,\omega)$ that propagates solely according to linear effects with the boundary condition $\luu(0,t) = \uu(0,t)$ at the input of the transmission link.
The \textsl{linear} channel transfer function and impulse response is then given by
\begin{align}
    H_\Ch(z,\omega) &\defeq \exp\left(\frac{\logG(z) - \jj \omega^2\accD(z)}{2}\right)\label{eqn:lin:impulse}\\
        &\reflectbox{\rotatebox[origin=c]{-90}{$\itrf$}}\nonumber\\
    h_\Ch(z, t) &= \frac{1}{\sqrt{2 \pi}} \frac{1}{\sqrt{\jj \accD(z)}}\exp\left(\frac{\logG(z) + \jj t^2 / \accD(z)}{2}\right),
	\label{eqn:lin:transfer}
\end{align}
which represents the joint effect of chromatic dispersion and the gain/loss variation along the link.
We finally have the linear channel relation in time-domain $\luu(z,t) = h_{\Ch}(z,t)\conv \luu(0,t)$ and frequency-domain $\lUU(z, \omega) = H_\Ch(z, \omega) \lUU(0, \omega)$, which will be used in the next section in the context of the first-order perturbation method.

\subsubsection{Receiver Frontend}
Again, we assume ideal \ac{O/E} and \ac{A/D} conversion.
The received continuous-time, optical signal $\uu(L,t)$ is first matched filtered w.r.t.~the \textsl{linear} channel response and transmit pulse, and then sampled at the symbol period $\Tsym$ to obtain the discrete-time receive symbols ${\ve{y}[k]}$, cf.~Fig.~\ref{fig:generic_model_submodel}~(c).
For linear channels this establishes the optimal transition from continuous-time to discrete-time \cite{Fischer2002}.
The receiver frontend hence also compensates for any residual link loss and performs perfect \ac{CD} compensation.
Note, that the analog frontend is usually realized using an oversampled digital representation.
E.g., \ac{CD} compensation is typically performed in the (oversampled) digital domain. 
Here, we prefer to conceptually incorporate it in the analog domain\footnote{When obeying the sampling theorem, both discrete- and continuous-time representations are equivalent.} since it significantly simplifies notation in the derivation of the end-to-end channel model.
The transfer function of the entire cascade of the receiver frontend is given by 
\begin{equation}
    H_\Rx(\omega) = \frac{\Tsym}{\Eg} H_\Ch^*(L, \omega) H_\Tx^*(\omega).
    \label{eqn:h_r-cascade}
\end{equation}
Due to $\PP(L) = 1$ and the pre-factor $\Tsym/\Eg$, the received signal is re-normalized to the variance of the constellation $\sigma_a^2$.
Since we only consider $\Tsym$-spaced sampling any fractional sampling phase-offset or timing synchronization\footnote{Note, that the time delay $L/\vg$ at $\omega_0$ and any initial phase $\beta_0$ has already been canceled from the propagation equation in (\ref{eqn:nlse}).} is already incorporated as suited delay in the receive filter $h_\Rx(t)$, s.t.~the transmitted and received sequence of the probe are perfectly aligned in time.

\section{First-Order Perturbation}\label{sec:III}
The principle philosophy of fiber-optical channel models based on the perturbation method is to assume that nonlinear distortions are \textsl{weak} compared to its source, i.e., the propagating signal. 
Starting from this premise, the well-known \ac{RP} ansatz for the \textsl{optical} end-to-end channel is written as \cite[Eq.~(14)]{Vannucci2002}\cite[Eq.~(3)]{Wei2006}\cite[Eq.~(2)]{Johannisson2013}
\begin{align}
    \uu(L, t) &= \luu(L,t) + \pert{\uu}(L,t)\\
              &\reflectbox{\rotatebox[origin=c]{-90}{$\trf$}}\nonumber\\
    \ve{U}(L, \omega) &= \lUU(L,\omega) + \pert{\UU}(L,\omega),
    \label{eqn:pert_ansatz}
\end{align}
where \( \luu(z,t) \trf \lUU(z, \omega)\) is the signal propagating according to the linear effects, i.e., according to (\ref{eqn:lin:impulse}), (\ref{eqn:lin:transfer}).
In this context, the nonlinear distortion $\pert{\uu}(z,t) \trf \Delta \ve{U}(z, \omega) \in \C^2$ is the accumulated \textsl{perturbation}, which is generated locally according to nonlinear signal-signal interaction and is then propagated linearly and independently of the signal $\luu(z,t)$ to the end of the optical channel at $z=L$.
We assume that the optical perturbation at $z = 0$ is zero, i.e., $\Delta \uu(0,t) = 0$.
The received signal is then given as the sum of linearly propagating signal and the accumulated perturbation.

The objective of this section is to develop the input/output relation of the equivalent discrete-time end-to-end channel in the form of
\begin{align}
    \label{eqn:ch-dise2e-regular}
    \ve{y}[k] &= \ve{a}[k] + \Delta \ve{a}[k]\\
              &\reflectbox{\rotatebox[origin=c]{-90}{$\trf$}}\nonumber\\
    \label{eqn:ch-dise2e-regular-freq}
    \ve{Y}(\ejwt) &= \ve{A}(\ejwt) + \Delta \ve{A}(\ejwt),
\end{align}
where the total \ac{NLI} is condensed into a single \textsl{discrete-time} perturbative term $\Delta \ve{a}[k] \trf \Delta \ve{A}(\ejwt)$, cf.~Fig.~\ref{fig:generic_model_submodel} (c). To that end, we start with a known \ac{RP} solution of the optical end-to-end relation in frequency-domain and successively incorporate the required components according to Fig.~\ref{fig:generic_model} and Fig.~\ref{fig:generic_model_submodel}.
The input/output relation in periodic frequency-domain according to (\ref{eqn:ch-dise2e-regular-freq}) is one of the key original results of this paper.

\begin{figure}[t]
    \centering
    \subfloat[Time-domain]{
        \input{\figpath/var_assignment/timeAssignment_1.tex}
        \includegraphics[width=0.23\textwidth]{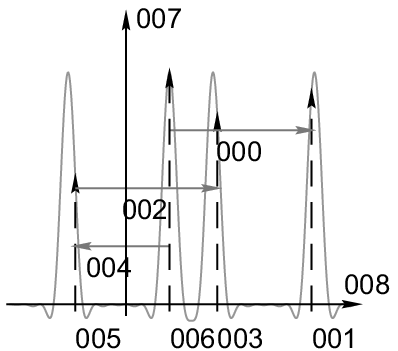}
    }
    \subfloat[Frequency-domain]{
        \input{\figpath/var_assignment/freqAssignment_1.tex}
        \includegraphics[width=0.23\textwidth]{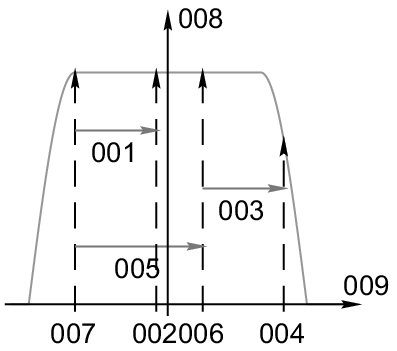}
    }
    \caption{Definitions of variables in the time- and frequency-domain. Note, that both $\tau_1$, $\tau_2$ and $\upsilon_1$, $\upsilon_2$ can take positive and negative values in $\R$.}
    \label{fig:var_assign}
\end{figure}

\subsection{Optical End-to-End Channel}
The solution to the optical perturbation after transmission at $z=L$ is given in frequency-domain by \cite[Eq.~(12)]{Vannucci2002a},\cite[Eq.~(2)]{Louchet2005},\cite[Eq.~(4)]{Wei2006},\cite[Eq.~(24)--(27)]{Liu2012a}
\begin{align}
    \pert{\UU}(L,\omega) &= -\jj \gamma \frac{8}{9} \frac{\Leff}{(2\pi)^2 } H_\Ch(L, \omega) \nonumber\\
                         &\times\int_{\R^2} \underline{\UU}(\omega, \upsilon_1, \upsilon_2)
    H_{\rm NL}(\upsilon_1, \upsilon_2)\, \dd^2 \ve{\upsilon}, \label{eqn:pertOpt2e2}
\end{align}
with the normalized \textsl{nonlinear transfer function} $H_{\rm NL}(\upsilon_1, \upsilon_2)$ and $\underline{\UU}(\omega, \upsilon_1, \upsilon_2) \defeq \UU(0, \omega + \upsilon_2) \UU^{\H}(0, \omega + \upsilon_1 + \upsilon_2) \UU(0, \omega + \upsilon_1)$, i.e., a term that depends on the optical field envelope at the \textsl{input} of the transmission system.
Note, that we made use of the common variable substitution 
\begin{align}
    \omega_1 &\defeq \omega + \upsilon_1\\
    \omega_2 &\defeq \omega + \upsilon_1 + \upsilon_2\\
    \omega_3 &\defeq \omega - \omega_1 + \omega_2 = \omega + \upsilon_2,
    \label{eqn:varDef}
\end{align}
to express the field $\ve{U}$ in terms of difference frequencies $\upsilon_1$ and $\upsilon_2$ relative to $\omega$.
Fig.~\ref{fig:var_assign} summarizes the definitions of the time and frequency variables that are used throughout this text\footnote{The integral over $\R^2$ in (\ref{eqn:pertOpt2e2}) can also be performed w.r.t.~$\omega_1$ and $\omega_2$.}.

\begin{figure}[!t]
    \centering
    \input{./figures/nonlinear_transfer_fun_2D/nonlinear_transfer_fun_2D_11.tex}
    \includegraphics{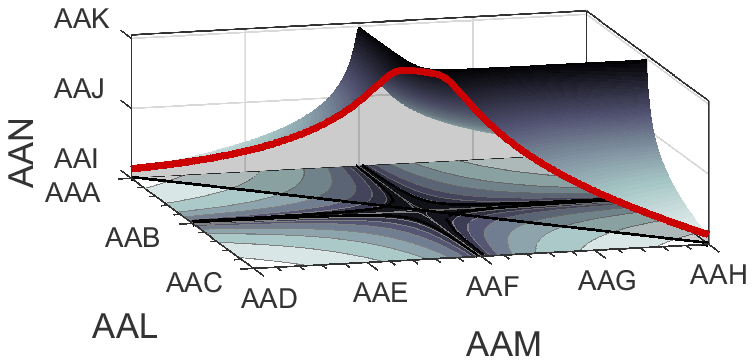}
    \caption{Magnitude in logarithmic scale of the \textsl{single-span} nonlinear transfer function for $\beta_2 = -21~\ps^2/\km$, $\accD_0 = 0~\ps^2$, $10 \log_{10}\ee^\alpha = 0.2~\dB/\km$ and $\Lsp = 100~\km$ over the \textsl{difference} frequencies $\upsilon_1$ and $\upsilon_2$ normalized to $\Rs = 64~\GBd$. The red line denotes $H_{\rm NL}(\dOm)$ which only depends on the scalar $\dOm=\upsilon_1 \upsilon_2$. (Part for $\upsilon_1 > \upsilon_2$ not shown).}
    \label{fig:nonlinear_transfer_fun_2d}
\end{figure}
\begin{figure}[!t]
    \centering
    \input{./figures/nonlinear_transfer_fun_Rs/nonlinear_transfer_fun_rs_01.tex}
    \includegraphics{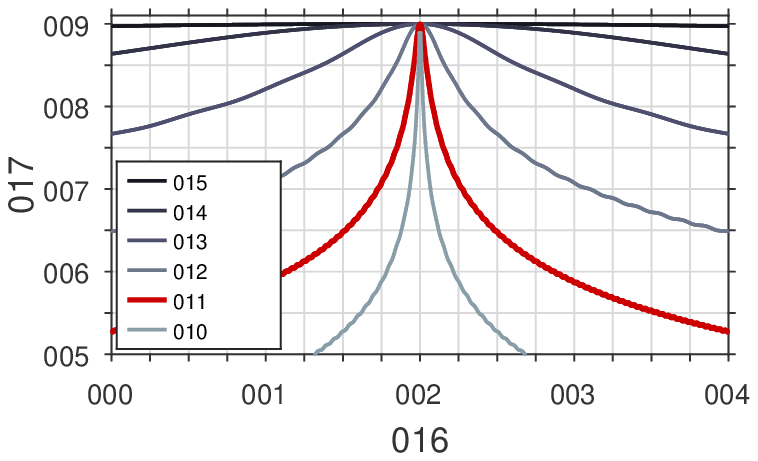}
    \caption{Magnitude in logarithmic scale of the \textsl{single-span} nonlinear transfer function for $\beta_2 = -21~\ps^2/\km$, $\accD_0 = 0~\ps^2$, $10 \log_{10}\ee^\alpha = 0.2~\dB/\km$ and $\Lsp = 100~\km$ over $\dOm=\upsilon_1\upsilon_2$. The normalization by $(2 \pi\Rs)^2$ relates $H_\NL(\dOm)$ to the probe's spectral width. The width of $|H_\NL(\dOm/\Rs^2)|^2$ is then proportional to the inverse map strength $1/\STp =\Ld/\Leff \propto \Rs^{-2}$, i.e., doubling $\Rs$ reduces the spectral width by a factor of $4$.}
    \label{fig:nonlinear_transfer_fun_Rs}
    \vspace{-0.5cm}
\end{figure}

The normalized nonlinear transfer function is a measure of the phase matching condition and defined as
\begin{align}
    \Hnl(\upsilon_1,\upsilon_2) &\defeq \frac{1}{\Leff} \int_0^L \exp\left(\logG(\zeta) + \jj \upsilon_1 \upsilon_2 \accD(\zeta)\right) \,\dd \zeta,
    \label{eqn:Hnl}
\end{align}
or equivalently in terms of \textsl{absolute frequencies} as \cite[Eq.~(10)]{Secondini2013}
\begin{align}
    \Hnl(\omega_1 - \omega&, \omega_2 - \omega_1) = \frac{1}{\Leff} \int_{0}^{L} H_\Ch(\zeta,\omega)^{-1} \nonumber\\
                                                   & \times H_\Ch(\zeta, \omega_1) H_\Ch^*(\zeta, \omega_2) H_\Ch(\zeta, \omega_3) \,\dd \zeta,
    \label{eqn:Hnl2}
\end{align}
which recovers the dependence on the \textsl{linear transfer function} evaluated at the four involved frequencies and \textsl{path-averaged} over the transmission link.
The pre-factor in (\ref{eqn:Hnl}), (\ref{eqn:Hnl2}) is the inverse of the effective length $\Leff$ and acts as a normalization constant s.t.~$H_{\rm NL}(0,0) = 1$.

The \textsl{phase mismatch} $\Delta \beta$, i.e., the difference in the (path-average) propagation constant due to dispersion, is defined as \cite[Eq.~(6.3.19)]{Agrawal2010}
\begin{align}
    \Delta \beta &\defeq \beta(\omega) - \beta(\omega_1) + \beta(\omega_2) - \beta(\omega_3)\nonumber\\
                 &=\frac{\beta_2}{2} (\omega^2-\omega_1^2+\omega_2^2 - (\omega - \omega_1 + \omega_2)^2)\nonumber\\
                 &=\beta_2 (\omega_1-\omega)(\omega_2 - \omega_1) = \beta_2 \upsilon_1 \upsilon_2,
    \label{eqn:phase_mismatch}
\end{align}
where the propagation constants at the four frequencies are developed in a second-order Taylor series according to (\ref{eqn:beta2_taylor}).
E.g., for transmission systems without inline dispersion compensation and zero pre-dispersion $\accD_0 = 0$, we have $\accD(z) = \beta_2 z$ and the phase mismatch $\Delta \beta$ can be found in the argument of the exponential in (\ref{eqn:Hnl}) with $\upsilon_1 \upsilon_2\, \accD(z) = \Delta \beta\, z$.

In the context of the equivalent ansatz following the regular \ac{VSTF} \cite{Peddanarappagari1997, Vannucci2002a, Liu2012a}, the nonlinear transfer function $\Hnl(\upsilon_1, \upsilon_2)$ is also referred to as $3^{\rm rd}$-order Volterra \textsl{kernel}.
Closed-form analytical solutions to (\ref{eqn:Hnl}) can be obtained for single-span or homogeneous multi-span systems \cite{Liu2012a,Poggiolini2012a}.

Fig.~\ref{fig:nonlinear_transfer_fun_2d} shows the magnitude of $H_{\rm NL}(\upsilon_1, \upsilon_2)$ (in logarithmic scale) exemplifying a single-span \ac{SSMF} link.
Note, that $H_{\rm NL}(\upsilon_1, \upsilon_2)$ depends in fact on the product $\dOm \defeq \upsilon_1 \upsilon_2$ and is hence a hyperbolic function in two dimensions \cite[Sec. VIII]{Poggiolini2012} (cf. the contour in Fig.~\ref{fig:nonlinear_transfer_fun_2d}).
The bold red line drawn into the diagonal cross section in Fig.~\ref{fig:nonlinear_transfer_fun_2d} is the corresponding nonlinear transfer function $H_{\rm NL}(\dOm)$ which only depends on the scalar variable $\dOm = \upsilon_1 \upsilon_2$.

Fig.~\ref{fig:nonlinear_transfer_fun_Rs} shows the magnitude of $H_{\rm NL}(\dOm)$ (in logarithmic scale) over the normalized variable $\dOm/(2\pi\Rs)^2$ to relate the nonlinear transfer function to the spectral width of the probe channel.
The spectral width of $|H_{\rm NL}(\dOm/(2 \pi \Rs)^2)|^2$ is proportional to the inverse dimensionless \textsl{map strength} $1/\STp \defeq \Ld /\Leff$, and to the \textsl{nonlinear diffusion bandwidth} defined in \cite{Louchet2005}. 
Conversely, the map strength $\STp$ quantifies the number of nonlinearly interacting pulses in time over the effective length $\Leff$ within the probe channel \cite{Bononi2008}. It is therefore a direct measure of \textsl{intra-channel} (i.e., \ac{SCI}) nonlinear effects \cite{Fischer2009}. 
The relevant quantity for \textsl{inter-channel} (i.e., \ac{XCI}) effects is given by $\smap_{\Tsym, \nu} \defeq \Leff / L_{\mathrm{wo}, \nu}$ (with $\nu \ne \probe$) where the temporal walk-off between wavelength channels is the relevant length scale.
In \cite{Wei2006} it was shown that $H_{\rm NL}(\dOm)$ is related to the \ac{PWDD} by a (one-dimensional) Fourier transformation (w.r.t.~the scalar variable $\dOm$) and has a time-domain counterpart which is discussed in the next paragraph.

\subsection{Electrical End-to-End Channel}
To derive the discrete-time end-to-end channel model the filter cascade of the linear receiver frontend is subsequently applied to $\Delta \ve{U}(L, \omega)$.
The perturbation $\Delta \ve{S}(\omega)$ (i.e., the perturbation in the analog \textsl{electrical} domain following our terminology, cf.~Fig.~\ref{fig:generic_model_submodel} (c)) is obtained by 
\begin{equation}
    \Delta \ve{S}(\omega) = H^*_\Ch(L, \omega) \Delta \ve{U}(L, \omega),
    \label{eqn:el_pert}
\end{equation}
which cancels out the leading term $H_\Ch(L, \omega)$ in (\ref{eqn:pertOpt2e2}) since $|H_\Ch(L, \omega)| = 1$. The result is shown in (\ref{eqn:ele2ebottompage_freq}) at the bottom of the page.
\begin{figure*}[!b]
    \vspace{-0.2cm}
    \noindent\rule{\textwidth}{0.25pt}
    \begin{align}
        \pert{\bS}(\omega) &= -\jj \gamma \frac{8}{9} \Leff \frac{1}{(2\pi)^2}\int_{\R^2}\UU(0,\underbrace{\omega+\upsilon_2}_{\mathclap{\omega_3 = \omega - \omega_1 + \omega_2}}) \UU^{\H}(0,\underbrace{\omega+\upsilon_1+\upsilon_2}_{\omega_2}) \UU(0,\underbrace{\omega+\upsilon_1}_{\omega_1}) H_{\rm NL}\tikzmark{c}(\upsilon_1, \upsilon_2) \;\dd^2\ve{\upsilon}
        \label{eqn:ele2ebottompage_freq}
        \\
        &\reflectbox{\rotatebox[origin=c]{-90}{$\itrf$}}\nonumber\\
        \pert{\bs}(t) &= -\jj \gamma \frac{8}{9} \Leff \int_{\R^2} \uu(0,\underbrace{t+\tau_1}_{t_1}) \uu^{\H}(0,\underbrace{t+\tau_1 +\tau_2}_{t_2}) \uu(0,\underbrace{t+\tau_2}_{\mathclap{t_3 = t - t_1 + t_2}}) h_{\rm NL}\tikzmark{d}(\tau_1,\tau_2)\; \dd^2 \ve{\tau}
        \label{eqn:ele2ebottompage_time}
    \end{align}
    \tikz[remember picture]
    \draw[overlay,<->] ([shift={(-0.3cm,-0.2cm)}]pic cs:c)to  [bend left] node[right]{$\mathcal{F}_{\ve{\tau} \leftrightarrow \ve{\upsilon}}\{\;\cdot\;\}$} ([shift={(0.2cm,0.4cm)}]pic cs:d);
\end{figure*}

Remarkably, there exists an equivalent time-domain representation $\Delta \ve{s}(t) \trf \Delta \ve{S}(\omega)$ shown in (\ref{eqn:ele2ebottompage_time}) where the Fourier relation is derived in Appendix \ref{apx:proof1}.
The time-domain perturbation $\Delta \ve{s}(t)$ has the same form as its frequency-domain counterpart, i.e., the integrand is constituted by the respective time-domain representation of the optical signal and the double integral is performed over the time variables $\tau_1$ and $\tau_2$ (cf.~Fig.~\ref{fig:var_assign} (a) and \cite{Ablowitz2002d, Wei2006}).

The frequency matching with $\omega_3 \defeq \omega - \omega_1 + \omega_2$ is translated to a \textsl{temporal matching}\footnote{Not to be confused with the phase matching condition in (\ref{eqn:Hnl}), (\ref{eqn:phase_mismatch}).} $t_3 \defeq t - t_1 + t_2$ (cf.~\cite{Ablowitz2000}), i.e., the selection rules of \ac{FWM} apply both in time and frequency.
Remarkably, the time-domain \textsl{kernel} $h_{\rm NL}(\tau_1, \tau_2)$ is related to $\Hnl(\upsilon_1, \upsilon_2)$ by an inverse \ac{2D} Fourier transform (cf.~\cite[Appx.]{Ablowitz2002d} and \cite[Eq.~(6)]{Bononi2008}) which can be written as
\begin{align}
    h_{\rm NL}(\tau_1,\tau_2) &= h_{\rm NL}(\ve{\tau}) = \mathcal{F}^{-1} \{H_{\rm NL}(\ve{\upsilon})\}
    \label{eqn:hNL}
    \\
    &= \frac{1}{\Leff} \int_0^{L} \frac{1}{2 \pi|\accD(\zeta)|} \exp\left(\logG(\zeta) - \jj \frac{\tau_1 \tau_2}{\accD(\zeta)} \right) \dd \zeta \nonumber,
\end{align}
with the tuples $\ve{\tau} = [\tau_1, \tau_2]^\T$ and $\ve{\upsilon} = [\upsilon_1, \upsilon_2]^\T$. The time-domain kernel maintains its hyperbolic form as it is a function of the product $\tau_1 \tau_2$.
Again, we can express $h_\NL(\tau_1, \tau_2)$ also in terms of \textsl{absolute time variables} as
\begin{align}
    h_{\rm NL}(t_1 - t,\,&t_2 - t_1) = \frac{1}{\Leff} \int_0^{L} h_\Ch(\zeta,t)^{-1} \nonumber\\
                                   &\times h_\Ch(\zeta,t_1) h_\Ch^*(\zeta,t_2) h_\Ch(\zeta,t_3) \,\dd \zeta.\label{eqn:hNLabstime}
\end{align}
Also note the duality to (\ref{eqn:Hnl2}), where in both representations the nonlinear transfer function can be understood as the \textsl{path-average} (cf. \cite{Gabitov1996}) over an expression related to the linear channel response $h_\Ch(z, t) \trf H_\Ch(z, \omega)$.

The next step is to dissect the perturbation $\Delta\ve{s}(t) \trf \Delta\ve{S}(\omega)$ into contributions originating from \ac{SCI}, \ac{XCI}, or \ac{MCI}.
We notice from Fig.~\ref{fig:nonlinear_transfer_fun_Rs} that, given $\Rs$ is sufficiently large, $|H_\NL(\dOm)|^2$ vanishes if $\dOm \gg (2 \pi \Rs)^2$, i.e., when the \textsl{phase matching condition} is not properly met.
Conversely, if the spectral width of $|H_{\rm NL}(\dOm/\Rs^2)|^2$ (or equivalently the inverse map strength $1/\STp$) is small enough, the integrand in (\ref{eqn:ele2ebottompage_freq}), (\ref{eqn:ele2ebottompage_time}) can be factored into a \ac{SCI} and \ac{XCI} term, i.e., mixing terms that originate either from within the probe channel (both $\upsilon_1<2 \pi \Rs$ and $\upsilon_2 < 2 \pi \Rs$) or from within the probe channel and a single interfering wavelength channel (either $\upsilon_1 <2 \pi \Rs$ or $\upsilon_2 < 2 \pi \Rs$).
Mixing terms originating from \ac{MCI} are only relevant for small $\Rs$.
We hence neglect any \ac{FWM} terms involving more than two wavelength channels.

The optical field envelope $\ve{u}(0,t) \trf \ve{U}(0, \omega)$ in (\ref{eqn:ele2ebottompage_freq}), (\ref{eqn:ele2ebottompage_time}) is now expanded according to (\ref{eqn:wdmsignal_time}), (\ref{eqn:wdmsignal_freq}).
By definition we have $\Delta \omega_\probe = 0$ and we can expand the triple product of $\ve{U}(0,\omega)$ in (\ref{eqn:ele2ebottompage_freq}) as
\begin{equation}
    \UU\UU^{\H}\UU = \underbrace{\UU_\probe \UU_\probe^{\H} \UU_\probe}_{\rm SCI} + \sum_{\nu \ne \probe} \underbrace{\left( \UU_\nu \UU_\nu^{\H} \UU_\probe + \UU_\probe \UU_\nu^{\H} \UU_\nu  \right) }_{\rm XCI}
    \label{eqn:opticalExpand}
\end{equation}
where the frequency-dependency of $\ve{U}(0,\omega)$ is omitted for short notation.
The \ac{XCI} term has two contributions---the first results from an interaction where $\omega_3$ and $\omega_2$ are from the $\nu^{\rm th}$ interfering wavelength channel and $\omega$ and $\omega_1$ are within the probe's support ($\upsilon_2 \to \Delta \omega_\nu$ in Fig.~\ref{fig:var_assign} (b)). The second involves an interaction where $\omega_2$ and $\omega_1$ are from the interfering wavelength channel and $\omega$ and $\omega_3$ are from the probe channel ($\upsilon_1 \to \Delta \omega_\nu$).

We can exploit the symmetry of the nonlinear transfer function $H_\NL(\upsilon_1, \upsilon_2) = H_\NL(\upsilon_2, \upsilon_1)$ to simplify the XCI expression in (\ref{eqn:opticalExpand}). 
We obtain with the definition of the electrical signal of each wavelength channel (cf.~(\ref{eqn:wdmsignal_time}), (\ref{eqn:wdmsignal_freq})) after rearranging some terms\footnote{Since $\UU_\nu^\H \UU_\nu$ is a scalar, we have $\UU_\probe \UU_\nu^\H \UU_\nu = \UU_\nu^\H \UU_\nu \UU_\probe$. The $2\!\times\!2$ identity matrix $\mathbf{I}$ is required to factor the XCI expression in a $\nu$- and $\probe$-dependent term.}
\begin{align}
    \UU(&0,\, \omega_3)\UU^{\H}(0, \omega_2)\UU(0, \omega_1)  H_\NL(\omega_2 - \omega_3, \omega_2 - \omega_1)\nonumber\\
    &=\ve{S}_\probe(\omega_1) \ve{S}_\probe^{\H}(\omega_2) \ve{S}_\probe(\omega_3) H_\NL(\omega_2 - \omega_1, \omega_2 - \omega_3)\nonumber \\
                         &\quad+ \sum_{\nu \ne \probe} \left( \ve{S}_\nu(\omega_1) \ve{S}_\nu^{\H}(\omega_2) + \ve{S}_\nu^{\H}(\omega_2) \ve{S}_\nu(\omega_1) \mathbf{I} \right) \ve{S}_\probe(\omega_3) \nonumber\\
                         &\quad\quad \times H_\NL(\underbrace{\omega_2 - \omega_1}_{\upsilon_2}, \underbrace{\omega_2 - \omega_3}_{\upsilon_1} - \Delta \omega_\nu),
    \label{eqn:opticalExpand_reorder}
\end{align}
which now corresponds to the case that $\omega_3$ always lays in the support of the probe%
\footnote{An alternative formulation with $\omega_1$ in the support of the probe is obtained by exchanging the subscripts of $\omega_1$ and $\omega_3$ in frequency-domain and $t_1$ and $t_3$ in time-domain.}%
.
The signals of the interfering wavelength channels are now represented in their respective \ac{ECB} and the relative frequency offset $\Delta \omega_\nu$ is accounted for via the modified argument of $H_\NL(\cdot, \cdot)$.


At this point, considering (\ref{eqn:ele2ebottompage_freq}) and (\ref{eqn:opticalExpand_reorder}), we formulated the relation between the perturbation at the probe $\Delta \ve{S}(\omega)$ after chromatic dispersion compensation and the transmit spectra $\ve{S}_\nu(\omega)$ of the probe and the interferers in their respective baseband.
The remaining operation in the receiver cascade is to perform matched filtering w.r.t.~the transmit pulse and then to perform $\Tsym$-spaced sampling.

\begin{figure*}[!b]
    \vspace{-0.2cm}
    \noindent\rule{\textwidth}{0.25pt}
    \begin{align}
        \pert{\ve{A}}^{\superSCI}(\ee^{\jj \omega \Tsym}) &= \frac{\Tsym}{\Eg} \alias{\Delta \ve{S}^{\superSCI}(\omega) \cdot H_\Tx^*(\omega)} =  -\jj \underbrace{\frac{8}{9} \frac{\Leff}{L_{\mathrm{NL}, \probe}}}_{\phinlp} \frac{\Tsym^2}{(2\pi)^2} \! \int_{\TT^2}\ve{A}(\ee^{\jj \omega_1 \Tsym}) \ve{A}^{\H}(\ee^{\jj \omega_2 \Tsym}) \ve{A}(\underbrace{\ee^{\jj \omega_3 \Tsym}}_{\mathclap{\omega_3 = \fold{\omega - \omega_1 + \omega_2} \; \in \, \TT ~ ~ ~ ~ ~ ~ ~ ~ ~ ~ ~ ~ ~ ~ ~  }}) H_{\probe}\tikzmark{a}(\ee^{\jj \ve{\omega} \Tsym}) \,\dd^2\ve{\omega}\label{eqn:discretee2ebottompage_freq}
        \\
        &\;\reflectbox{\rotatebox[origin=c]{90}{$\trf$}}\nonumber\\
        \Delta \ve{a}^{\superSCI}[k] &=\frac{\Tsym}{\Eg} \Delta \bs^{\superSCI}(t) \conv h^*_\Tx(-t) \Big|_{t =k\Tsym}
        = -\jj \frac{8}{9} \frac{\Leff}{L_{\mathrm{NL}, \probe}} \sum_{\ve{\kappa}\in \Z^3}\ve{a}[k+\kappa_1] \ve{a}^\H[k+\kappa_2] \ve{a}[k+\kappa_3] h_\probe\tikzmark{b}[\ve{\kappa}]
        \label{eqn:discretee2ebottompage_time}
    \end{align}
    \tikz[remember picture]
    \draw[overlay,<->] ([shift={(0cm,-0.2cm)}]pic cs:a)to  [bend left] node[right]{$\hat{\mathcal{F}}_{\ve{\kappa} \leftrightarrow \ve{\omega}}\{\;\cdot\;\}$} ([shift={(0cm,0.4cm)}]pic cs:b);
\end{figure*}
\subsection{Discrete-Time End-to-End Channel}
We recap that the periodic spectrum $X(\ee^{\jj \omega \Tsym})$ of the sampled signal $x[k] \defeq x(k\Tsym)$ is related to the \textsl{aliased} spectrum of the continuous-time signal $x(t)$ over the Nyquist interval $\TT$ by
\begin{equation}
    X(\ee^{\jj \omega \Tsym}) \defeq \alias{X(\omega)} = \frac{1}{\Tsym} \sum_{m \in \Z} X(\omega - \frac{2 \pi m}{\Tsym}).
    \label{eqn:specFolding}
\end{equation}
The matched filter $H_\Tx^*(\omega)$ and the \textsl{aliasing operator} are used to translate (\ref{eqn:ele2ebottompage_freq}), (\ref{eqn:ele2ebottompage_time}) to the equivalent discrete-time form in (\ref{eqn:discretee2ebottompage_freq}), (\ref{eqn:discretee2ebottompage_time}), shown at the bottom of the page, again exemplarily for the \ac{SCI} contribution $\Delta \ve{a}^{\superSCI}$.
The total perturbation inflicted on the probe channel is $\Delta \ve{a}[k] = \Delta \ve{a}^{\superSCI}[k] + \Delta \ve{a}^{\superXCI} [k]$.

\begin{figure*}[t]
    \centering
    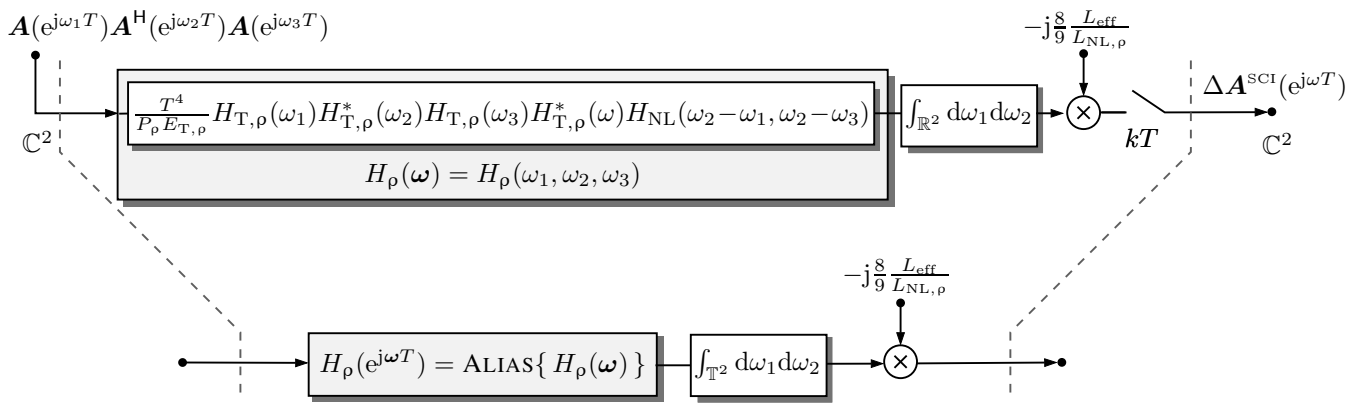
    \caption{A block diagram representation of the frequency-domain continuous-time (single-channel, i.e., $\nu = \probe$) perturbation model (top), and the deduced discrete-time end-to-end equivalent where $T$-spaced sampling is included via the \textsl{aliased} kernel representation and integration bounds that coincide with the Nyquist interval (bottom).}%
    \label{fig:end_to_end_pert_model} %
\end{figure*}
\begin{figure*}[t]
    \centering
    \input{./figures/nonlinear_transfer_fun_alias/nonlinear_transfer_aliased_kern.tex}
    \psfrag{AAI}[cl][cl]{\footnotesize $10 \log_{10}(|H_\probe(\ve{\omega})|^2)\;\; [\dB]$}%
    \includegraphics[height=0.32\figwidth]{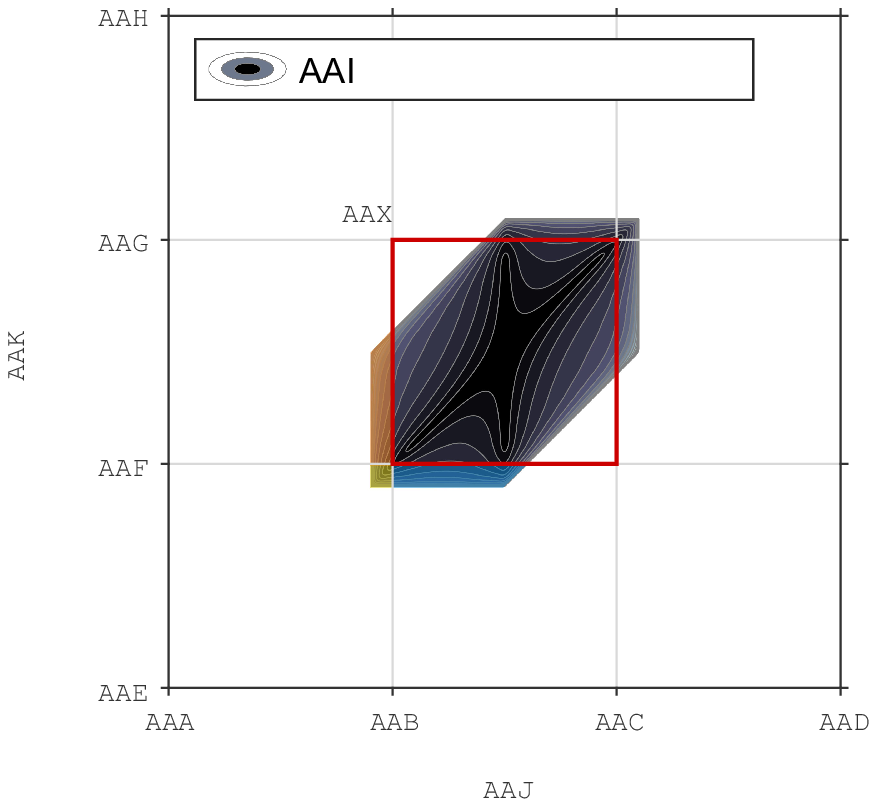}
    \hspace*{0.8cm}
    \psfrag{AAI}[cl][cl]{\footnotesize $10 \log_{10}(|H_\probe(\ejwwt)|^2)\;\; [\dB]$}%
    \includegraphics[height=0.32\figwidth]{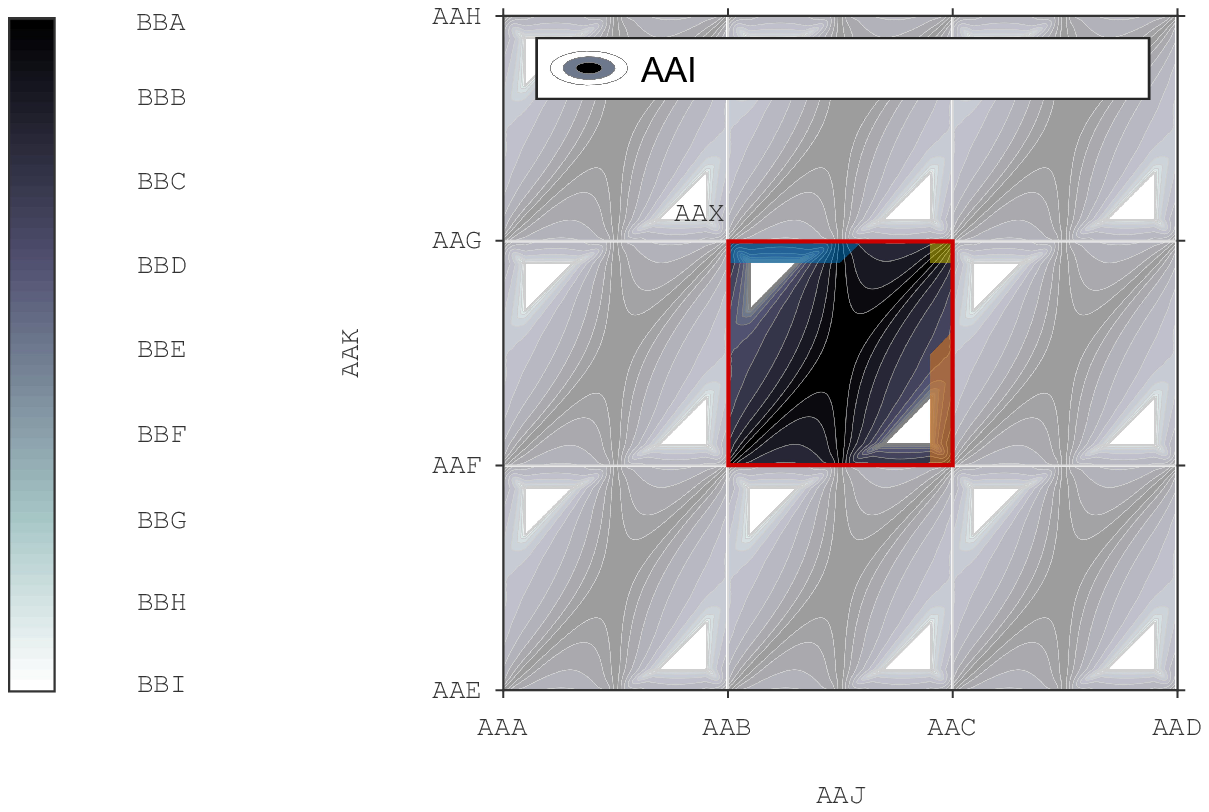}
    \caption{Contour plot of a 2D cut ($\omega_3 = 0$) of the 3D (single-channel, i.e., $\nu = \probe$) frequency-domain kernel $H_\probe(\ve{\omega})$ (left). The parameters are the same as in Fig.~\ref{fig:nonlinear_transfer_fun_2d}, and the basic pulse has RRC shape with roll-off factor $\rho = 0.2$. The kernel exhibits the well-known polygon-shape, compare e.g.~with \cite[Fig.~4]{Carena2014} or \cite[Fig.~2]{Johannisson2014}. The projection of the Nyquist cube $\TT^3$ into two dimensions is highlighted by the red boundaries. The discrete-time end-to-end nonlinear transfer function $H_\probe(\ejwwt)$ (right) is obtained by aliasing the kernel $H_\probe(\ve{\omega})$ into the Nyquist cube over all three dimensions $[\omega_1, \omega_2, \omega_3]^\T$. The three color-coded regions indicate where the spectral components outside the Nyquist region appear after the aliasing operation. As the notation implies, $H_\probe(\ejwwt)$ has a $1/T$-periodic structure into all three dimensions---shown here by the transparent continuations in the $\omega_1$-$\omega_2$-plane.}%
    \label{fig:end_to_end_pert_alias} %
\end{figure*}

In (\ref{eqn:discretee2ebottompage_freq}), (\ref{eqn:discretee2ebottompage_time}) we use the $1/\Tsym$-periodic spectrum $\ve{A}(\ee^{\jj \omega \Tsym})$ which is related to the discrete-time sequence $\sequence{\ve{a}[k]}$ by a \ac{DTFT} $\ve{A}(\ee^{\jj \omega \Tsym}) = \dtft{\ve{a}[k]}$.
The channel-dependent \textsl{nonlinear length} is $L_{\mathrm{NL},\nu} \defeq 1/(\gamma \Pow_\nu)$ and $\Pow_\nu = \frac{\sigma_{b,\nu}^2}{T} E_{\Tx,\nu}$ is the optical launch power of the $\nu^{\rm th}$ wavelength channel.

The normalized \textsl{nonlinear} end-to-end transfer function $H_\nu(\ve{\omega}) = H_\nu(\omega_1, \omega_2, \omega_3)$ characterizes the nonlinear cross-talk from the $\nu^{\rm th}$ wavelength channel to the probe channel. In particular, $H_\probe(\ve{\omega})$ describes \ac{SCI} and $H_\nu(\ve{\omega})$ with $\nu \ne \probe$ describes \ac{XCI}.
It is defined as \cite[Eq.~(12)]{Dar2015a}\cite[Eq.~(14)]{Dar2017}
\begin{align}
    H_\nu(\ve{\omega}) \defeq 
    &\,\Tsym \cdot H_{\Tx, \nu}(\omega_1) \; \Tsym \cdot H_{\Tx, \nu}^{*}(\omega_2) \;/\,\Pow_\nu\nonumber\\
    \times&\,\Tsym \cdot H_{\Tx, \probe}(\omega_3) \; \Tsym \cdot H^*_{\Tx, \probe}(\omega_1-\omega_2+\omega_3) \;/ \,E_{\Tx,\probe} \nonumber\\
    \times &\,H_{\rm NL}(\omega_2 -\omega_1,\omega_2-\omega_3 - \Delta \omega_\nu).
    \label{eqn:H_dt_e2e}
\end{align}
A block diagram of the end-to-end relation is shown at the top of Fig.~\ref{fig:end_to_end_pert_model}.
Here, the nonlinear transfer function $H_\probe(\ve{\omega})$ relates the periodic spectrum of the transmit \textsl{symbol} sequence to the received signal \textsl{before} sampling.
The nonlinear end-to-end transfer function in (\ref{eqn:H_dt_e2e}) depends on the characteristics of the transmission link, comprised by $H_{\rm NL}(\cdot, \cdot)$, the characteristics of the pulse-shapes of the probe and interfering wavelength channel (assuming the matched filter receiver frontend) and the frequency offset $\Delta \omega_\nu$ between probe and interferer.

We now take the sampling operation with period $T$ into account by considering the \textsl{periodic continuation}, i.e., the \textsl{aliased} discrete-time equivalent of $H_\nu(\ve{\omega})$. It is given by
\begin{equation}
    H_\nu(\ee^{\jj \ve{\omega} \Tsym}) = \frac{1}{\Tsym^3} \sum_{\ve{m}\in \Z^3} H_\nu(\ve{\omega} - \frac{2 \pi \ve{m}}{\Tsym}),
    \label{eqn:H_dt_e2e_folded}
\end{equation}
where the three-fold aliasing is done along each frequency dimension with $\ve{\omega} = [\omega_1, \omega_2, \omega_3]^\T$ and $\ve{m} = [m_1, m_2, m_3]^\T \in \Z^3$.
The normalization\footnote{Note, that by definition the optical launch power $P_\nu$ of the $\nu^{\rm th}$ wavelength channel is related to the pulse energy of $H_{\Tx, \nu}(\omega)$ in (\ref{eqn:pulseEnergy}), (\ref{eqn:transmitPower}).} in (\ref{eqn:H_dt_e2e}) is done s.t. $H_\probe(\ee^{\jj \ve{0} \Tsym}) = 1$ and dimensionless.
Fig.~\ref{fig:end_to_end_pert_model} (bottom) shows the deduced discrete-time end-to-end model%
\footnote{This representation is in analogy with linear, dispersive channels \cite[Fig.~2.3]{Fischer2002}.}
which now includes $T$-spaced sampling via the \textsl{aliased} frequency-domain kernel $H_\probe(\ejwwt)$.
The integration is now performed over the Nyquist interval $\TT^2$ instead of $\R^2$.
The difference to prior continuous-time models is subtle but fundamental, and one of the central results of this work.

In Fig.~\ref{fig:end_to_end_pert_alias}, the contour plot of a 2D cut from $H_\probe(\ve{\omega})$, i.e., before aliasing, and the corresponding 2D cut from $H_\probe(\ejwwt)$ after aliasing to the Nyquist interval is shown.
The kernel $H_\probe(\ve{\omega})$ exhibits the well-known polygon-shape of the \ac{SCI} contribution, compare, e.g., with the integration islands displayed in \cite[Fig.~4]{Carena2014} and \cite[Fig.~2]{Johannisson2014}.
The aliased kernel is obtained by the operation defined in (\ref{eqn:H_dt_e2e_folded}) and can be visually understood as \textsl{folding} the frequency components from outside the Nyquist interval into the Nyquist interval (indicated by the red square in Fig.~\ref{fig:end_to_end_pert_alias}).
The folded spectrum $H_\probe(\ejwwt)$ has a $1/T$-periodic structure in all three dimensions $[\omega_1, \omega_2, \omega_3]^\T$, indicated in Fig.~\ref{fig:end_to_end_pert_alias} by the transparent continuations.
Here, only the cut $\omega_3 = 0$ is shown, but the periodicity also extents to the third dimension $\omega_3$ which is not displayed.

It is remarkable that the integration in (\ref{eqn:discretee2ebottompage_freq}) is over the two-fold tuple $[\omega_1, \omega_2] ^\T \in \TT^2$ while the time-domain summation in (\ref{eqn:discretee2ebottompage_time}) is over three independent variables $\ve{\kappa} = [\kappa_1, \kappa_2, \kappa_3]^\T \in \Z^3$.
This is a consequence of the time-frequency relation between convolution and element-wise multiplication.
The temporal matching required for the \textsl{optical field} in (\ref{eqn:ele2ebottompage_time}) is now canceled in (\ref{eqn:discretee2ebottompage_time}) due to the convolution with the matched filter $h_\Tx^*(-t)$, i.e., $\kappa_3$  does not depend on $\kappa_1$ and $\kappa_2$ unlike $t_3 \defeq t - t_1 + t_2$.
Note, that the frequency variable $\omega_3$ in (\ref{eqn:discretee2ebottompage_freq}) still complies with the frequency matching $\omega_3 = \omega - \omega_1 + \omega_2$ which may be outside the Nyquist interval $\TT$. 
Due to the $1/T$-periodicity of the spectrum $\ve{A}(\ejwt)$ any frequency component outside $\TT$ is effectively \textsl{folded} back into the Nyquist interval by addition of integer multiples of $\omega_{\rm Nyq}$ (denoted by the $\fold{\cdot}$ operation in (\ref{eqn:discretee2ebottompage_freq})).

The \ac{XCI} complement to (\ref{eqn:discretee2ebottompage_freq}) reads
\begin{align}
    \Delta &\ve{A}^{\superXCI}(\ee^{\jj \omega \Tsym}) = - \jj \sum_{\nu \ne \probe} \frac{8}{9} \frac{\Leff}{L_{\mathrm{NL},\nu}} \frac{\Tsym^2}{(2\pi)^2} \int_{\TT^2}  \nonumber\\
                                                      &\times \Big(\ve{B}_\nu(\ee^{\jj \omega_1 \Tsym}) \ve{B}_\nu^\H(\ee^{\jj \omega_2 \Tsym})+ \ve{B}_\nu^{\H}(\ee^{\jj \omega_2 \Tsym}) \ve{B}_\nu(\ee^{\jj \omega_1 \Tsym}) \mathbf{I} \Big) \nonumber\\
                                                      &\times\ve{A}(\ee^{\jj \omega_3 \Tsym}) H_\nu(\ee^{\jj \ve{\omega} \Tsym}) \; \dd^2 \ve{\omega}.
        \label{eqn:model_freq_xci}
\end{align}
The time-domain description of the $\Tsym$-spaced channel model in (\ref{eqn:discretee2ebottompage_time}) is equivalent to the \textsl{pulse-collision picture} (cf.~\cite[Eq.~(3-4)]{Dar2015b} and \cite[Eq.~(3-4)]{Dar2016c}) and the \ac{XCI} result is repeated here for completeness
\begin{align}
    \Delta \ve{a}^{\superXCI}[k] &= - \jj \sum_{\nu \ne \probe} \frac{8}{9} \frac{\Leff}{L_{\mathrm{NL},\nu}} \sum_{\ve{\kappa}\in \Z^3}  \Big(\ve{b}_\nu[k+\kappa_1] \ve{b}_\nu^\H[k+\kappa_2] \nonumber\\
                                 &+ \ve{b}_\nu^{\H}[k+\kappa_2] \ve{b}_\nu[k+\kappa_1] \mathbf{I} \Big) \ve{a}[k+\kappa_3] h_\nu[\ve{\kappa}].
        \label{eqn:pulse_collision_xci}
\end{align}
The time-domain and \textsl{aliased} frequency-domain kernel are related by a \ac{3D} \ac{DTFT} according to
\begin{equation}
    h_\nu[\ve{\kappa}] = \idtft{H_\nu(\ee^{\jj \ve{\omega} \Tsym})}.
    \label{eqn:h_dt_e2e}
\end{equation}
The kernel $h_\nu[\ve{\kappa}] = h_\nu[\kappa_1, \kappa_2, \kappa_3]$ is equivalent to the kernel derived in the seminal paper by A.~Mecozzi and R.-J.~Essiambre in \cite[Eq.~(61), (62)]{Mecozzi2012a} and later used in \cite[Eq.~(6), (7)]{Dar2013} via an alternative approach that instead involves an integration over time and space\footnote{The spatial integration in \cite{Mecozzi2012a} is similar to (\ref{eqn:hNLabstime}) where the power profile (i.e., $f(z)$ in the original source) is already included in $h_\Ch(z,t)$.}.

\subsection{Relation to the GN-Model and System Design Rules}\label{sec:gn}
Parseval's theorem applied to (\ref{eqn:h_dt_e2e}) yields
\begin{equation}
E_{h,\nu} \defeq \sum_{\ve{\kappa} \in \Z^3} |h_\nu[\ve{\kappa}]|^2 = \left(\frac{T}{2 \pi}\right)^3 \int_{\TT^3} |H_\nu(\ee^{\jj \ve{\omega} \Tsym})|^2 \;\dd^3 \ve{\omega},
    \label{eqn:parseval}
\end{equation}
where the right-hand side can be interpreted as an alternative formulation of the (frequency-domain) \ac{GN}-model%
\footnote{Compared to the GN/EGN-model, the common pre-factor $(\frac{8}{9} \frac{\Leff}{L_{\NL, \nu}})^2$ in (\ref{eqn:parseval}) is omitted, and the energy in time- and frequency-domain is calculated over the whole Nyquist interval of the probe channel, whereas the GN/EGN-model \cite[Eq.~(1)]{Poggiolini2012} is often only evaluated at the probe's center-frequency. Beyond that, to include all \ac{SCI} and \ac{XCI} contributions in (\ref{eqn:parseval}) one needs to sum over all $\nu$---the GN/EGN-model in its standard form also includes \ac{MCI}.}%
~\cite{Poggiolini2012}.
In contrast to (\ref{eqn:parseval}), the GN-model essentially performs the integration over $|H_\nu(\ve{\omega})|^2$, i.e., the \textsl{unaliased} kernel, and the integration bounds correspond to the \textsl{bandwidth}\footnote{Another common approximation of the GN/EGN-model is to limit the integration bounds of $\omega = \omega_3 + \omega_1 - \omega_2$ to the Nyquist region whereas the inner integrals over $\omega_1$ and $\omega_2$ are still in $\R$, cf. \cite[Eq.~(2-3)]{Poggiolini2017} and \cite[Sec.~V]{Poggiolini2012a}. This approximation is, however, only valid if no aliasing occurs, i.e., for zero roll-off factor. Then, the analog and $T$-spaced formulation are equivalent.} of the wavelength channel, cf.~ the integration islands in \cite[Fig.~4]{Carena2014}.
The energies computed by both models are in general not equivalent, since in (\ref{eqn:parseval}) the squared magnitude is taken \textsl{after} the kernel is folded to the Nyquist interval.
In this respect, we follow that the GN-model calculates the \textsl{strength} of the \textsl{optical} end-to-end nonlinear distortion, i.e., \textsl{before} sampling, whereas (\ref{eqn:parseval}) calculates the same quantity for the discrete-time end-to-end channel on a \textsl{per-symbol} basis, i.e., after $T$-spaced sampling.

Equation (\ref{eqn:parseval}) is also the dual representation to the original \ac{GN}-model where the optical signal is constructed---for numerical convenience---as a continuous-time signal with period $T_0$ and \textsl{discrete} frequency components (c.f.~the Karhunen-Lo\`eve formula in \cite{Poggiolini2012a,Carena2012}).
In other words, the discretization in one domain and the periodicity in the other is exchanged in (\ref{eqn:parseval}) compared to the GN-model.
In this view, the \textsl{kernel energy} $E_{h,\nu}$ of the corresponding end-to-end channel (summed over all $\nu$ and weighted with $\phinlnu^2$) corresponds to the system-relevant strength of the nonlinear distortion \textsl{after} $T$-spaced sampling.
For Nyquist-shaped wavelength channels with roll-off factor $\rho=0$ both models produce the same numerical results, because---in that case---the signal band coincides with the Nyquist interval obeying the sampling theorem.

At the same time, the variance of the perturbation $\sigma^2_{\Delta \ve{a}} \defeq \expval{\norm{\Delta \ve{a}}^2}$ depends as well on the properties of the modulation format $\mathcal{A}$ which in turn is a problem addressed by the \ac{EGN}-model \cite{Carena2012}, cf. also the discussion in \cite[Sec.~F and Appx.]{Ghazisaeidi2017a}.
Note, that the derivation of (\ref{eqn:parseval}) does not require any assumptions on the signal (albeit its pulse-shape)---in particular no Gaussian assumption.

We can identify three relevant system parameters that characterize the nonlinear response: Firstly, the map strength $\STp = \Leff / \Ld$ (or equivalently the $\nu$-dependent $\STnu = \Leff / L_{\mathrm{wo}, \nu}$) which is a measure of the temporal extent, i.e., the memory of the nonlinear interaction. 
Secondly, the ($\nu$-dependent) nonlinear phase shift $\phinlnu \defeq \frac{8}{9} \frac{\Leff}{L_{\NL, \nu}}$ that depends via $L_{\NL, \nu}$ linearly on the launch power $\Pow_\nu$ and essentially acts as a scaling factor to the nonlinear distortion $\Delta \ve{a}[k]$.
Finally, the total kernel energy $E_{h,\nu}$ which charactarizes the strength of the nonlinear interaction---independent of the launch power.

\subsection{Algorithmic Implementation in Discrete Frequency-Domain}\label{sec:app}

In this section, we present an algorithmic implementation of the proposed end-to-end model in $1/T$-periodic frequency-domain.
To that end, the algorithm is exemplarily derived for intra-channel (i.e., SCI) contributions corresponding to the continuous-frequency relation in (\ref{eqn:discretee2ebottompage_freq}), and shown at the bottom of Fig.~\ref{fig:end_to_end_pert_model}.
Generalization to \ac{XCI} contributions is straightforward.

In order to realize the frequency-domain processing, the periodic spectrum of the transmit sequence $\ve{A}(\ejwt)$ and the frequency-domain kernel $H_\probe(\ejwwt)$ are \textsl{discretized}.
Then, the point-wise multiplication in frequency-domain results in a \textsl{cyclic} convolution in time-domain, and we have to resort to block-wise processing using the \textsl{overlap-and-save} method \cite{Shynk1992}.

\begin{algorithm}[t]
\SetKw{Break}{break}
{
    $\ve{a}_{\lambda}[k] = \mathop{\mathrm{overlapSaveSplit}}(\sequence{\ve{a}[k]}, \Nfft, K)$\\
$k, \mu, \mu_1, \mu_2 \in \set{0,1, \dots, \Nfft-1}$\\
$H_\probe[\mu_1, \mu_2, \mu_3] = H_{\probe}[\ve{\mu}] =  H_\probe(\ee^{\jj \frac{2 \pi}{\Nfft} \ve{\mu}})$\\
\ForAll{$\lambda$}{
    $\ve{A}_{\lambda}[\mu] = \dft{\ve{a}_{\lambda}[k]}$\\
    \ForAll{$\mu$}{
        $\mu_3 = \mathop{\mathrm{mod}}_{\Nfft}(\mu - \mu_1 + \mu_2)$\\[1.5mm]
        $\Delta \ve{A}_{\lambda}^{\superSCI}[\mu] = -\jj \frac{\phinlp}{\Nfft^2}$\\
        \nonl $\;\;\times \sum_{\mu_1, \mu_2} \ve{A}_\lambda[\mu_1] \ve{A}_\lambda^{\H}[\mu_2] \ve{A}_\lambda[\mu_3] H_\probe[\mu_1, \mu_2, \mu_3]$\\[2mm]
        $\ve{Y}_\lambda^{\PERT}[\mu] = \ve{A}_\lambda[\mu] + \Delta \ve{A}_\lambda^{\superSCI}[\mu]$\\
    }
    $\ve{y}_\lambda^{\PERT}[k] = \idft{\ve{Y}_\lambda^{\PERT}[\mu]}$\\
  }
  $\sequence{\!\ve{y}^{\PERT}[k]\!} = \!\mathop{\mathrm{overlapSaveAppend}}(\ve{y}^{\PERT}_\lambda[k], \Nfft, K)$\\
}
\caption{\label{IR} $\rm\scriptstyle REG$-$\rm\scriptstyle PERT$-$\rm\scriptstyle FD$ \small for the \ac{SCI} contribution}
\end{algorithm}

Algorithm \ref{IR} realizes the regular perturbation ($\REGs$-$\PERTs$) procedure in $1/T$-periodic \textsl{discrete} frequency-domain ($\FDs$).
Here, the overlap-save algorithm is used to split the sequence $\sequence{\ve{a}[k]}$ into overlapping blocks $\ve{a}_\lambda[k] \trf \ve{A}_\lambda[\mu]$ of size $\Nfft$ enumerated by the subindex $\lambda \in \NN$.
The block size is equal to the size of the \ac{DFT}\footnote{The one-dimensional DFT is performed on each vector component of $\ve{a}_\lambda[k]$ and always relates the \textsl{whole} blocks of length $\Nfft$.}
and the overlap between successive blocks is $K$.
The \textsl{aliased} frequency-domain kernel is discretized to obtain the coefficients 
\begin{equation}
    H_\probe[\mu_1, \mu_2, \mu_3] = H_{\probe}[\ve{\mu}] \defeq  H_\probe(\ee^{\jj \frac{2 \pi}{\Nfft} \ve{\mu}})
    \label{eqn:H_discretized}
\end{equation}
where $\Nfft$ is the number of discrete-frequency samples per dimension\footnote{Note, that the frequency discretization of the kernel must not necessarily coincide with the transformation length $\Nfft$.}.
The discrete-frequency indices $\mu_1$ and $\mu_2$ are elements of the set $\set{0, 1, \dots, \Nfft-1}$ whereas $\mu_3$ must be (modulo) \textsl{reduced} to the same number set, cf.~Line 7.
This is due to the frequency matching constraint in (\ref{eqn:varDef}) which may result in a component $\omega_3 = \omega - \omega_1 + \omega_2$ outside the Nyquist interval.
But due to the periodicity of both the discrete spectrum $\ve{A}_\lambda[\mu]$ and the kernel $H_\probe[\ve{\mu}]$, the frequency index $\mu_3$ can be \textsl{folded} back into the Nyquist interval via the modulo reduction.

Line 8 of the algorithm realizes equation (\ref{eqn:discretee2ebottompage_freq}) where the (double) sum is performed over all $\mu_1$ and $\mu_2$.
After frequency-domain processing the blocks of \textsl{perturbed} receive symbols $\ve{Y}_\lambda^{\PERT}[\mu] \itrf \ve{y}_\lambda^{\PERT}[k]$ are transformed back to time-domain where the $\Nfft - K$ desired output symbols of each block are \textsl{appended} to obtain the perturbed sequence $\sequence{\ve{y}^{\PERT}[k]}$.

The number of coefficients can be controlled by pruning the kernel, similar to techniques already applied to (oversampled) \ac{VSTF} models \cite{Guiomar2013b}. 
However, note that in contrast to \ac{VSTF} models the proposed algorithm operates on the $1/T$-periodic spectrum of blocks of transmit symbols $\ve{a}_\lambda[k]$ and the filter coefficients are taken from the \textsl{aliased} frequency-domain kernel which also includes the matched filter on the receiver-side%
\footnote{For fiber nonlinearity compensation, \ac{VSTF} models typically operate on the oversampled signal, e.g., at \textsl{two samples-per-symbol}, in place of bulk chromatic dispersion compensation prior to polarization demultiplexing and matched filtering. Hence, \ac{VSTF} models compute the per-sample distortion based on $H_\NL(\ve{\upsilon})$ or $H_\probe(\ve{\omega})$, cf.~Fig.~\ref{fig:end_to_end_pert_alias} (left), while the proposed model calculates the distortion on a per-symbol basis after matched filtering and $T$-spaced sampling, cf.~Fig.~\ref{fig:end_to_end_pert_alias} (right).}%
.

The time- and frequency-domain picture of the \textsl{regular} perturbation approach are equivalent due to the \ac{DTFT} in (\ref{eqn:discretee2ebottompage_freq}), (\ref{eqn:discretee2ebottompage_time}) which interrelates both representations. 
Algorithm 1 represents a practical realization in discrete-frequency which produces the same (numerical) results as the discrete-time model as long as $\Nfft$ and $K$ are chosen sufficiently large for a given system scenario.

In terms of computational efficiency a frequency-domain implementation \textsl{can} be superior to the time-domain implementation, in particular, for cases where the number of nonlinear interacting pulses is large.
This is typically the case if the system memory is large, i.e., for large map strengths $\STp$ or $\STnu$, or large relative frequency offsets $\Delta \omega_\nu$, or pulse shapes $h_\Tx(t)$ that extend over multiple symbol durations, e.g., a \ac{RRC} shape with small roll-off factor $\rolloff$.
Then, the number of coefficients of the time-domain kernel $h_\nu[\ve{\kappa}]$ exceeding a relevant energy level grows very rapidly leading to a large number of multiplications and summations.
Vice-versa, we can conclude from Fig.~\ref{fig:nonlinear_transfer_fun_Rs} and Fig.~\ref{fig:end_to_end_pert_alias} that for increasing system memory, the energy of the kernel coefficients is confined in a smaller \textsl{volume} within the Nyquist cube, i.e., more coefficients can be pruned.
This is in analogy with linear systems, where a large-memory system is represented by a narrow-banded transfer function.
Moreover, the frequency-domain picture comprises only a double sum per frequency index $\mu$ instead of a triple sum for each $k$ in the time-domain model---this is again in analogy with linear systems where time-domain convolution is dual to frequency-domain point-wise multiplication.

A thorough complexity analysis is, however, beyond the scope of this work, as it heavily depends on the specific application and system scenario in mind. 
Section \ref{sec:IIII} will hence focus on the validity and accuracy of the proposed model---deliberately using a very low pruning level of the coefficients to provide a benchmark performance of the discussed schemes.
To that end, Section \ref{sec:IIII} will compare the regular discrete-time and -frequency model to the reference channel model implemented via the \ac{SSFM}.
In the next section, the regular model is extended to a combined regular-logarithmic model where a subset of the perturbations are considered as multiplicative, i.e., perturbations that cause a rotation in \textsl{phase} or in the \textsl{\ul{s}tate \ul{o}f \ul{p}olarization} (SOP). \acused{SOP}

\subsection{Regular-Logarithmic Model in Discrete-Time Domain}
It was already noted in \cite{Xu2002} that the regular \ac{VSTF} approach (or the equivalent \ac{RP} method) in (\ref{eqn:pert_ansatz}) reveals an energy-divergence problem if the optical launch power $\Pow$ is too high---or more precisely if the nonlinear phase shift $\phinl$ is too large.
Using a first-order \ac{RP} approach, a pure phase rotation is approximated by $\exp(\jj \phi) \approx 1 + \jj \phi$. 
While multiplication with $\exp(\jj \phi)$ is an energy conserving transformation (i.e., the norm is invariant under phase rotation), the \ac{RP} approximation is obviously \textsl{not} energy conserving (cf.~also the discussion in \cite[Sec.~II B.]{Tao2011} and \cite[Sec.~VIII]{Dar2015b}).
In the context of optical transmission, already a trivial (time-constant) average phase rotation due nonlinear interaction is not well modeled by the \ac{RP} method.

This inconsistency was first addressed in the early 2000s \cite{Xu2001a, Vannucci2002a} and years later revived in the context of intra-channel fiber nonlinearity mitigation.
In \cite[Sec.~VI]{Vannucci2002a}, the \ac{RP} method is derived in a reference system rotated by the time-average nonlinear phase, called \ac{eRP} method.
It was shown in \cite{Vannucci2002a, Serena2013, Serena2015}, that the \ac{eRP} model provides a significant improvement over pure \ac{RP} models.
A similar correction formula was proposed for \ac{VSTF} methods in \cite{Xu2002}.
In \cite{Forestieri2005, Secondini2012, Secondini2013}, a \ac{LP} model is derived which is exact in the limit of zero-dispersion links.
On the other hand in \ac{DU} links, as pointed out in \cite{Serena2013}, the \ac{LP} method yields a log-normal distribution of the nonlinear distortion which is inconsistent with observations from simulations and experiments.

In the \textsl{additive-multiplicative} (A-M) model derived in \cite{Fan2012, Tao2014}, it turned out that a certain subset of symbol combinations in the time-domain \ac{RP} model deterministically creates a perturbation oriented into the $-\jj$-direction from the transmit symbol $\ve{a}[k]$.
Similarly, in the pulse-collision picture \cite{Dar2013, Dar2014b, Dar2015b} a subset of \textsl{degenerate}\footnote{in the sense that not all four interacting pulses are distinct.} cross-channel pulse collisions were properly associated to distortions exhibiting a \textsl{multiplicative} nature.
In the same series of contributions, these subsets of degenerate distortions were first termed \textsl{two}- and \textsl{three-pulse collisions}, i.e., symbol combinations $\ve{\kappa}\in \Z^3$ in (\ref{eqn:pulse_collision_xci}) with $\kappa_3 = 0$ in our terminology.
While the pulse collision picture covers mainly cross-channel effects, we will extend the discussion on separating additive and multiplicative terms also to intra-channel effects.

In this context, we review some properties of the kernel coefficients relevant for inter-channel ($\nu \ne \probe$) two- and three-pulse collisions \cite{Dar2015b}
\begin{align}
    h_\nu[\kappa_1, \kappa_2, 0] &\in \R, \quad&\text{if } \kappa_1 = \kappa_2\;
    \label{eqn:kernelSymmetry1}
    \\
    h_\nu[\kappa_1, \kappa_2, 0] &= h_\nu^*[\kappa_2, \kappa_1, 0] \in \C \quad&\text{if } \kappa_1 \ne \kappa_2,
    \label{eqn:kernelSymmetry2}
\end{align}
where two-pulse collisions with $\kappa_1 = \kappa_2$ in (\ref{eqn:kernelSymmetry1}) are doubly degenerate and the kernel is real-valued\footnote{The transmit pulse-shape $h_\Tx(t)$ is assumed to be a real-valued (root) raised-cosine.}.
In case of three-pulse collisions, the kernel is generally complex-valued but due to its symmetry property in (\ref{eqn:kernelSymmetry2}) and the double sum over all (nonzero) pairs of $[\kappa_1, \kappa_2]^\T$ in (\ref{eqn:pulse_collision_xci}) the overall effect is still multiplicative.

Additionally, for intra-channel contributions ($\nu = \probe$) we find the following symmetry properties of the kernel
\begin{align}
    h_\probe[\kappa_1, \kappa_2, \kappa_3] &= h_\probe[\kappa_3, \kappa_2, \kappa_1]
    \label{eqn:kernelSymmetry3}
    \\
    h_\probe[\kappa_1, \kappa_2, \kappa_3] &= h_\probe[-\kappa_1, -\kappa_2, -\kappa_3],
    \label{eqn:kernelSymmetry4}
\end{align}
and we identify a second degenerate case with $\kappa_1 = 0$ as source for multiplicative distortions, cf. the symmetric form of (\ref{eqn:discretee2ebottompage_time}) w.r.t.~$\kappa_1$ and $\kappa_3$.

In the following, the original RP solution is modified such that perturbations originating from certain degenerate mixing products are associated with a multiplicative perturbation.
Similar to \cite{Secondini2009,Tao2014,Dar2015b}, we extend the previous \ac{RP} model to a combined regular-logarithmic model.
It takes the general form of%
\footnote{Note, that the order, in which the additive and multiplicative perturbation is applied, matters. We chose the same order as in the original \textsl{additive-multiplicative (A-M) model} from \cite{Tao2014}, but we have no proof that this is the optimal order of how to combine the two operations.}
\begin{equation}
    \ve{y}[k] = \exp\left(\jj \phiT[k] + \jj \vec{\ve{s}}[k] \cdot \vec{\ve{\sigma}}\right) \left(\ve{a}[k] + \Delta \ve{a}[k]\right).
    \label{eqn:pert_ansatz_hybrid}
\end{equation}
In addition to the regular, additive perturbation $\Delta \ve{a}[k]$ we now also consider a \textsl{phase} rotation by $\exp(\jj \phiT[k])$ and a rotation in the \textsl{state of polarization} by $\exp(\jj \vec{\ve{s}}[k] \cdot \vec{\ve{\sigma}})$.
Here, $\exp(\cdot)$ denotes the \textsl{matrix exponential}.
All perturbative terms combine both \ac{SCI} and \ac{XCI} effects, i.e., the \textsl{additive} perturbation $\Delta \ve{a}[k] \in \C^2$ is the sum of \ac{SCI} and \ac{XCI} contributions.
The \textsl{time-dependent} phase rotation is given by $\exp(\jj {\phiT}[k])$ with the diagonal matrix ${\phiT}[k] \in \R^{2\times2}$ defined as
\begin{align}
    {\phiT}[k] &\defeq 
    {\phi}^{\superSCI}[k]\, \mathbf{I} + {\phi}^{\superXCI}[k] \,\mathbf{I},
    \label{eqn:diagphase}
\end{align}
i.e., we find a \textsl{common} phase term for both polarizations originating from intra- and inter-channel effects.

The combined effect of intra- and inter-channel \ac{XPolM} is expressed by the \textsl{Pauli matrix expansion} $\vec{\ve{s}}[k] \cdot \vec{\ve{\sigma}} \in \C^{2\times2}$ using (\ref{eqn:pauliexpand}), with the notation adopted from \cite{Gordon2000} and \cite{Winter2009a}.
The expansion defines a unitary rotation in Jones space of the perturbed vector $\ve{a}[k] + \Delta\ve{a}[k]$ around the \textsl{time-dependent} Stokes vector $\vec{\ve{s}}[k]$ and is explained in more detail in the subsequent subsection.

\subsubsection{SCI Contribution}
To discuss the \ac{SCI} contribution we first introduce the following symbol sets
\begin{align}
    \setk^{\superSCI} &= \set{[\kappa_1, \kappa_2, \kappa_3]^\T \in \Z^3 \mid |h_{\probe}[\ve{\kappa}]/h_{\probe}[\ve{0}]|^2 > \Gamma^{\superSCI}}
    \label{eqn:setDef1}
    \\
     \setk_\phi^\oplus &\defeq \set{\setk^{\superSCI} \mid \kappa_1 = 0 \wedge \kappa_2 \ne 0 \wedge \kappa_3 \ne 0}
    \label{eqn:setDef2}
    \\
     \setk_\phi^\ominus &\defeq \set{\setk^{\superSCI} \mid \kappa_3 = 0 \wedge \kappa_2 \ne 0 \wedge \kappa_1 \ne 0}
    \label{eqn:setDef3}
    \\
     \setk_\phi^\superSCI &\defeq \setk_\phi^{\oplus} \cup \setk_\phi^\ominus \cup \set{\ve{\kappa} = \ve{0}} 
    \label{eqn:setDef4}
    \\
     \setk_\Delta^\superSCI &\defeq \setk^{\superSCI} \setminus \setk_\phi^\superSCI,
    \label{eqn:setDef5}
\end{align}
where (\ref{eqn:setDef1}) defines the \textsl{base} set including all possible symbol combinations that exceed a certain energy  (clipping) level $\Gamma^{\superSCI}$ normalized to the energy of the \textsl{center} tap at $\ve{\kappa} = \ve{0}$.
In (\ref{eqn:setDef2}), (\ref{eqn:setDef3}) the joint set of degenerate two- and three-pulse collisions for \ac{SCI} are defined which follow directly from the kernel properties in (\ref{eqn:kernelSymmetry1}), (\ref{eqn:kernelSymmetry2}) for $\kappa_3 =0$, and (\ref{eqn:kernelSymmetry3}), (\ref{eqn:kernelSymmetry4}) for $\kappa_1 = 0$.
The set of indices for \textsl{multiplicative} distortions $\setk_\phi^{\superSCI}$ in (\ref{eqn:setDef4}) also includes the singular case $\ve{\kappa} = \ve{0}$.
Then, the additive set is simply the complementary set of $\setk^{\superSCI}_\phi$ w.r.t.~the base set $\setk^{\superSCI}$.

We start with the additive perturbation from the previous section in (\ref{eqn:discretee2ebottompage_time}) which now reads
\begin{equation}
    \Delta \ve{a}^{\superSCI}[k] = -\jj \phinlp \sum_{\setk^\superSCI_\Delta} \ve{a}[k\!+\!\kappa_1] \ve{a}^\H[k\!+\!\kappa_2] \ve{a}[k\!+\!\kappa_3] h_\probe[\ve{\kappa}],
\end{equation}
where the triple sum is now restricted to the set $\setk^\superSCI_\Delta$ excluding all combinations which result in a multiplicative distortion, cf. (\ref{eqn:setDef5}).

To calculate the common phase $\phi^{\superSCI}[k]$ and the intra-channel Stokes rotation vector $\vec{\ve{s}}^{\,\superSCI}[k]$ we first analyse the expression $\ve{a}[k+\kappa_1] \ve{a}^\H[k+\kappa_2] \ve{a}[k+\kappa_3]$ from the original equation in (\ref{eqn:discretee2ebottompage_time}).
For the set $\setk_\phi^\oplus$ with $\kappa_1=0$ the triple product factors into the respective transmit symbol $\ve{a}[k] $ and a scalar value $\ve{a}^\H[k+\kappa_2]\ve{a}[k+\kappa_3]$.
After multiplication with $h_\probe[0, \kappa_2, \kappa_3]$ and summation of all $\ve{\kappa} \in \setk_\phi^\oplus$ the perturbation is strictly imaginary-valued (cf. symmetry properties in (\ref{eqn:kernelSymmetry3}), (\ref{eqn:kernelSymmetry4})).

On the other hand, for $\setk_\phi^\ominus$ with $\kappa_3=0$ we have to rearrange the triple product using the matrix expansion from (\ref{eqn:pauliproject}) to factor the expression accordingly as\footnote{multiplication with $h_\probe[\ve{\kappa}]$ and summation over $\ve{\kappa}\in \setk_\phi^{\superSCI}$ are implied.}
\begin{align}
    \ve{a} \ve{a}^\H \ve{a} = \frac{1}{2} \left( \ve{a}^\H \ve{a}\,\mathbf{I} + (\ve{a}^\H \vec{\ve{\sigma}} \ve{a}) \cdot \vec{\ve{\sigma}} \right) \ve{a}.
    \label{eqn:pauliA}
\end{align}
The first term $\ve{a}^\H \ve{a} \,\mathbf{I}$ also contributes to a common phase term, whereas the second term $(\ve{a}^\H \vec{\ve{\sigma}} \ve{a}) \cdot \vec{\ve{\sigma}} \in \C^{2\times2}$ is a traceless and Hermitian matrix s.t.~$\exp(\jj (\ve{a}^\H \vec{\ve{\sigma}} \ve{a}) \cdot \vec{\ve{\sigma}})$ is a unitary polarization rotation\footnote{Since the Pauli expansion $\vec{\ve{u}}\cdot \vec{\ve{\sigma}}$ in (\ref{eqn:pauliexpand}) is Hermitian, the expression $\exp(\jj\, \vec{\ve{u}}\cdot\vec{\ve{\sigma}})$ is unitary.}.

The multiplicative perturbation $\exp(\jj \phi^\superSCI[k])$ with $\phi^\superSCI[k] \in \R$ is then given by 
\begin{align}
    \phi^{\superSCI}[k] = &-\phinlp \sum_{\setk_\phi^\oplus}\! \ve{a}^\H[k\!+\!\kappa_2] \ve{a}[k\!+\!\kappa_3] h_\probe[\ve{\kappa}]\nonumber\\
     &-\frac{1}{2} \phinlp \sum_{\setk_\phi^\ominus}\! \ve{a}^\H[k\!+\!\kappa_2] \ve{a}[k\!+\!\kappa_1] h_\probe[\ve{\kappa}]\nonumber\\
                          &-\phinlp \norm{\ve{a}[k]}^2 h_\probe[\ve{0}]\\
     = &-\frac{3}{2} \phinlp \sum_{\setk_\phi^\ominus}\! \ve{a}^\H[k\!+\!\kappa_2] \ve{a}[k\!+\!\kappa_1] h_\probe[\ve{\kappa}]\nonumber\\
                          &-\phinlp \norm{\ve{a}[k]}^2 h_\probe[\ve{0}].
\end{align}
Given a wide-sense stationary transmit sequence $\sequence{\ve{a}[k]}$, the induced nonlinear phase shift has a \textsl{time-average} value $\bar{\phi}^{\superSCI}$, around which the instantaneous phase $\phi^{\superSCI}[k]$ may fluctuate (cf. also \cite{Serena2015}).

The instantaneous rotation of the \ac{SOP} due to the expression $\exp(\jj \vec{\ve{s}}^{\,\superSCI}[k] \cdot \vec{\ve{\sigma}}) \in \C^{2\times2}$ causes intra-channel \ac{XPolM} \cite{Mecozzi2012b}. It is given by
\begin{align}
    \vec{\ve{s}}^{\,\superSCI}[k] \cdot \vec{\ve{\sigma}} &= -\frac{1}{2} \phinlp \sum_{\setk_{\phi}^{\ominus}}\! \Big(2\,\ve{a}[k+\kappa_1] \ve{a}^\H[k+\kappa_2] \nonumber\\
                                 &\quad\quad- \ve{a}^{\H}[k+\kappa_2] \ve{a}[k+\kappa_1] \mathbf{I} \Big) h_\probe[\ve{\kappa}],
                                 \label{eqn:intraXPolM}
\end{align}
where we made use of the relation in (\ref{eqn:pauliexpand}). The rotation matrix $\exp(\jj \vec{\ve{s}}^{\, \superSCI}[k]\cdot \vec{\ve{\sigma}})$ is unitary and $\vec{\ve{s}}^{\, \superSCI}[k] \cdot \vec{\ve{\sigma}}$ is Hermitian and traceless.
The physical meaning of the transformation described in (\ref{eqn:intraXPolM}) is as follows: The perturbed transmit vector $(\ve{a}[k] + \Delta \ve{a}[k])$ in (\ref{eqn:pert_ansatz_hybrid}) is transformed into the polarization eigenstate $\vec{\ve{s}}^{\,\superSCI}[k]$ (i.e., into the basis defined by the eigenvectors of $\vec{\ve{s}}^{\, \superSCI}[k] \cdot \vec{\ve{\sigma}}$).
There, both vector components receive equal but opposite phase shifts and the result is transformed back to the $\X/\Y$-basis of the transmit vector.
In Stokes space, the operation can be understood as a precession of $(\vec{\ve{a}}[k] + \Delta \vec{\ve{a}}[k])$ around the Stokes vector $\vec{\ve{s}}^{\, \superSCI}[k] $ by an angle equal to its length $\norm{\vec{\ve{s}}^{\, \superSCI}[k]}$.
The intra-channel Stokes vector $\vec{\ve{s}}^{\, \superSCI}[k]$ depends via the nonlinear kernel $h_\probe[\ve{\kappa}]$ on the transmit symbols within the memory of the nonlinear interaction $\STp$ around $\ve{a}[k]$.
Similar to the nonlinear phase shift---for a wide-sense stationary input sequence---the Stokes vector $\vec{\ve{s}}^{\, \superSCI}[k]$ has a time-constant average value around which it fluctuates over time.

\subsubsection{XCI Contribution}
The same methodology is now applied to cross-channel effects.
The symbol set definitions for \ac{XCI} follow from the considerations in the previous section.
\begin{align}
    \setk_\nu^{\superXCI} &= \set{[\kappa_1, \kappa_2, \kappa_3]^\T \in \Z^3 \mid |h_{\nu}[\ve{\kappa}]/h_{\nu}[\ve{0}]|^2 > \Gamma^{\superXCI}_\nu}
    \label{eqn:setDefXCI1}
    \\
    \setk_{\phi,\nu}^\superXCI &\defeq \set{\setk_\nu^{\superXCI} \mid \kappa_3 = 0 \wedge \kappa_2 \ne 0 \wedge \kappa_1 \ne 0} \nonumber\\
                           &\cup \set{\ve{\kappa} = \ve{0}}
    \label{eqn:setDefXCI2}
    \\
    \setk_{\Delta,\nu}^\superXCI &\defeq \setk_\nu^{\superXCI} \setminus \setk_{\phi,\nu}^\superXCI,
    \label{eqn:setDefXCI4}
\end{align}
where the subscript $\nu$ indicates the channel number of the respective interfering channel.
For $\setk_{\phi,\nu}^\superXCI$, only the degenerate case $\kappa_3 = 0$ has to be considered\footnote{due to the kernel properties of $h_\nu[\kappa_1, \kappa_2, 0]$ in (\ref{eqn:kernelSymmetry1}), (\ref{eqn:kernelSymmetry2}).}.
Similar to (\ref{eqn:pauliA}), the expression $\ve{b} \ve{b}^\H + \ve{b}^{\H} \ve{b} \,\mathbf{I}$
from (\ref{eqn:pulse_collision_xci}) is rearranged to obtain
\begin{align}
    \frac{3}{2} \underbrace{\begin{bmatrix}
        b_\X b_\X^*\! +\! b_\Y b_\Y^*\!&\!0\\
        0\!&\!b_\Y b_\Y^* \!+\!b_\X b_\X^*
\end{bmatrix}}_{\ve{b}^\H \ve{b}\, \mathbf{I}}
+\frac{1}{2} \underbrace{\begin{bmatrix}
        b_\X b_\X^* \!-\! b_\Y b_\Y^*&2b_\X b_\Y^*\\
       2 b_\Y b^*_\X&b_\Y b_\Y^* \!-\! b_\X b_\X^*
   \end{bmatrix}}_{2\, \ve{b} \ve{b}^\H - \ve{b}^\H \ve{b}\, \mathbf{I}\; =\; (\ve{b}^\H \vec{\ve{\sigma}} \ve{b}) \cdot \vec{\ve{\sigma}}},
\end{align}
where the argument and subscript $\nu$ is omitted for concise notation.
The multiplicative cross-channel contribution is again split into a common phase shift in both polarizations and an equal but opposite phase shift in the basis given by the instantaneous Stokes vector of the $\nu^{\rm th}$ interferer.

We define the total, common phase shift due to cross-channel interference as
\begin{equation}
    {\phi}^{\superXCI}[k] = -\sum_{\nu \ne \probe}\frac{3}{2} \phinlnu  \sum_{\setk_{\phi,\nu}^\superXCI}\! \ve{b}_{\nu}^{\H}[k\!+\!\kappa_1] \ve{b}_{\nu}[k\!+\!\kappa_2]\; h_\nu[{\ve{\kappa}}]
    \label{eqn:phi_inter}
\end{equation}
which depends on the instantaneous sum over all interfering channels and the sum of $\ve{b}_\nu^{\H} \ve{b}_\nu$ over $[\kappa_1, \kappa_2]^\T$.
The effective, instantaneous cross-channel Stokes vector $\vec{\ve{s}}^\superXCI[k]$ is given by
\begin{align}
    \vec{\ve{s}}^{\,\superXCI}[k] \cdot \vec{\ve{\sigma}} &= - \sum_{\nu \ne \probe}\frac{1}{2}\phinlnu  \sum_{\setk_{\phi, \nu}^\superXCI}\! \Big(2\,\ve{b}_\nu[k+\kappa_1] \ve{b}_\nu^\H[k+\kappa_2] \nonumber\\
                                 &- \ve{b}_\nu^{\H}[k+\kappa_2] \ve{b}_\nu[k+\kappa_1] \mathbf{I} \Big) h_\nu[\ve{\kappa}].
                                 \label{eqn:stokes_inter}
\end{align}
Note, that the expressions in (\ref{eqn:phi_inter}), (\ref{eqn:stokes_inter}) include both contributions from two- and three pulse collisions (cf. \cite[Eq.~(10)--(13)]{Dar2015b}).

\subsubsection{Energy of Coefficients in Discrete-Time Domain}
The energy of the kernel coefficients is defined for the subsets given in (\ref{eqn:setDef1})--(\ref{eqn:setDef5}). We find for the different symbol sets
\begin{align}
    E^{\superSCI}_{h} &\defeq \sum_{\setk^{\superSCI}} |h_\probe[\ve{\kappa}]|^2
    \label{eqn:kernelEnergy_time_tot}
    \\
    E^{\superSCI}_{h, \Delta} &\defeq \sum_{\setk_\Delta^{\superSCI}} |h_\probe[\ve{\kappa}]|^2\\
    E^{\superSCI}_{h, \phi} &\defeq \sum_{\setk_\phi^{\superSCI}} |h_\probe[\ve{\kappa}]|^2,
    \label{eqn:kernelEnergy_time}
\end{align}
with the clipping factor $\Gamma^{\superSCI}$ in (\ref{eqn:setDef1}) equal to zero. 
The energy for cross-channel effects is defined accordingly with the sets from (\ref{eqn:setDefXCI1})--(\ref{eqn:setDefXCI4}).
Since the subsets for additive and multiplicative effects are always disjoint we have $E_h^\superSCI = E^\superSCI_{h, \Delta} + E^\superSCI_{h, \phi}$.

\subsection{Regular-Logarithmic Model in Frequency-Domain}
Similar to the previous section, we first review some kernel properties of the aliased frequency-domain kernel 
\begin{align}
    H_\nu(\ejwwt) &\in \R, &\text{if }\, \omega_2 = \omega_1\; \Leftrightarrow \;\omega_3 = \omega,
    \label{eqn:kernelFreqSymmetry1}
    \\
    H_\probe(\ejwwt) &\in \R, &\text{if } \,\omega_2 = \omega_1\;\Leftrightarrow\; \omega_3 = \omega\hphantom{,}
    \label{eqn:kernelFreqSymmetry2}
    \\
    &~&\vee~ \,\omega_2 = \omega_3\;\Leftrightarrow\; \omega_1 = \omega, \nonumber
\end{align}
where the two (doubly) degenerate cases $\omega_1 = \omega_2$ and $\omega_3 = \omega_2$ correspond to classical inter- and intra-channel \ac{XPM}.
The first degenerate case with $\omega_1 = \omega_2$ corresponds to the \textsl{diagonal} of the Nyquist region in Fig.~\ref{fig:end_to_end_pert_alias}, shown here for $\omega_3 = 0$. 
For the special case that the transmit pulse $h_{\Tx}(t)$ has a \ac{RRC} shape with roll-off factor $\rho = 0$ (i.e., no spectral support outside the Nyquist region), we find that the folded kernel always takes the value $1$ on the diagonal---independent of $\omega_3$.
In general, the elements on the diagonal are only \textsl{nearly} (frequency-) flat, and the flatness depends on the amount of spectral support which is folded into the Nyquist region, cf.~the yellow region in Fig.~\ref{fig:end_to_end_pert_alias}.
This approximation will be used in the remainder to simplify the expression for the average phase- and polarization rotation.

We also find the following symmetry properties of the intra-channel kernel
\begin{align}
    H_\probe(\ee^{\jj\, [\omega_1, \omega_2, \omega_3]^\T T}) &= H_\probe(\ee^{\jj\, [\omega_3, \omega_2, \omega_1]^\T T}),
    \label{eqn:kernelFreqSymmetry3}
    \\
    H_\probe(\ee^{\jj\, [\omega_1, \omega_2, \omega_3]^\T T}) &= H_\probe(\ee^{\jj\, [-\omega_1, -\omega_2, -\omega_3]^\T T}),
    \label{eqn:kernelFreqSymmetry4}
\end{align}
and we can conclude that the cut of the kernel shown in Fig.~\ref{fig:end_to_end_pert_alias} is equivalent to the cut with $\omega_1 = 0$, shown in the $\omega_3$-$\omega_2$-plane instead.

The frequency-domain model is now modified such that the degenerate contributions will be associated with multiplicative distortions.
Due to the multiplicative nature, average effects can be straightforwardly incorporated into the frequency-domain model as they are both treated as constant pre-factors in the time- and frequency-domain representation.
We will see in the next section that this already leads to significantly improved results compared to the regular model.
Note that, in contrast to the regular models, the regular-logarithmic model in time and frequency are no longer equivalent.

The general form of the combined regular-logarithmic model in frequency is given by
\begin{align}
    \ve{Y}(\ejwt) &= \exp\left(\jj {\aPhi} + \jj \vec{\ve{S}} \cdot \vec{\ve{\sigma}}\right)\nonumber\\
                  &\times \left(\ve{A}(\ejwt) + \Delta \ve{A}(\ejwt)\right),
    \label{eqn:pert_ansatz_hybrid_freq}
\end{align}
where the phase- and polarization-term take on a \textsl{frequency}-constant value, i.e., independent of $\ejwt$ (indicated here by the lack of argument for $\aPhi$ and $\vec{\ve{S}}$).
Following the same terminology as before, we introduce the \textsl{average} multiplicative perturbation of the common phase term
\begin{align}
    {\aPhi} &\defeq 
    {\aphi}^{\superSCI}\, \mathbf{I} + {\aphi}^{\superXCI} \,\mathbf{I},
    \label{eqn:diagphase_freq}
\end{align}
as the sum of the intra-channel contribution $\aphi^\superSCI\in \R$ and the inter-channel contribution $\aphi^\superXCI \in \R$.
Similarly, for the \textsl{average} polarization rotation we have 
\begin{align}
    {\vec{\ve{S}} \cdot \vec{\ve{\sigma}}} &\defeq 
    {\vec{\ve{S}}}^{\,\superSCI} \cdot \vec{\ve{\sigma}} + {\vec{\ve{S}}}^{\superXCI} \cdot \vec{\ve{\sigma}},
    \label{eqn:polrot_freq}
\end{align}
where $\vec{\ve{S}} \cdot \vec{\ve{\sigma}}$ is again Hermitian and traceless.

\subsubsection{SCI Contribution}
The two degenerate frequency conditions in (\ref{eqn:kernelFreqSymmetry2}) are used in the expression (\ref{eqn:discretee2ebottompage_freq}) to obtain the \textsl{average}, intra-channel phase distortion.
To that end, the triple product $\ve{A}\ve{A}^\H\ve{A}$ in (\ref{eqn:discretee2ebottompage_freq}) is rearranged similar to (\ref{eqn:pauliA}).
First, the general \textsl{frequency-dependent} expression $\phi^\superSCI(\ejwt)$ is given by
\begin{align}
    {\phi}^{\superSCI} \!= &\!-\!\hphantom{\frac{1}{2}}\phinlp \frac{T}{(2\pi)^2} \!\int_{\TT}\! \norm{\ve{A}(\ee^{\jj \omega_2 T})}^2 H_\probe( \ee^{\jj [\omega, \omega_2, \omega_2]^\T T} ) \dd \omega_2 \nonumber\\
                          &\!-\!\frac{1}{2} \phinlp \frac{T}{(2\pi)^2} \! \int_{\TT}\! \norm{\ve{A}(\ee^{\jj \omega_1 T})}^2 H_\probe( \ee^{\jj [\omega_1, \omega_1, \omega]^\T T} ) \dd \omega_1,
                                                  \label{eqn:phase_freq_sci_freqdep}
\end{align}
where the first term on the right-hand side in (\ref{eqn:phase_freq_sci_freqdep}) corresponds to the degeneracy $\omega_2 = \omega_3 \Leftrightarrow \omega_1 = \omega$ and the second term corresponds to $\omega_2 = \omega_1 \Leftrightarrow \omega_3 = \omega$.
We simplify the expression using the \ac{RRC} $\rho=0$ approximation (i.e., the kernel on the diagonal is equal to $1$) and the symmetry property in (\ref{eqn:kernelFreqSymmetry3}) to obtain the average, intra-channel phase distortion
\begin{align}
    {\aphi}^{\superSCI} = &-\frac{3}{2}\phinlp \frac{\Tsym}{(2 \pi)^2} \int_\TT \norm{\ve{A}(\ee^{\jj \omega \Tsym})}^2\; \dd \omega,
                                                  \label{eqn:phase_freq_sci}
\end{align}
which does no longer depend on the power or dispersion profile of the transmission link (given a fixed $\Leff$).

Similarly, the average intra-channel \ac{XPolM} contribution can be simplified to
\begin{align}
    \vec{\ve{S}}^{\,\superSCI} \cdot \vec{\ve{\sigma}} &= -\frac{1}{2}\phinlp \frac{\Tsym}{(2 \pi)^2}  \int_\TT \Big(2\,\ve{A}(\ee^{\jj \omega\Tsym}) \ve{A}^\H(\ee^{\jj \omega\Tsym}) \nonumber\\
                                                  &- \ve{A}^{\H}(\ee^{\jj \omega\Tsym}) \ve{A}(\ee^{\jj \omega\Tsym}) \mathbf{I} \Big) \dd \omega.
                                                  \label{eqn:xpolm_freq_sci}
\end{align}
In Algorithm \ref{IR:2} the required modifications to the regular perturbation model ($\REGs$-$\PERTs$) are highlighted to arrive at the regular-logarithmic perturbation model ($\REGLOGs$-$\PERTs$)---again exemplarily for the \ac{SCI} contribution.
Lines 6,7 of Algorithm 2 translate Eq.~(\ref{eqn:phase_freq_sci}), (\ref{eqn:xpolm_freq_sci}) to the discrete-frequency domain where the integral over all $\omega \in \TT$ becomes a sum over all $\mu$ of the $\lambda^{\rm th}$ processing block.
The average values, here, are always associated to the average values of the $\lambda^{\rm th}$ block.
In Lines 10, 11, the double sum to obtain $\Delta \ve{A}_\lambda^{\superSCI}[\mu]$ is restricted to all combinations $\mathcal{U}$ of the discrete frequency pair $[\mu_1, \mu_2]^\T$ excluding the degenerate cases corresponding to Eq.~(\ref{eqn:kernelFreqSymmetry1}), (\ref{eqn:kernelFreqSymmetry2}).
The perturbed receive vector $\ve{Y}_\lambda^{\PERT}$ is then calculated according to (\ref{eqn:pert_ansatz_hybrid_freq}) before it is transformed back to the discrete-time domain.
\subsubsection{XCI Contribution}
The cross-channel contributions follow from the considerations in the previous sections and we obtain for the degenerate case in (\ref{eqn:kernelFreqSymmetry1}) the total, average \ac{XCI} phase shift
\begin{equation}
    {\aphi}^{\superXCI} = -\sum_{\nu \ne \probe} \frac{3}{2} \phinlnu  \frac{\Tsym}{(2 \pi)^2} \int_\TT \norm{\ve{B}_{\nu}(\ee^{\jj \omega\Tsym})}^2 \; \dd \omega
\end{equation}
and analogously for the total, average \ac{XCI} Stokes vector we find
\begin{align}
    \vec{\ve{S}}^{\,\superXCI} \cdot \vec{\ve{\sigma}} &= -\sum_{\nu \ne \probe} \frac{1}{2} \phinlnu \frac{\Tsym}{(2 \pi)^2}  \int_\TT \Big(2\,\ve{B}_\nu(\ee^{\jj \omega\Tsym}) \ve{B}_\nu^\H(\ee^{\jj \omega\Tsym}) \nonumber\\
                                                  &- \ve{B}_\nu^{\H}(\ee^{\jj \omega\Tsym}) \ve{B}_\nu(\ee^{\jj \omega\Tsym}) \mathbf{I} \Big) \dd \omega.
\end{align}
\begin{algorithm}[t]
\SetKw{Break}{break}
{
    \color{mygray}
    $\ve{a}_{\lambda}[k] = \mathop{\mathrm{overlapSaveSplit}}(\sequence{\ve{a}[k]}, \Nfft, K)$\\
$k, \mu, \mu_1, \mu_2 \in \set{0,1, \dots, \Nfft-1}$\\
$H_\probe[\mu_1, \mu_2, \mu_3] = H_{\probe}[\ve{\mu}] =  H_\probe(\ee^{\jj \frac{2 \pi}{\Nfft} \ve{\mu}})$\\
\ForAll{$\lambda$}{
    $\ve{A}_{\lambda}[\mu] = \dft{\ve{a}_{\lambda}[k]}$\\
    \textcolor{black}{$\aphi^{\superSCI}_\lambda = -\frac{3}{2}\frac{\phinlp}{\Nfft^2} \;\sum_\mu \norm{\ve{A}_\lambda[\mu]}^2$\\}
    \textcolor{black}{$\vec{\ve{S}}^{\,\superSCI}_\lambda\!\! \cdot \vec{\ve{\sigma}} = -\frac{1}{2}\frac{\phinlp}{\Nfft^2} \sum_\mu 2\ve{A}_\lambda[\mu] \ve{A}_\lambda^\H[\mu] - \norm{\ve{A}_\lambda[\mu]}^2 \mathbf{I}$\\}
    \ForAll{$\mu$}{
        $\mu_3 = \mathop{\mathrm{mod}}_{\Nfft}(\mu - \mu_1 + \mu_2)$\\[1.5mm]
        $\textcolor{black}{\mathcal{U} = \set{[\mu_1, \mu_2]^\T \mid \mu_2 \ne \mu_1 \wedge \mu_2 \ne \mu_3}}$\\[1.5mm]
        $\Delta \ve{A}_{\lambda}^{\superSCI}[\mu] = -\jj \frac{\phinlp}{\Nfft^2}$\\
        \nonl $\;\;\times \sum_{\textcolor{black}{\mathcal{U}}} \ve{A}_\lambda[\mu_1] \ve{A}_\lambda^{\H}[\mu_2] \ve{A}_\lambda[\mu_3] H_\probe[\mu_1, \mu_2, \mu_3]$\\[2mm]
        $\textcolor{black}{\ve{Y}_\lambda^{\PERT}[\mu] = \mathrm{exp}(\jj \aphi^{\superSCI}_\lambda\mathbf{I} + \jj  \vec{\ve{S}}^{\, \superSCI}_\lambda \!\!\cdot \vec{\ve{\sigma}})}$\\ 
        \nonl $\textcolor{black}{\quad \quad \quad \quad \quad \times (\ve{A}_\lambda[\mu] + \Delta \ve{A}_\lambda^{\superSCI}[\mu])}$\\
    }
    $\ve{y}_\lambda^{\PERT}[k] = \idft{\ve{Y}_\lambda^{\PERT}[\mu]}$\\
  }
  $\sequence{\!\ve{y}^{\PERT}[k]\!} = \! \mathop{\mathrm{overlapSaveAppend}}(\ve{y}^{\PERT}_\lambda[k], \Nfft, K)$\\
}
\caption{\label{IR:2} $\rm\scriptstyle REGLOG$-$\rm\scriptstyle PERT$-$\rm\scriptstyle FD$ \small for the \ac{SCI} contribution}
\end{algorithm}

\subsubsection{Energy of Coefficients in Discrete-Frequency Domain}
With the notation of the discrete-frequency kernel from (\ref{eqn:H_discretized}) we have according to Parseval's theorem in (\ref{eqn:parseval}) the following definitions
\begin{align}
    E^{\superSCI}_{H} &\defeq \frac{1}{\Nfft^3} \sum_{\mathcal{U}^\superSCI} |H_\probe[\ve{\mu}]|^2\\
    E^{\superSCI}_{H, \Delta} &\defeq \frac{1}{\Nfft^3} \sum_{\mathcal{U}_\Delta^{\superSCI}} |H_\probe[\ve{\mu}]|^2\\
    E^{\superSCI}_{H, \phi} &\defeq \frac{1}{\Nfft^3} \sum_{\mathcal{U}_\phi^{\superSCI}} |H_\probe[\ve{\mu}]|^2  \stackrel{\rho=0}{\approx} \frac{2}{\Nfft},
    \label{eqn:kernelEnergy_freq}
\end{align}
with the sets according to (\ref{eqn:kernelFreqSymmetry2})
\begin{align}
    \mathcal{U}^{\superSCI} &= \set{ \ve{\mu} = [\mu_1, \mu_2, \mu_3]^\T \in \set{0,1,\dots,\Nfft-1}^3 }\\
    \mathcal{U}^{\superSCI}_\Delta &= \set{ \mathcal{U}^\superSCI \mid \mu_2 \ne \mu_1 \wedge \mu_2 \ne \mu_3}\\
    \mathcal{U}^{\superSCI}_\phi &= \set{ \mathcal{U}^\superSCI \mid \mu_2 = \mu_1 \vee \mu_2 = \mu_3}.
\end{align}
Note, that we have again $E^\superSCI_H = E^\superSCI_{H, \Delta} + E^\superSCI_{H, \phi}$ and due to Parseval's theorem $E_{h}^\superSCI = E_H^\superSCI$ for $\Nfft \to \infty$.
The cardinalities of the sets are $|\mathcal{U}^\superSCI| = \Nfft^3$, $|\mathcal{U}_\phi^\superSCI| = 2\Nfft^2-\Nfft$ and $|\mathcal{U}_\Delta^\superSCI| = |\mathcal{U}^\superSCI| - |\mathcal{U}_\phi^\superSCI|$.
With the \ac{RRC} pulse-shape and $\rho=0$ we find again that $H_\probe(\ve{\mu}) = 1$ with $\ve{\mu} \in \mathcal{U}_{\phi}^\superSCI$, and with that the kernel energy is simplified to $E_{H, \phi}^{\superSCI} = (2\Nfft-1)/\Nfft^2\approx 2/{\Nfft}$.

The cross-channel sets are defined according to (\ref{eqn:kernelFreqSymmetry1}) with only a single degeneracy.

\section{Numerical Results}\label{sec:IIII}
This section complements the theoretical considerations of the previous sections by numerical simulations.
To this end, we compare the simulated received symbol sequence $\sequence{\ve{y}[k]}$ obtained by the perturbation-based ($\rm \scriptstyle PERT$) end-to-end channel models to the sequence obtained by numerical evaluation via the \ac{SSFM} (in the following indicated by the superscript $\rm \scriptstyle SSFM$).

The evaluated metric is the \textsl{normalized} \ac{MSE} between the two $T$-spaced output sequences for a given input symbol sequence $\sequence{\ve{a}[k]}$, i.e., we have
\begin{equation}
    \sigma^2_{\rm e} \defeq \expval{\norm{\ve{y}^{\SSFM} - \ve{y}^{\PERT}}^2},
    \label{eqn:model_mse}
\end{equation}
where the expectation takes the form of a statistical average of the received sequence over the discrete time index $k$. The \ac{MSE} is already normalized due to the fixed variance $\sigma_a^2=1$ of the symbol alphabet and the receiver-side re-normalization in (\ref{eqn:h_r-cascade}), s.t.~the received sequence has (approximately\footnote{In the numerical simulation via \ac{SSFM} \textsl{signal depletion} takes place due to an energy transfer from signal to \ac{NLI}. For simplicity, this additional signal energy loss is not accounted for by additional receiver-side re-normalization.}) the same fixed variance as the transmit sequence.

\begin{figure*}[b]
    \centering
    \subfloat[$\REGs$-$\PERTs$-$\TDs$, single-channel, single-span, lossless fiber]{
        \input{./figures/results_reg-pert_td_rs_vs_p/rs_vs_p_lossy_02.tex}
        \includegraphics{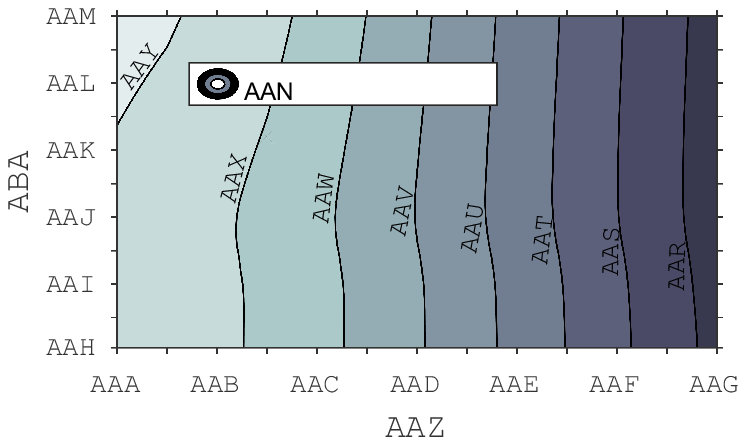}
        \label{fig:1_reg_pert_td}
    }
    \hfil
    \subfloat[$\REGLOGs$-$\PERTs$-$\TDs$, single-channel, single-span, lossless fiber]{
        \input{./figures/results_reglog-pert_td_rs_vs_p/rs_vs_p_lossy_02.tex}
        \includegraphics{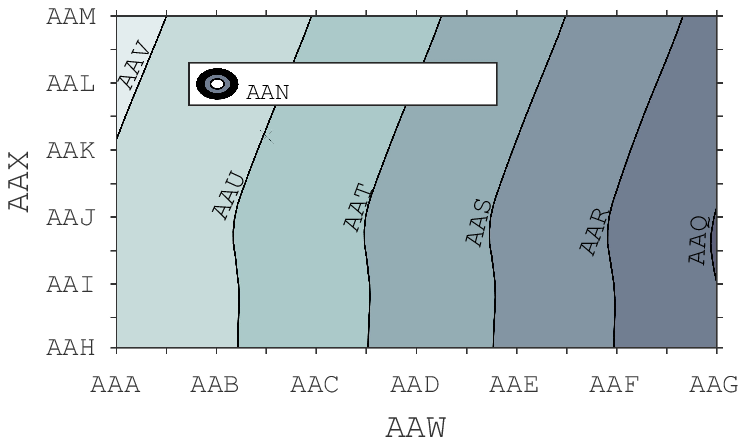}
        \label{fig:1_reglog_pert_td}
    }
    \caption{Contour plot of the normalized mean-square error $\sigma_{\rm e}^2 = \expval{\norm{\ve{y}^{\rm \scriptscriptstyle SSFM} - \ve{y}^{\rm \scriptscriptstyle PERT}}^2}$ in $\dB$ between the perturbation-based ($\rm \scriptstyle PERT$) end-to-end model and the split-step Fourier method ($\rm \scriptstyle SSFM$). The results are shown w.r.t. the symbol rate $\Rs$ and the optical launch power of the probe $\Pow_\probe$ in $\dBm$. Parameters as in Table \ref{tab:sim} with roll-off factor $\rolloff = 0.2$, $\Nsp = 1$, $10 \log_{10} \ee^\alpha = 0\,\dB/\km$ and $\Lsp = 21.71\,\km$. In (a) the \ul{reg}ular ($\REGs$) \ul{t}ime-\ul{d}omain ($\TDs$) model is carried out as in (\ref{eqn:discretee2ebottompage_time}) and in (b) the \ul{reg}ular-\ul{log}arithmic ($\REGLOGs$) model is carried out as in (\ref{eqn:pert_ansatz_hybrid}).}
    \label{fig:1}
\end{figure*}

\begin{table}[!t]
    \renewcommand{\arraystretch}{1.3}
    \caption{Simulation Parameters}
    \label{tab:sim}
    \centering
    \setlength{\tabcolsep}{.4em}
        \begin{tabular}{c||c|c}
            \hline
            \hline
            $\ve{a}$, $\ve{b} \in \mathcal{A}$ & \multicolumn{2}{c}{PDM 64-QAM}\\
            \hline
            $M$ & \multicolumn{2}{c}{4096 ($\equiv$ 64-QAM per polarization)}\\
            \hline
            $~ ~ ~h_\Tx(t)$~ ~ ~& \multicolumn{2}{c}{~ ~ ~$h_{\rm RRC}(t)$ with roll-off factor $\rolloff$}~ ~ ~\\
            \hline
            \hline
            $\gamma$ & \multicolumn{2}{c}{$1.1\,\watt^{-1}\mathrm{km}^{-1}$}\\
            \hline
            $\beta_2$ & \multicolumn{2}{c}{$-21\,\mathrm{ps}^2/\mathrm{km}$}\\
            \hline
            $\accD_0$ & \multicolumn{2}{c}{$0\,\mathrm{ps}^2$}\\
            \hline
            $\accD(z)$ & \multicolumn{2}{c}{$\beta_2 z$}\\
            \hline
            $10 \log_{10} \ee^\alpha$ &~ ~ ~ ~$0~\dB/\km$~ ~ ~ ~& $0.2\,\mathrm{dB/km}$\\
            \cline{1-1}
            $\Lsp$ & $21.71\,\km$ & $100\,\km$\\
            \cline{1-1}
            $\logG(z)$ & $0$ & $-\alpha z + \alpha \Lsp \sum_{i = 1}^{\Nsp} \delta(z-i\Lsp)$\\
            \hline
            \hline
            $\Nsym$ & \multicolumn{2}{c}{$2^{16}$}\\
            \hline
            $\Nfft$ & \multicolumn{2}{c}{${\max} \,(\;2^{\lceil \log_2 \STnu \rceil + 2}\;,\; 64\,)$}\\
            \hline
            $10\log_{10}\Gamma$ & \multicolumn{2}{c}{$-60\,\dB$}\\
            \hline
            \hline
        \end{tabular}
\end{table}

The simulation parameters are summarized in Table \ref{tab:sim}. A total number of $\Nsym = 2^{16}$ transmit symbols $\sequence{\ve{a}[k]}$ are randomly drawn from a \ac{PDM} $64$-ary \ac{QAM} symbol alphabet $\mathcal{A}$ with (4D) cardinality $M = |\mathcal{A}| = 4096$, i.e., 64-QAM per polarization.
The transmit pulse shape $h_\Tx(t)$ is a \ac{RRC} with roll-off factor $\rho$ and energy $\Eg$ to vary the optical launch power\footnote{In the previous sections, signals are always treated as dimensionless entities, but by convention we will still associate the optical launch power $P$ with units of $[\watt]$ and the nonlinearity coefficient $\gamma$ with $[1/(\watt\met)]$.} $\Pow$.

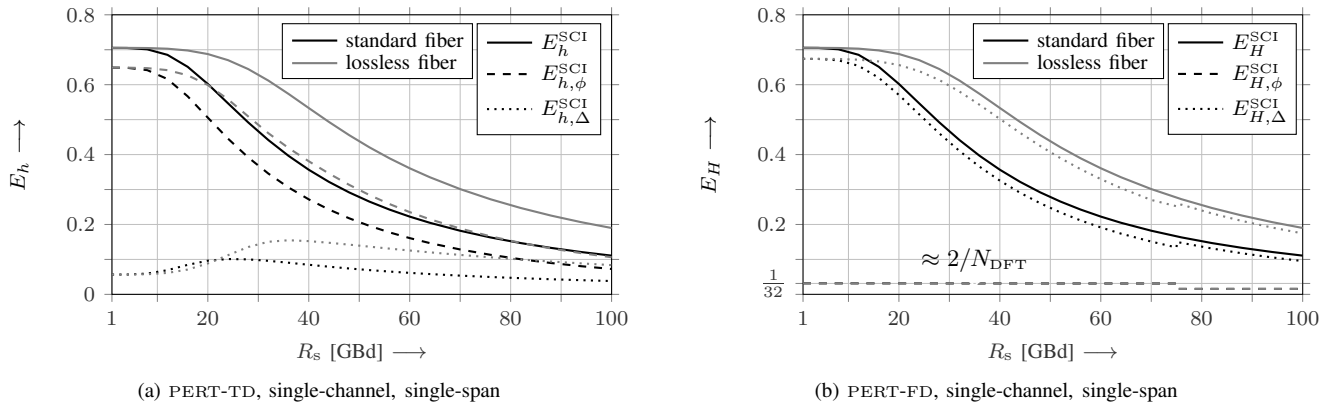
\begin{figure*}[t]
    \centering
    \subfloat[$\PERTs$-$\TDs$, single-channel, single-span]{
        \centering
        \input{./figures/E_td_sc_ss_lossy.tex}
    }
    \hfil
    \subfloat[$\PERTs$-$\FDs$, single-channel, single-span]{
        \centering
        \input{./figures/E_fd_sc_ss_lossy.tex}
    }
    \caption{Energy of the kernel coefficients in time-domain $E_h$ (a) and in frequency-domain $E_H$ (b) over the symbol rate $\Rs$. The kernel coefficients are obtained from the \ul{reg}ular-\ul{log}arithmic ($\REGLOGs$) model for a single-channel ($\rho =0.2$) over a standard single-mode fiber ($10 \log_{10} \ee^\alpha = 0.2\,\dB/\km$ and $\Lsp = 100\,\km$) or a lossless fiber ($10\log_{10} \ee^{\alpha} = 0\,\dB/\km$ and $\Lsp = 21.71\,\km$). The subscript $\Delta$ denotes the subset of all coefficients associated with additive perturbations and the subscript $\phi$ denotes the subset of all coefficients with multiplicative perturbations.}
    \label{fig:1_energy}
\end{figure*}

Two different optical amplification schemes are considered: ideal distributed Raman amplification (i.e., lossless transmission) and \textsl{transparent} end-of-span lumped amplification (i.e., lumped amplification where the effect of \textsl{signal-gain depletion} \cite[Sec. II B.]{Ghazisaeidi2017a} is neglected in the derivation of the perturbation model).
For lumped amplification we consider homogeneous spans of \ac{SSMF} with fiber attenuation $10\log_{10}\ee^\alpha = 0.2~\dB/\km$ and a span length of $\Lsp = 100~\km$.
In case of lossless transmission we have $10\log_{10}\ee^\alpha = 0~\dB/\km$ and span length $\Lsp= 21.71~\km$ corresponding to the asymptotic effective length $\Leffa \defeq 1/\alpha$ of a fictitious fiber with infinite length and attenuation $10 \log_{10} \ee^\alpha = 0.2~\dB/\km$.
The dispersion profile $\accD(z) = \beta_2 z$ conforms with modern \ac{DU} links, i.e., without optical inline dispersion compensation and bulk compensation at the receiver-side (typically performed in the digital domain).
Dispersion pre-compensation at the transmit-side can be easily incorporated via $\accD_0$ but is not considered in this work.
The dispersion coefficient is $\beta_2 = -21~\ps^2/\km$ and the nonlinearity coefficient is $\gamma = 1.1~\watt^{-1}\km^{-1}$, both constant over $z$ and $\omega$.
Additive noise due to \ac{ASE} and laser \ac{PN} are neglected since we only focus on deterministic signal-signal \ac{NLI}.

\begin{figure*}[b]
    \centering
    \subfloat[$\REGs$-$\PERTs$-$\FDs$, single-channel, single-span, lossless fiber]{
        \input{./figures/results_reg-pert_fd_rs_vs_p/rs_vs_p_lossy_02.tex}
        \includegraphics{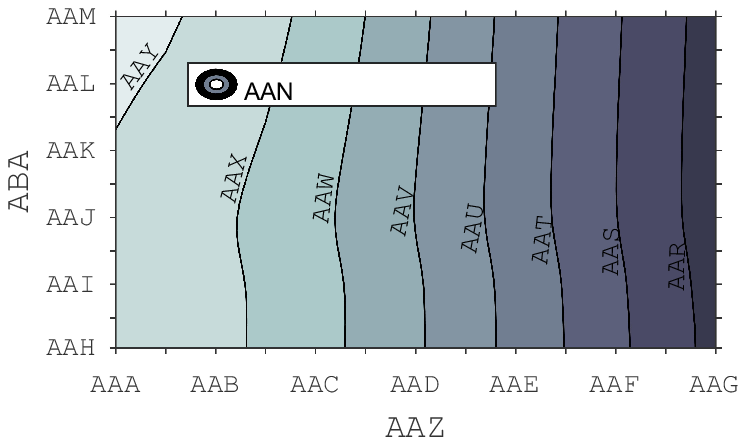}
        \label{fig:2_reg-pert-fd}
    }
    \hfil
    \subfloat[$\REGLOGs$-$\PERTs$-$\FDs$, single-channel, single-span, lossless fiber]{
        \input{./figures/results_reglog-pert_fd_rs_vs_p/rs_vs_p.tex}
        \includegraphics{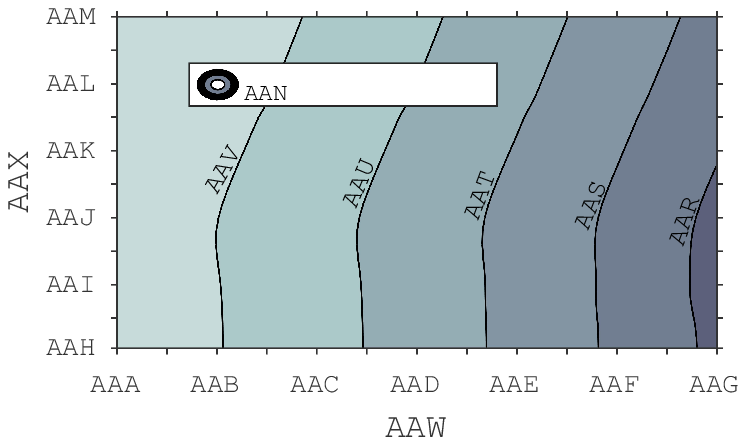}
        \label{fig:2_reglog-pert-fd}
    }
    \caption{Contour plot of the normalized mean-square error $\sigma_{\rm e}^2$ in $\dB$. The results are shown w.r.t. the symbol rate $\Rs$ and the optical launch power of the probe $\Pow_\probe$ in $\dBm$. Parameters as in Table \ref{tab:sim} with roll-off factor $\rolloff = 0.2$, $\Nsp = 1$, $10 \log_{10} \ee^\alpha = 0\,\dB/\km$ and $\Lsp = 21.71\,\km$. In (a) the \ul{reg}ular ($\REGs$) \ul{f}requency-\ul{d}omain ($\FDs$) model is carried out as in Algorithm \ref{IR} and in (b) the \ul{reg}ular-\ul{log}arithmic ($\REGLOGs$) model is carried out as in Algorithm \ref{IR:2}.}
    \label{fig:2}
\end{figure*}

The numerical reference simulation is a full-vectorial field simulation implemented via the \textsl{symmetric split-step Fourier method} \cite{Sinkin2003} with adaptive step size and a maximum nonlinear phase-rotation per step of $\phi_{\NL}^{\rm max} = 3.5\times 10^{-4}~{\rm rad}$. The simulation bandwidth is $B_{\rm \scriptscriptstyle SIM} = 8\Rs$ for single-channel and $16\Rs$ for dual-channel transmission.
All filter operations in the \ac{SSFM} reference simulation (i.e., pulse-shaping, linear step in the \ac{SSFM}, linear channel matched filter) are performed at the full simulation bandwidth via fast convolution and regarding periodic boundary conditions.

\subsection{Discussion of the Results}
In Fig.~\ref{fig:1} (a), we start our evaluation with the most simple scenario, i.e., single-channel, single-span, and lossless fiber. 
The \ac{MSE} is shown in logarithmic scale $10\log_{10}\sigma_{\rm e}^2$ in $\dB$ over the symbol rate $\Rs$ and the launch power of the probe $10 \log_{10}(\Pow_\probe /\mwatt)$ in $\dBm$.
The results are obtained from the \ul{reg}ular ($\REGs$) \ul{pert}urbation-based ($\PERTs$) end-to-end channel model in discrete \ul{t}ime-\ul{d}omain ($\TDs$), corresponding to (\ref{eqn:discretee2ebottompage_time}).
For the given effective length $\Leff$ and dispersion parameter $\beta_2$, the range of the symbol rate between $1\,\GBd$ and $100\,\GBd$ corresponds to a map strength $\STp$ between $0.003$ and $28.7$.
This amounts to virtually no memory of the intra-channel nonlinear interaction for small symbol rates (hence only very few coefficients $h_\probe[\ve{\kappa}]$ exceeding the minimum energy level of $10 \log_{10} \Gamma^{\superSCI} = -60\,\dB$) to a very broad intra-channel nonlinear memory for high symbol rates (with coefficients $h_\probe[\ve{\kappa}]$ covering a large number of symbols).
Likewise, the launch power of the probe $\Pow_\probe$ spans a nonlinear phase shift $\phinlp$ from $0.02$ to $0.34\,{\rm rad}$.
We can observe a gradual increase in $\sigma_\ee^2$ of about $5\,\dB$ per $1.5\,\dBm$ launch power in the nonlinear transmission regime.
We deliberately consider a \ac{MSE} $10 \log_{10}\sigma_\ee^2 > -30\,\dB$ as a \textsl{poor} match between the perturbation-based model and the full-field simulation, i.e., here for $\Pow_\probe$ larger than $9\,\dBm$ ($\equiv 0.168\,{\rm rad} \approx 10^{\circ}$) independent of $\Rs$.

In Fig.~\ref{fig:1} (b) the same system scenario is considered but instead of the regular model, now, the \ul{reg}ular-\ul{log}arithmic ($\REGLOGs$) model is employed according to (\ref{eqn:pert_ansatz_hybrid}).
The gradual increase in $\sigma_{\ee}^2$ with increasing $\Pow_\probe$ is now considerably relaxed to about $5\,\dB$ per $2.5\,\dBm$ launch power.
The region of poor model match with $10 \log_{10}\sigma_\ee^2 > -30\,\dB$ is now only approached for launch powers larger than $12\,\dBm$.
We can also observe that $\sigma_\ee^2$ improves with increasing symbol rate $\Rs$, in particular for rates $\Rs > 40\,\GBd$.
This is explained by the fact that the kernel energy $E_{h}^{\superSCI}$ in (\ref{eqn:kernelEnergy_time_tot}) depends on the symbol rate $\Rs$ s.t.~$\sigma_\ee^2$ is reduced for higher symbol rates.

\begin{figure*}[t]
    \centering
    \subfloat[$\REGLOGs$-$\PERTs$-$\TDs$, single-channel, single-span, standard fiber]{
        \input{./figures/results_reglog-pert_td_rs_vs_p_lossy/rs_vs_p.tex}
        \includegraphics{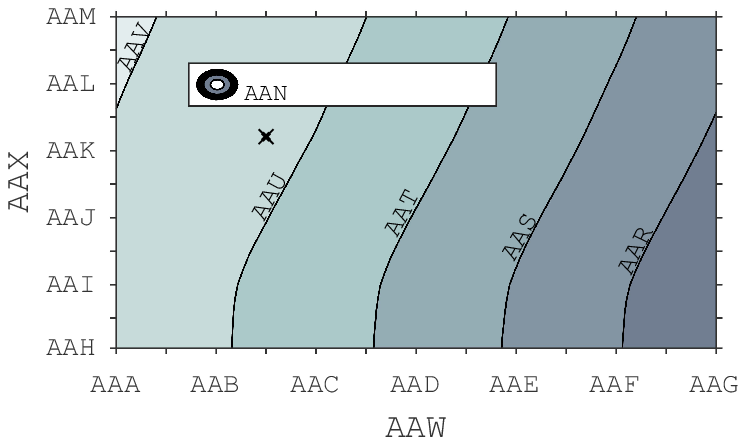}
        \label{fig:3_reglog-pert-td:single-span}
    }
    \hfil
    \subfloat[$\REGLOGs$-$\PERTs$-$\TDs$, single-channel, multi-span, standard fiber]{
        \input{./figures/results_reglog-pert_td_rho_vs_Nsp_lossy/rho_vs_nsp.tex}
        \includegraphics{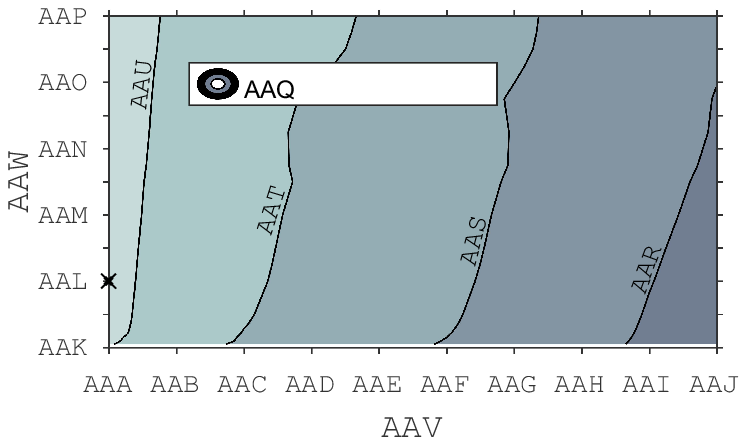}
        \label{fig:3_reglog-pert-td:multi-span}
    }
    \caption{Contour plot of the normalized mean-square error $\sigma_{\rm e}^2$ in $\dB$. The results are obtained from the \ul{reg}ular-\ul{log}arithmic ($\REGLOGs$) \ul{t}ime-\ul{d}omain ($\TDs$) model over a standard single-mode fiber ($10 \log_{10} \ee^\alpha = 0.2\,\dB/\km$ and $\Lsp = 100\,\km$) with end-of-span lumped amplification. In (a) the symbol rate $\Rs$ and the optical launch power $\Pow_\probe$ are varied for single-span ($\Nsp = 1$) transmission and fixed roll-off factor ($\rolloff = 0.2$). In (b) the roll-off factor $\rolloff$ and number of spans $\Nsp$ are varied with fixed symbol rate ($\Rs = 64\,\GBd$) and fixed launch power ($10\log_{10}(\Pow_\probe/\mwatt) = 3\,\dBm$). The black marker indicates the joint reference point with the same absolute value of $\sigma_\ee^2 = -51.4\,\dB$ but different gradient over the sweep parameter.}
    \label{fig:3}
\end{figure*}

Fig.~\ref{fig:1_energy} (a) shows the energy of the (time-domain) kernel coefficients $E_h^\superSCI$ over $\Rs$ for a single-span \ac{SSMF} with $\Lsp = 100\,\km$ and for a lossless fiber with $\Lsp = 21.71\,\km$. 
Generally, we see that $E_{h}^\superSCI$ is constant for small $\Rs$ and then curves into a transition region towards smaller energies for increasing $\Rs$.
For transmission over \ac{SSMF} this transition region is shifted to smaller $\Rs$, e.g., $E_h^\superSCI$ drops from $0.7$ to $0.6$ around $33\,\GBd$ for lossless transmission and at around $20\,\GBd$ for transmission over \ac{SSMF}.
We also present the kernel energies $E_{h, \Delta}^\superSCI$ associated with additive perturbations, and $E_{h,\phi}^{\superSCI}$ associated with multiplicative perturbations.
For this single-span scenario, most of the energy is concentrated in $E_{h, \phi}^{\superSCI}$, i.e., corresponding to the degenerate symbol combinations with $\kappa_1 = 0$ or $\kappa_3 = 0$ defined in (\ref{eqn:setDef2})--(\ref{eqn:setDef4}).
Interestingly, while the total energy $E_h^\superSCI$ decreases monotonically with $\Rs$, the additive contribution $E_{h, \Delta}^\superSCI$ increases in the transition region and then decreases again for large $\Rs$.
This behaviour is also visible in the results presented in Fig.~\ref{fig:1} (a) and (b).

Fig.~\ref{fig:1_energy} (b) shows the energy of the kernel coefficients $E_H^\superSCI$ in frequency-domain for the same system scenario as in (a).
The total energies are the same, i.e., $E_h^\superSCI = E_H^\superSCI$ (cf.~Parseval's theorem), however, the majority of the energy is now contained in the regular (additive) subset of coefficients.
The energy of the degenerate, i.e., multiplicative, subset of coefficients $E_{H,\phi}^\superSCI$ depends on the frequency discretization (which coincides here with the transformation length $\Nfft$) and is approximately $2/\Nfft$. The exact value $(2\Nfft-1)/\Nfft^2$ would be achieved for $\rho = 0$. 
For $\Rs>75.1\,\GBd$ we have $\STp > 16$ and it can be seen that $E_{H,\phi}^{\superSCI}$ drops from $1/32$ to $1/64$ and $E^\superSCI_{H, \Delta}$ jumps up by an equal amount because $\Nfft$ increases from $64$ to $128$ (cf. the set of simulation parameters in Table \ref{tab:sim}).
The $\REGLOGs$ frequency-domain model is hence pre-dominantly a regular model, where only the average multiplicative effects are truly treated as such.

In Fig.~\ref{fig:2} (a) and (b), the respective results on $\sigma_\ee^2$ using the discrete \ul{f}requency-\ul{d}omain ($\FDs$) model according to Algorithm 1 and 2 are shown.
We can confirm our previous statement that the regular perturbation model in time and frequency are equivalent considering that the results shown in Fig.~\ref{fig:1} (a) and Fig.~\ref{fig:2} (a) are (virtually) the same.
We also observe that the $\REGLOGs$-$\FDs$ performs very similar to the corresponding $\TDs$ model despite the fact that only \textsl{average} terms are considered as multiplicative distortions.
We conclude that---in the considered system scenario---$\REGLOGs$ models benefit from the fact that the \textsl{average} phase and polarization rotations are properly represented compared to pure $\REGs$ models.
The \textsl{time-variant} phase and polarization rotations that fluctuate around the average can to some extent also be represented by an additive perturbation without significant loss in performance.
This observation is in line with the \ac{eRP} method introduced in \cite[Sec.~VI]{Vannucci2002a}.
In the \ac{eRP} view, the perturbation expansion is performed in a ``SPM-rotated reference system'' \cite{Serena2013, Serena2015} where the time-average phase rotation is a priori included or a posteriori removed from the regular solution, cf.~\cite[Eq.~(33)]{Serena2013}.

Fig.~\ref{fig:3} (a) shows $\sigma_\ee^2$ for a single-channel over standard single-mode fiber ($\Lsp = 100\,\km$ and $10 \log_{10} \ee^\alpha = 0.2\,\dB/\km$) and lumped end-of-span amplification.
In the full-field simulation, the lumped amplifier is operated in \textsl{constant-gain} mode compensating for the exact span-loss of $20\,\dB$.
The results over a single-span in Fig.~\ref{fig:3} (a) are very similar in the low symbol rate regime compared to the lossless case in Fig.~\ref{fig:1} (b).
For $\Rs$ larger than $20\,\GBd$, the \ac{MSE} starts to decrease at a higher rate compared to the lossless case. 
This is in line with the energy of the kernel coefficients $E_h^{\superSCI}$ for the standard fiber shown in Fig.~\ref{fig:1_energy} (a).

\begin{figure*}[b]
    \centering
    \subfloat[$\PERTs$-$\TDs$, single-channel, multi-span, standard fiber]{
        \centering
        \input{./figures/E_td_sc_multspan_lossy.tex}
    }
    \hfil
    \subfloat[$\PERTs$-$\TDs$, dual-channel, single-span, lossless fiber]{
        \centering
        \input{./figures/E_td_2chan_ss_lossless.tex}
    }
    \caption{In (a), the energy of the kernel coefficients (black lines, bullet markers, left y-axis) in time-domain $E_h$ is shown over $\Nsp$ spans of standard single-mode fiber ($10 \log_{10} \ee^\alpha = 0.2\,\dB/\km$ and $\Lsp = 100\,\km$, $\rho = 0.2$). Additionally, the kernel energies are shown scaled with $\Nsp^2 \propto \phinlp^2$ (gray lines, cross markers, right y-axis) to indicate the general growth of nonlinear distortions with increasing $\Nsp$ (similar to the \ac{GN}-model). In (b), kernel energies $E_h$ are shown for cross-channel interference (XCI) imposed by a single wavelength channel spaced at $\Delta \omega_1/(2 \pi)\, \GHz$ over a single span of lossless fiber. Both probe and interferer have $\Rs = 64\,\GBd$ and $\rho = 0.2$.}
    \label{fig:3_energy}
\end{figure*}
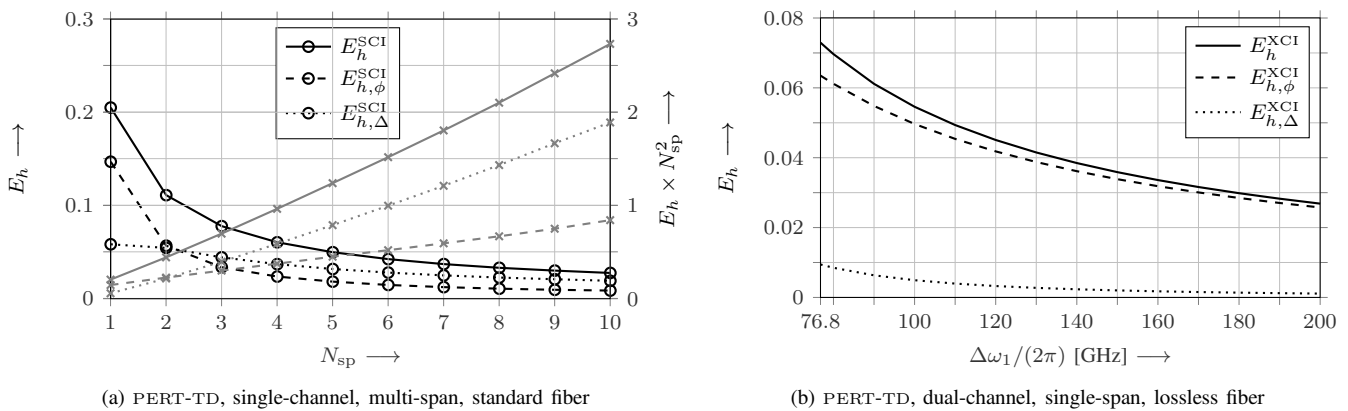

In Fig.~\ref{fig:3} (b), $\sigma_\ee^2$ is shown over the roll-off factor $\rolloff$ and the number of spans $\Nsp$ for a fixed symbol rate of $\Rs = 64\,\GBd$ and a fixed launch power of $10 \log_{10}(\Pow_\probe/\mwatt) = 3\,\dBm$.
The black cross in Fig.~\ref{fig:3} (a) and (b) indicates the point with a common set of parameters.
We can see a dependency on the roll-off factor $\rho$ which is due to a dependency of $E_h^\superSCI$ on $\rho$ (not shown here).
With increasing $\rho$ the kernel energy $E_h^\superSCI$ decreases and hence does $\sigma_\ee^2$ too. 

The scaling laws of $\sigma_\ee^2$ with $\Nsp$ are complemented in Fig.~\ref{fig:3_energy} (a) by the energy of the kernel coefficients $E_h^\superSCI$ for the same system scenario as in Fig.~\ref{fig:3} (b) (with $\rho = 0.2$).
It is interesting to see that (for this particular system scenario) $E_{h,\Delta}^\superSCI$ and $E_{h, \phi}^\superSCI$ intersect at $\Nsp = 2$. 
We can conclude that after the second span more energy is comprised within the additive subset of coefficients than in the multiplicative one.
With increasing $\Nsp$ the relative contribution of $E_{h, \Delta}^\superSCI$ to the total energy $E_h^\superSCI$ is increasing.
Note, while $E_h^\superSCI$ is actually monotonically decreasing with $\Nsp$, the common pre-factor $\phinlp$ has to be factored in as it effectively scales the nonlinear distortion. 
Since for heterogeneous spans we have $\phinlp \propto \Leff \propto \Nsp$, the same traces are shown scaled by $\Nsp^2$ to illustrate how the energy of the total distortion accumulates with increasing transmission length.
In this respect, similar results can be obtained from the presented channel model as from the \ac{GN}/\ac{EGN}-model (given proper scaling with $\phinlp^2$ instead of just $\Nsp^2$, and similarly taking all other wavelength channels into account).

In particular, the model correctly predicts the strength of the nonlinear distortion when the roll-off factor is larger than zero. Then, aliasing of frequency components from nonlinear distortions is properly included.

Additionally, qualitative statements can be derived, e.g., whether the nonlinear distortion is pre-dominantly additive or multiplicative. 
From the energy spread of the kernel coefficients one can also deduce the time scale over which nonlinear distortions are still correlated.

\begin{figure*}[b]
    \centering
    \subfloat[$\REGLOGs$-$\PERTs$-$\TDs$, dual-channel, single-span, lossless fiber]{
        \input{./figures/results_reglog-pert_td_rs_vs_p_2chan/chsp_vs_p1.tex}
        \includegraphics{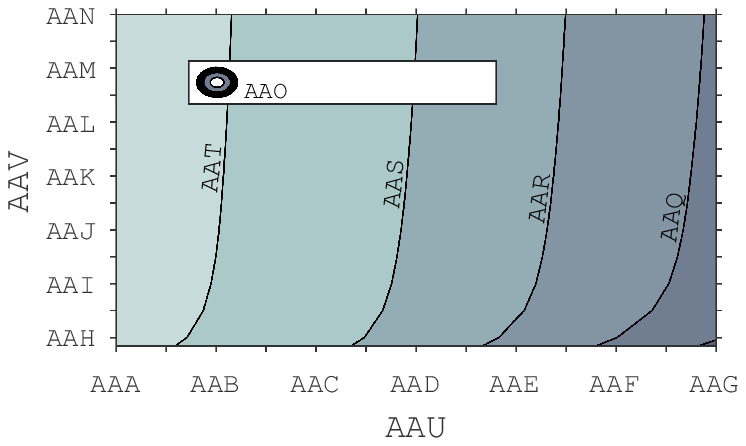}
        \label{fig:4_reglog-pert-td}
    }
    \hfil
    \subfloat[$\REGLOGs$-$\PERTs$-$\FDs$, dual-channel, single-span, lossless fiber]{
        \input{./figures/results_reglog-pert_fd_rs_vs_p_2chan/chsp_vs_p1.tex}
        \includegraphics{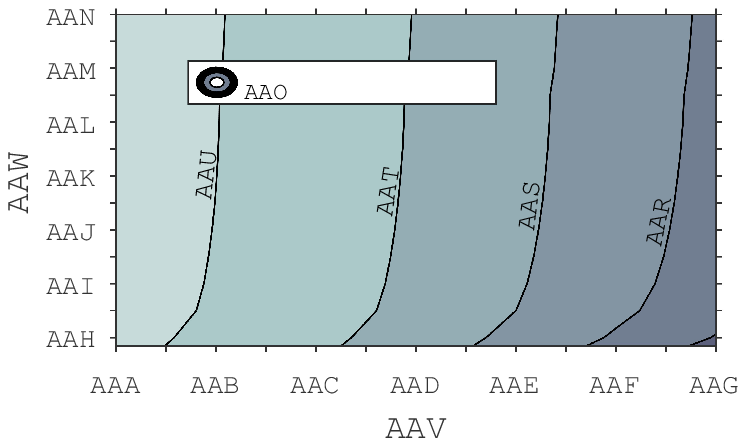}
        \label{fig:4_reglog-pert-fd}
    }
    \caption{Contour plot of the normalized mean-square error $\sigma_{\rm e}^2$ in $\dB$. The results are obtained from two co-propagating wavelength channels with PDM $64$-QAM and a symbol rate of $64\,\GBd$ and roll-off factor $\rho = 0.2$. The launch power of the probe is fixed at $10\log_{10}(\Pow_\probe/\mwatt) = 0\,\dBm$ while the power of the interferer $\Pow_1$ and the relative frequency offset $\Delta\omega_1$ are varied. In (a) the \ul{reg}ular-\ul{log}arithmic ($\REGLOGs$) \ul{t}ime-\ul{d}omain ($\TDs$) model for both \ac{SCI} and \ac{XCI} is carried out as in (\ref{eqn:pert_ansatz_hybrid}) and in (b) the $\REGLOGs$ \ul{f}requency-\ul{d}omain ($\FDs$) model is carried out as in Algorithm \ref{IR:2} and (\ref{eqn:pert_ansatz_hybrid_freq}) for both \ac{SCI} and \ac{XCI}.}
    \label{fig:4}
\end{figure*}

Fig.~\ref{fig:4} shows the $\sigma_\ee^2$ for dual-channel transmission using either the $\REGLOGs$ time-domain (a) or frequency-domain model (b).
The transmit symbols of the interferer $\sequence{\ve{b}[k]}$ are drawn from the same symbol set $\mathcal{A}$, i.e., $64$-\ac{QAM} per polarization.
For both wavelength channels, the symbol rate is fixed to $\Rs = 64\,\GBd$ and the roll-off factor of the \ac{RRC} shape is $\rolloff = 0.2$.
The transmit power of the probe is set to $10 \log_{10}(\Pow_\probe/\mwatt) = 0\,\dBm$ while the transmit power of the interferer $\Pow_\nu$ with channel number $\nu = 1$ is varied together with the relative frequency offset $\Delta \omega_1/(2 \pi)$ ranging from $76.8\,\GHz$ (i.e., no guard interval with $(1+\rho)\times 64\,\GHz$) to $200\,\GHz$.
In the numerical simulation via \ac{SSFM} we use an ideal channel combiner and both wavelength channels co-propagate at the full simulation bandwidth $B_{\rm \scriptscriptstyle SIM} = 16\Rs$.
In case of the end-to-end channel model both contributions from intra- and inter-channel distortions are combined into a single perturbative term (cf. (\ref{eqn:pert_ansatz_hybrid}) and (\ref{eqn:pert_ansatz_hybrid_freq})).
The baseline error $\sigma_\ee^2$ is therefore approximately $-55\,\dB$ considering the respective case with $\Rs = 64\,\GBd$ and $\Pow_\probe = 0\,\dBm$ in Fig.~\ref{fig:1} (b).
It is seen that the time- and frequency-domain model perform very similar. 

The dependency on the channel spacing $\Delta \omega_1$ is explained considering Fig.~\ref{fig:3_energy} (b).
Here, the energy of the cross-channel coefficients $h_1[\ve{\kappa}]$ is shown over $\Delta \omega_1$.
Generally, with increasing $\Delta \omega_1$, $E_h^\superXCI$ decreases and additionally the relative contribution of the degeneracy at $\kappa_3 = 0$, i.e., $E_{h, \phi}^\superSCI$, is growing.
Ultimately, the main distortion caused by an interferer spaced far away from the probe channel is a distortion in phase and state of polarization.

\section{Conclusion and Outlook}\label{sec:IIIII}
In this paper, a comprehensive analysis of end-to-end channel models for fiber-optic transmission based on a perturbation approach is presented.
The existing view on nonlinear interference following the \textsl{pulse collision picture} is described in a unified framework with a novel frequency-domain perspective that incorporates the time-discretization via an aliased frequency-domain kernel.
The relation between the time- and frequency-domain representation is elucidated and we show that the kernel coefficients in both views are related by a \ac{3D} discrete-time Fourier transform.
The energy of the \textsl{unaliased} kernel can be directly related to the GN-model.
The energy of the \textsl{aliased} kernel also takes the $T$-spaced sampling on the receiver into account.

While the pulse collision picture addresses the importance of separating additive and multiplicative terms for \textsl{inter}-channel nonlinear interactions, a generalization to \textsl{intra}-channel nonlinear interactions is presented.
An intra-channel phase distortion term and an intra-channel \ac{XPolM} term are introduced and both correspond to a subset of degenerate intra-channel pulse collisions.
In analogy to the time-domain model, the frequency-domain model is modified to treat certain degenerate mixing products as multiplicative distortions.
As a result, we have established a complete formulation of strictly \textsl{regular} (i.e., additive) models, and \textsl{regular-logarithmic} (i.e., mixed additive and multiplicative) models---both in time- and in frequency-domain, both for intra- and inter-channel nonlinear interference.

Derived from the frequency-domain description, a novel class of algorithms is proposed which effectively computes the end-to-end relation between transmit and receive sequences over discrete frequencies from the Nyquist interval.
One potential application of the frequency-domain model can be in fiber nonlinearity compensation.
Here, the model can be applied in a reverse manner at the transmit-side before pulse-shaping or on the receive-side after matched filtering.
Moreover, while the time-domain implementation requires a triple summation per time-instance, the frequency-domain implementation involves only a double summation per frequency index.
Similar as for linear systems, this characteristic allows for an efficient implementations using the fast Fourier transform when the time-domain kernel comprises many coefficients.

The derived algorithms were compared to the (oversampled and inherently sequential) split-step Fourier method based on the mean-squared error between both output sequences.
We show that, in particular, the regular-logarithmic models have good agreement with the split-step Fourier method over a wide range of system parameters.
The presented results are further supported by a qualitative analysis involving the kernel energies to quantify the relative contributions of either additive or multiplicative distortions.

Future work will address a thorough complexity analysis of the presented models.

\appendices
\section{Proof of the relation in (\ref{eqn:ele2ebottompage_freq}), (\ref{eqn:ele2ebottompage_time})}\label{apx:proof1}
In this appendix we compute the Fourier transform of $\Delta \ve{s}(t)$ in (\ref{eqn:ele2ebottompage_time}) similar to \cite[Appx.]{Ablowitz2002d}.

We start our derivation by expressing the optical field envelope $\ve{u}(0,t)$ by its inverse Fourier transform of $\ve{U}(0, \omega)$ to obtain
\begin{align}
    \pert{{\ve{s}}}(t) = & - \jj\gamma \frac{8}{9} \Leff \int_{\R^2} h_{\NL}(\tau_1, \tau_2)\\
                        &\times \ve{u}(0,t + \tau_1) \ve{u}^{\H}(0,t + \tau_1 + \tau_2) \ve{u}(t+\tau_2) \dd^2 \ve{\tau}\nonumber\\
    = &-\jj\gamma \frac{8}{9} \Leff \frac{1}{(2 \pi)^3} \iint_{-\infty}^{+\infty} \dd \tau_1 \dd \tau_2 \; h_{\NL}(\tau_1,\tau_2)\nonumber\\
    &\times \int_{-\infty}^{\infty} \dd \omega_3 \ve{U}(0,\omega_3) \exp(\jj \omega_3 \tau_1) \nonumber\\
    &\times \int_{-\infty}^{\infty} \dd \omega_2 \ve{U}^{\H}(0,\omega_2) \exp(-\jj \omega_2 (\tau_1 + \tau_2)) \nonumber\\
    &\times \int_{-\infty}^{\infty} \dd \omega_1 \ve{U}(0,\omega_1) \exp(\jj \omega_1 \tau_2) \nonumber\\
    &\times \exp(\jj (\omega_3 - \omega_2 + \omega_1) t).\nonumber
\end{align}
The Fourier transform of the former expression yields
\begin{align}
    \pert{{\ve{S}}}(\omega) = &-\jj\gamma \frac{8}{9} \Leff \frac{1}{(2 \pi)^3} \iiint_{-\infty}^{+\infty} \dd t \dd \tau_1 \dd \tau_2 \; h_{\NL}(\tau_1,\tau_2)\nonumber\\ 
                              &\times \int_{-\infty}^{\infty} \dd \omega_3 \ve{U}(0,\omega_3) \exp(\jj \omega_3 \tau_1) \nonumber\\ 
                              &\times \int_{-\infty}^{\infty} \dd \omega_2 \ve{U}^{\H}(0,\omega_2) \exp(-\jj \omega_2 (\tau_1 + \tau_2)) \nonumber\\ 
                              &\times \int_{-\infty}^{\infty} \dd \omega_1 \ve{U}(0,\omega_1) \exp(\jj \omega_1 \tau_2) \nonumber\\ 
                              &\times \exp(\jj (\omega_3 - \omega_2 + \omega_1 - \omega) t).
\end{align}
We now use the identity $\int_{-\infty}^{\infty} \exp(\jj (\omega_3 - \omega_2 + \omega_1 - \omega) t)\dd t  = 2 \pi \delta(\omega_3 - \omega_2 + \omega_1 - \omega)$ to obtain
\begin{align}
    \pert{{\ve{S}}}(\omega) = &-\jj\gamma \frac{8}{9} \Leff \frac{1}{(2 \pi)^2} \iint_{-\infty}^{+\infty} \dd \tau_1 \dd \tau_2 \; h_{\NL}(\tau_1,\tau_2)\nonumber\\
                  &\times \ve{U}(0,\omega -\omega_1+\omega_2) \exp(\jj (\omega-\omega_1+\omega_2) \tau_1)\nonumber\\
                  &\times \int_{-\infty}^{\infty} \dd \omega_2 \ve{U}^{\H}(0,\omega_2) \exp(-\jj \omega_2 (\tau_1 + \tau_2))\nonumber\\
                  &\times \int_{-\infty}^{\infty} \dd \omega_1 \ve{U}(0,\omega_1) \exp(\jj \omega_1 \tau_2).
    \label{eqn:A3}
\end{align}
After re-arranging the order of integration, we have
\begin{align}
    \pert{{\ve{S}}}(\omega) = &-\jj\gamma \frac{8}{9} \Leff \frac{1}{(2 \pi)^2} \iint_{-\infty}^{+\infty} \dd \omega_1 \dd \omega_2 \\
                              &\times \ve{U}(0, \omega - \omega_1 + \omega_2) \ve{U}^{\H}(0, \omega_2) \ve{U}(0,\omega_1)\nonumber\\
                    &\times \iint_{-\infty}^{\infty} \dd \tau_1 \dd \tau_2 h_{\NL}(\tau_1, \tau_2) \exp(\jj \omega_1 \tau_2) \nonumber\\
                    &\times \exp(-\jj \omega_2 (\tau_1 + \tau_2)) \exp(\jj (\omega-\omega_1 +\omega_2) \tau_1).\nonumber
\end{align}
And finally a change of variables with $\upsilon_1 = \omega_1 - \omega$ and $\upsilon_2 = \omega_2 - \omega_1$ yields
\begin{align}
    \pert{{\ve{S}}}(\omega) = &-\jj\gamma \frac{8}{9} \Leff \frac{1}{(2 \pi)^2} \iint_{-\infty}^{+\infty} \dd \upsilon_1 \dd \upsilon_2 \\
                    &\times \ve{U}(0, \omega + \upsilon_2) \ve{U}^{\H}(0, \omega + \upsilon_1 + \upsilon_2) \ve{U}(0,\omega + \upsilon_1)\nonumber\\
                    &\times \underbrace{\iint_{-\infty}^{\infty} \dd \tau_1 \dd \tau_2 h_{\NL}(\tau_1, \tau_2) \exp(-\jj \upsilon_1 \tau_1 - \jj \upsilon_2 \tau_2)}_{H_{\NL}(\upsilon_1, \upsilon_2) = \fourier{h_{\NL}(\tau_1, \tau_2)}}, \nonumber
\end{align}
which is equivalent to the expression in (\ref{eqn:ele2ebottompage_freq}). \QEDB

\bibliographystyle{IEEEtran}

\end{document}

%% file: figures/var_assignment/timeAssignment_1.tex
%
%
\providecommand\matlabtextA{\color[rgb]{0.000,0.000,0.000}\fontsize{8}{8}\selectfont\strut}%
\psfrag{000}[tc][tc]{\matlabtextA $\tau_2$}%
\psfrag{001}[tl][tl]{\matlabtextA $t_3$}%
\psfrag{002}[tc][tc]{\matlabtextA $\tau_2$}%
\psfrag{003}[tl][tl]{\matlabtextA $t_2$}%
\psfrag{004}[tc][tc]{\matlabtextA $\tau_1$}%
\psfrag{005}[tl][tl]{\matlabtextA $t_1$}%
\psfrag{006}[tl][tl]{\matlabtextA $t$}%
\psfrag{007}[bl][bl]{\matlabtextA $u(z,t)$}%
\psfrag{008}[bl][bl]{\matlabtextA $t$}%
%

%% file: figures/var_assignment/freqAssignment_1.tex
%
%
\providecommand\matlabtextA{\color[rgb]{0.000,0.000,0.000}\fontsize{8}{8}\selectfont\strut}%
\psfrag{001}[tc][tc]{\matlabtextA $\upsilon_2$}%
\psfrag{002}[tc][tc]{\matlabtextA $\omega_3$}%
\psfrag{003}[tc][tc]{\matlabtextA $\upsilon_2$}%
\psfrag{004}[tc][tc]{\matlabtextA $\omega_2$}%
\psfrag{005}[tc][tc]{\matlabtextA $\upsilon_1$}%
\psfrag{006}[tc][tc]{\matlabtextA $\omega_1$}%
\psfrag{007}[tc][tc]{\matlabtextA $\omega$}%
\psfrag{008}[bl][bl]{\matlabtextA $U(z,\omega)$}%
\psfrag{009}[bl][bl]{\matlabtextA $\omega$}%
\providecommand\matlabtextB{\color[rgb]{1.000,1.000,1.000}\fontsize{8}{8}\selectfont\strut}%
\psfrag{000}[bl][bl]{\matlabtextB $x$}%
%

%% file: figures/nonlinear_transfer_fun_2D/nonlinear_transfer_fun_2D_11.tex
%
%
\providecommand\matlabtextA{\color[rgb]{0.200,0.200,0.200}\fontsize{8}{8}}%
\psfrag{AAL}[cc][cc]{\matlabtextA $\upsilon_1 / (2\pi \Rs)$}%
\psfrag{AAM}[cc][cc]{\matlabtextA $\upsilon_2 / (2\pi \Rs)$}%
\psfrag{AAN}[cc][cc]{\matlabtextA $10 \log_{10} |H_{\rm NL}(\upsilon_1,\upsilon_2)|^2 \rightarrow$}%
%
%
%
\def\matlabfragNegXTick{\mathord{\makebox[0pt][r]{$-$}}}
\providecommand\matlabtextB{\color[rgb]{0.200,0.200,0.200}\fontsize{8}{8}}%
\psfrag{AAA}[cc][cc]{\matlabtextB $\matlabfragNegXTick 1$}%
\psfrag{AAB}[cc][cc]{\matlabtextB $0$}%
\psfrag{AAC}[cc][cc]{\matlabtextB $1$}%
%
%
%
\psfrag{AAD}[cc][cc]{\matlabtextB $-1$}%
\psfrag{AAE}[cc][cc]{\matlabtextB $-0.5$}%
\psfrag{AAF}[cc][cc]{\matlabtextB $0$}%
\psfrag{AAG}[cc][cc]{\matlabtextB $0.5$}%
\psfrag{AAH}[cc][cc]{\matlabtextB $1$}%
%
%
%
\psfrag{AAI}[cc][cc]{\matlabtextB $-40$}%
\psfrag{AAJ}[cc][cc]{\matlabtextB $-20$}%
\psfrag{AAK}[cc][cc]{\matlabtextB $0$}%
%

%% file: figures/nonlinear_transfer_fun_Rs/nonlinear_transfer_fun_rs_01.tex
%
%
\providecommand\matlabtextA{\color[rgb]{0.000,0.000,0.000}\fontsize{6}{6}\selectfont\strut}%
\psfrag{010}[bl][bl]{\matlabtextA $128~\GBd$}%
\psfrag{011}[bl][bl]{\matlabtextA $64~\GBd$}%
\psfrag{012}[bl][bl]{\matlabtextA $32~\GBd$}%
\psfrag{013}[bl][bl]{\matlabtextA $16~\GBd$}%
\psfrag{014}[bl][bl]{\matlabtextA $8~\GBd$}%
\psfrag{015}[bl][bl]{\matlabtextA $4~\GBd$}%
\providecommand\matlabtextB{\color[rgb]{0.200,0.200,0.200}\fontsize{10}{10}\selectfont\strut}%
\psfrag{016}[tc][tc]{\matlabtextB $\dOm / (2\pi \Rs)^2 \longrightarrow$}%
\psfrag{017}[bc][bc]{\matlabtextB $10 \log_{10} |H_\NL(\dOm)|^2 \longrightarrow$}%
%
%
%
\def\matlabfragNegXTick{\mathord{\makebox[0pt][r]{$-$}}}
\providecommand\matlabtextC{\color[rgb]{0.200,0.200,0.200}\fontsize{8}{8}\selectfont\strut}%
\psfrag{000}[ct][ct]{\matlabtextC $\matlabfragNegXTick 1$}%
\psfrag{001}[ct][ct]{\matlabtextC $\matlabfragNegXTick 0.5$}%
\psfrag{002}[ct][ct]{\matlabtextC $0$}%
\psfrag{003}[ct][ct]{\matlabtextC $0.5$}%
\psfrag{004}[ct][ct]{\matlabtextC $1$}%
%
%
%
\psfrag{005}[rc][rc]{\matlabtextC $-40$}%
\psfrag{006}[rc][rc]{\matlabtextC $-30$}%
\psfrag{007}[rc][rc]{\matlabtextC $-20$}%
\psfrag{008}[rc][rc]{\matlabtextC $-10$}%
\psfrag{009}[rc][rc]{\matlabtextC $0$}%
%

%% file: figures/end_to_end_pert_model/end_to_end_pert_model_v02.pstex_t
\begin{picture}(0,0)%
\includegraphics{end_to_end_pert_model_v02.ps}%
\end{picture}%
%
%
\setlength{\unitlength}{3158sp}%
\begingroup\makeatletter\ifx\SetFigFont\undefined%
\gdef\SetFigFont#1#2#3#4#5{%
  \reset@font\fontsize{#1}{#2pt}%
  \fontfamily{#3}\fontseries{#4}\fontshape{#5}%
  \selectfont}%
\fi\endgroup%
\begin{picture}(9809,3109)(8373,-2512)
\put(18167,-531){\makebox(0,0)[b]{\smash{{\SetFigFont{10}{12.0}{\familydefault}{\mddefault}{\updefault}{\color[rgb]{0,0,0}$\mathbb{C}^2$}%
}}}}
\put(12083,-747){\makebox(0,0)[b]{\smash{{\SetFigFont{10}{12.0}{\familydefault}{\mddefault}{\updefault}{\color[rgb]{0,0,0}$H_{\probe}(\ve{\omega}) = H_{\probe}(\omega_1, \omega_2, \omega_3)$}%
}}}}
\put(17096,-476){\makebox(0,0)[b]{\smash{{\SetFigFont{10}{12.0}{\familydefault}{\mddefault}{\updefault}{\color[rgb]{0,0,0}$k T$}%
}}}}
\put(11926,-2241){\makebox(0,0)[b]{\smash{{\SetFigFont{10}{12.0}{\familydefault}{\mddefault}{\updefault}{\color[rgb]{0,0,0}$H_{\probe}(\ee^{\jj \ve{\omega} T}) = \text{\scshape Alias} \{\, H_{\probe} (\ve{\omega}) \, \}$}%
}}}}
\put(15178,-1533){\makebox(0,0)[b]{\smash{{\SetFigFont{10}{12.0}{\familydefault}{\mddefault}{\updefault}{\color[rgb]{0,0,0}$-\mathrm{j}\frac{8}{9} \frac{L_\mathrm{eff}}{L_{\mathrm{NL},\probe}}$}%
}}}}
\put(17096,-476){\makebox(0,0)[b]{\smash{{\SetFigFont{10}{12.0}{\familydefault}{\mddefault}{\updefault}{\color[rgb]{0,0,0}$k T$}%
}}}}
\put(12071,-271){\makebox(0,0)[b]{\smash{{\SetFigFont{10}{12.0}{\familydefault}{\mddefault}{\updefault}{\color[rgb]{0,0,0}$\frac{T^4}{P_{\probe} E_{\mathrm{T},\probe}} H_{\mathrm{T},\probe}(\omega_1) H_{\mathrm{T},\probe}^*(\omega_2) H_{\mathrm{T},\probe}(\omega_3) H_{\mathrm{T},\probe}^*(\omega) H_{\mathrm{NL}}(\omega_2\! -\! \omega_1, \omega_2\! -\! \omega_3)$}%
}}}}
\put(9458,406){\makebox(0,0)[b]{\smash{{\SetFigFont{10}{12.0}{\familydefault}{\mddefault}{\updefault}{\color[rgb]{0,0,0}$\ve{A}(\ee^{\jj \omega_1 T}) \ve{A}^\H (\ee^{\jj \omega_2 T}) \ve{A}(\ee^{\jj \omega_3 T})$}%
}}}}
\put(8431,-514){\makebox(0,0)[b]{\smash{{\SetFigFont{10}{12.0}{\familydefault}{\mddefault}{\updefault}{\color[rgb]{0,0,0}$\mathbb{C}^2$}%
}}}}
\put(16592,414){\makebox(0,0)[b]{\smash{{\SetFigFont{10}{12.0}{\familydefault}{\mddefault}{\updefault}{\color[rgb]{0,0,0}$-\mathrm{j}\frac{8}{9} \frac{L_\mathrm{eff}}{L_{\mathrm{NL},\probe}}$}%
}}}}
\put(18143,-60){\makebox(0,0)[b]{\smash{{\SetFigFont{10}{12.0}{\familydefault}{\mddefault}{\updefault}{\color[rgb]{0,0,0}$\Delta\ve{A}^{\rm \scriptscriptstyle SCI}(\ee^{\jj \omega T})$}%
}}}}
\put(14081,-2227){\makebox(0,0)[b]{\smash{{\SetFigFont{10}{12.0}{\familydefault}{\mddefault}{\updefault}{\color[rgb]{0,0,0}$\int_{\TT^2} \mathrm{d}\omega_1 \mathrm{d}\omega_2$}%
}}}}
\put(15741,-263){\makebox(0,0)[b]{\smash{{\SetFigFont{10}{12.0}{\familydefault}{\mddefault}{\updefault}{\color[rgb]{0,0,0}$\int_{\mathbb{R}^2} \mathrm{d}\omega_1 \mathrm{d}\omega_2$}%
}}}}
\end{picture}%

%% file: figures/nonlinear_transfer_fun_alias/nonlinear_transfer_aliased_kern.tex
%
%
\providecommand\matlabtextA{\color[rgb]{0.000,0.000,0.000}\fontsize{7}{7}\selectfont\strut}%
%
\providecommand\matlabtextB{\color[rgb]{0.200,0.200,0.200}\fontsize{10}{10}\selectfont\strut}%
\psfrag{AAJ}[tc][tc]{\matlabtextB $\omega_2 \,/\, (2\pi \Rs) \longrightarrow$}%
\psfrag{AAK}[bc][bc]{\matlabtextB $\omega_1 \,/\, (2\pi \Rs) \longrightarrow$}%
\psfrag{AAX}[cc][cc]{\footnotesize \color{myred}$\TT^2$}%
%
%
%
%
\providecommand\matlabtextC{\color[rgb]{0.200,0.200,0.200}\fontsize{8}{8}\selectfont\strut}%
\psfrag{AAA}[ct][ct]{\matlabtextC $-1.5$}%
\psfrag{AAB}[ct][ct]{\matlabtextC $-0.5$}%
\psfrag{AAC}[ct][ct]{\matlabtextC $0.5$}%
\psfrag{AAD}[ct][ct]{\matlabtextC $1.5$}%
%
%
%
\psfrag{AAE}[rc][rc]{\matlabtextC $-1.5$}%
\psfrag{AAF}[rc][rc]{\matlabtextC $-0.5$}%
\psfrag{AAG}[rc][rc]{\matlabtextC $0.5$}%
\psfrag{AAH}[rc][rc]{\matlabtextC $1.5$}%

\providecommand\matlabtextD{\color[rgb]{0.200,0.200,0.200}\fontsize{6}{6}\selectfont\strut}%
\psfrag{BBA}[rc][rc]{\matlabtextD $0\,\dB$}%
\psfrag{BBB}[rc][rc]{\matlabtextD $-10\,\dB$}%
\psfrag{BBC}[rc][rc]{\matlabtextD $-20\,\dB$}%
\psfrag{BBD}[rc][rc]{\matlabtextD $-30\,\dB$}%
\psfrag{BBE}[rc][rc]{\matlabtextD $-40\,\dB$}%
\psfrag{BBF}[rc][rc]{\matlabtextD $-50\,\dB$}%
\psfrag{BBG}[rc][rc]{\matlabtextD $-60\,\dB$}%
\psfrag{BBH}[rc][rc]{\matlabtextD $-70\,\dB$}%
\psfrag{BBI}[rc][rc]{\matlabtextD $-80\,\dB$}%
%

%% file: figures/results_reg-pert_td_rs_vs_p/rs_vs_p_lossy_02.tex
%
%
\providecommand\matlabtextA{\color[rgb]{0.000,0.000,0.000}\fontsize{8}{8}\selectfont\strut}%
\psfrag{AAN}[bl][bl]{\matlabtextA $~10\log_{10}(\sigma_{\rm e}^2)$ [dB]}%
\providecommand\matlabtextB{\color[rgb]{0.000,0.000,0.000}\fontsize{8}{8}\selectfont\strut}%
\psfrag{AAO}[cc][cc]{\matlabtextB 5}%
\psfrag{AAP}[cc][cc]{\matlabtextB 10}%
\psfrag{AAQ}[cc][cc]{\matlabtextB 15}%
\psfrag{AAR}[cc][cc]{\matlabtextB $\ve{-20}$}%
\psfrag{AAS}[cc][cc]{\matlabtextB $\ve{-25}$}%
\psfrag{AAT}[cc][cc]{\matlabtextB $\ve{-30}$}%
\psfrag{AAU}[cc][cc]{\matlabtextB $\ve{-35}$}%
\psfrag{AAV}[cc][cc]{\matlabtextB $\ve{-40}$}%
\psfrag{AAW}[cc][cc]{\matlabtextB $\ve{-45}$}%
\psfrag{AAX}[cc][cc]{\matlabtextB $\ve{-50}$}%
\psfrag{AAY}[cc][cc]{\matlabtextB $\ve{-55}$}%
\providecommand\matlabtextC{\color[rgb]{0.200,0.200,0.200}\fontsize{8}{8}\selectfont\strut}%
\psfrag{AAZ}[tc][tc]{\matlabtextC $10 \log_{10}(\Pow_\probe/\mwatt)$ [dBm] $\longrightarrow$}%
\psfrag{ABA}[bc][bc]{\matlabtextC $\Rs$ [GBd] $\longrightarrow$}%
%
%
%
\def\matlabfragNegXTick{\mathord{\makebox[0pt][r]{$-$}}}
\providecommand\matlabtextD{\color[rgb]{0.200,0.200,0.200}\fontsize{8}{8}\selectfont\strut}%
\psfrag{AAA}[ct][ct]{\matlabtextD $0$}%
\psfrag{AAB}[ct][ct]{\matlabtextD $2$}%
\psfrag{AAC}[ct][ct]{\matlabtextD $4$}%
\psfrag{AAD}[ct][ct]{\matlabtextD $6$}%
\psfrag{AAE}[ct][ct]{\matlabtextD $8$}%
\psfrag{AAF}[ct][ct]{\matlabtextD $10$}%
\psfrag{AAG}[ct][ct]{\matlabtextD $12$}%
%
%
%
\psfrag{AAH}[rc][rc]{\matlabtextD $1$}%
\psfrag{AAI}[rc][rc]{\matlabtextD $20$}%
\psfrag{AAJ}[rc][rc]{\matlabtextD $40$}%
\psfrag{AAK}[rc][rc]{\matlabtextD $60$}%
\psfrag{AAL}[rc][rc]{\matlabtextD $80$}%
\psfrag{AAM}[rc][rc]{\matlabtextD $100$}%
%

%% file: figures/results_reglog-pert_td_rs_vs_p/rs_vs_p_lossy_02.tex
%
%
\providecommand\matlabtextA{\color[rgb]{0.000,0.000,0.000}\fontsize{8}{8}\selectfont\strut}%
\psfrag{AAN}[bl][bl]{\matlabtextA $~10\log_{10}(\sigma_{\rm e}^2)$ [dB]}%
\providecommand\matlabtextB{\color[rgb]{0.000,0.000,0.000}\fontsize{8}{8}\selectfont\strut}%
\psfrag{AAO}[cc][cc]{\matlabtextB $\ve{-20}$}%
\psfrag{AAP}[cc][cc]{\matlabtextB $\ve{-25}$}%
\psfrag{AAQ}[cc][cc]{\matlabtextB $\ve{-30}$}%
\psfrag{AAR}[cc][cc]{\matlabtextB $\ve{-35}$}%
\psfrag{AAS}[cc][cc]{\matlabtextB $\ve{-40}$}%
\psfrag{AAT}[cc][cc]{\matlabtextB $\ve{-45}$}%
\psfrag{AAU}[cc][cc]{\matlabtextB $\ve{-50}$}%
\psfrag{AAV}[cc][cc]{\matlabtextB $\ve{-55}$}%
\providecommand\matlabtextC{\color[rgb]{0.200,0.200,0.200}\fontsize{8}{8}\selectfont\strut}%
\psfrag{AAW}[tc][tc]{\matlabtextC $10 \log_{10}(\Pow_\probe/\mwatt)$ [dBm] $\longrightarrow$}%
\psfrag{AAX}[bc][bc]{\matlabtextC $\Rs$ [GBd] $\longrightarrow$}%
%
%
%
\def\matlabfragNegXTick{\mathord{\makebox[0pt][r]{$-$}}}
\providecommand\matlabtextD{\color[rgb]{0.200,0.200,0.200}\fontsize{8}{8}\selectfont\strut}%
\psfrag{AAA}[ct][ct]{\matlabtextD $0$}%
\psfrag{AAB}[ct][ct]{\matlabtextD $2$}%
\psfrag{AAC}[ct][ct]{\matlabtextD $4$}%
\psfrag{AAD}[ct][ct]{\matlabtextD $6$}%
\psfrag{AAE}[ct][ct]{\matlabtextD $8$}%
\psfrag{AAF}[ct][ct]{\matlabtextD $10$}%
\psfrag{AAG}[ct][ct]{\matlabtextD $12$}%
%
%
%
\psfrag{AAH}[rc][rc]{\matlabtextD $1$}%
\psfrag{AAI}[rc][rc]{\matlabtextD $20$}%
\psfrag{AAJ}[rc][rc]{\matlabtextD $40$}%
\psfrag{AAK}[rc][rc]{\matlabtextD $60$}%
\psfrag{AAL}[rc][rc]{\matlabtextD $80$}%
\psfrag{AAM}[rc][rc]{\matlabtextD $100$}%
%

%% file: figures/E_td_sc_ss_lossy.tex
%
%

\begin{tikzpicture}[font=\footnotesize]
\begin{axis}[%
width=\fwidth,
height=\fheight,
scale only axis,
xmin=1,
xmax=100,
major tick length=.1cm,
xtick={1,10,20,...,100},
xticklabels={{$1$},{},{$20$},{},{$40$},{},{$60$},{},{$80$},{},{$100$}},
xminorticks=true,
xlabel={\color[rgb]{0.2,0.2,0.2}$\Rs$ [GBd] $\longrightarrow$},
ylabel={$E_h$ $\longrightarrow$},
ytick={0,0.1,0.2,...,0.8},
yticklabels={{$0$},{},{$0.2$},{},{$0.4$},{},{$0.6$},{},{$0.8$}},
xmajorgrids,
tick align=outside,
ymin=0,
ymax=0.8,
yminorticks=true,
axis background/.style={fill=white},
xmajorgrids,
ymajorgrids,
yminorgrids,
grid style={solid,very thin},
legend style={legend cell align=left, align=left, draw=white!15!black}
]
\addplot [color=black,thick]
  table[]{data/E_td_sc_ss_lossy-1.tsv};
  \label{p1_t}

\addplot [color=black,thick,dashed]
  table[]{data/E_td_sc_ss_lossy-2.tsv};
  \label{p2_t}

\addplot [color=black,thick,dotted]
  table[]{data/E_td_sc_ss_lossy-3.tsv};
  \label{p3_t}

\addplot [color=gray,thick]
  table[]{data/E_td_sc_ss_lossless-1.tsv};
  \label{p4_t}

\addplot [color=gray,thick,dashed]
  table[]{data/E_td_sc_ss_lossless-2.tsv};

\addplot [color=gray,thick,dotted]
  table[]{data/E_td_sc_ss_lossless-3.tsv};

\node [draw,fill=white,anchor=north west] at (rel axis cs: 0.73,0.97) {\shortstack[l]{
\ref{p1_t} $E_{h}^{\superSCI}$ \\
\ref{p2_t} $E_{h,\phi}^{\superSCI}$\\
\ref{p3_t} $E_{h,\Delta}^{\superSCI}$}};

\node [draw,fill=white,anchor=north west] at (rel axis cs: 0.34,0.97) {\shortstack[l]{
\ref{p1_t} standard fiber \\
\ref{p4_t} lossless fiber}};

\end{axis}
\end{tikzpicture}%

%% file: figures/E_fd_sc_ss_lossy.tex
%
%
\pgfdeclarelayer{background layer}
\pgfdeclarelayer{foreground layer}
\pgfsetlayers{background layer,main,foreground layer}
\begin{tikzpicture}[font=\footnotesize]
\begin{axis}[%
width=\fwidth,
height=\fheight,
scale only axis,
xmin=1,
xmax=100,
major tick length=.1cm,
xtick={1,10,20,...,100},
xticklabels={{$1$},{},{$20$},{},{$40$},{},{$60$},{},{$80$},{},{$100$}},
every outer x axis line/.append style={white!5!black},
xminorticks=true,
xlabel={\color[rgb]{0.2,0.2,0.2}$\Rs$ [GBd] $\longrightarrow$},
ylabel={$E_H$ $\longrightarrow$},
ytick={0,0.03125,0.1,0.2,...,0.8},
yticklabels={{$~$},{$\frac{1}{32}$},{},{$0.2$},{},{$0.4$},{},{$0.6$},{},{$0.8$}},
xmajorgrids,
tick align=outside,
ymin=0,
ymax=0.8,
yminorticks=true,
axis background/.style={fill=white},
xmajorgrids,
ymajorgrids,
yminorgrids,
grid style={solid,very thin},
legend style={legend cell align=left, align=left, draw=white!15!black}
]
\addplot [color=black,thick]
  table[]{data/E_fd_sc_ss_lossy-1.tsv};
\label{p1}

\addplot [color=black,thick,dashed]
  table[]{data/E_fd_sc_ss_lossy-2.tsv};
\label{p2}

\addplot [color=black,thick,dotted]
  table[]{data/E_fd_sc_ss_lossy-3.tsv};
\label{p3}

\addplot [color=gray,thick]
  table[]{data/E_fd_sc_ss_lossless-1.tsv};
\label{p4}

\addplot [color=gray,thick,dashed]
  table[]{data/E_fd_sc_ss_lossless-2.tsv};

\addplot [color=gray,thick,dotted]
  table[]{data/E_fd_sc_ss_lossless-3.tsv};


\begin{pgfonlayer}{foreground layer}
    \draw[<->,thick,color=white,font=\fontsize{4}{5}] (axis cs:35, 0.02) -- node[above=0.5mm] {\color{black} \small $\approx2/\Nfft$} (axis cs:35, 0.03);
\end{pgfonlayer}
\node [draw,fill=white,anchor=north west] at (rel axis cs: 0.73,0.97) {\shortstack[l]{
\ref{p1} $E_{H}^{\superSCI}$ \\
\ref{p2} $E_{H,\phi}^{\superSCI}$\\
\ref{p3} $E_{H,\Delta}^{\superSCI}$}};

\node [draw,fill=white,anchor=north west] at (rel axis cs: 0.34,0.97) {\shortstack[l]{
\ref{p1} standard fiber \\
\ref{p4} lossless fiber}};

\end{axis}
\end{tikzpicture}%

%% file: figures/results_reg-pert_fd_rs_vs_p/rs_vs_p_lossy_02.tex
%
%
\providecommand\matlabtextA{\color[rgb]{0.000,0.000,0.000}\fontsize{8}{8}\selectfont\strut}%
\psfrag{AAN}[bl][bl]{\matlabtextA $~10\log_{10}(\sigma_{\rm e}^2)$ [dB]}%
\providecommand\matlabtextB{\color[rgb]{0.000,0.000,0.000}\fontsize{8}{8}\selectfont\strut}%
\psfrag{AAO}[cc][cc]{\matlabtextB 5}%
\psfrag{AAP}[cc][cc]{\matlabtextB 10}%
\psfrag{AAQ}[cc][cc]{\matlabtextB 15}%
\psfrag{AAR}[cc][cc]{\matlabtextB $\ve{-20}$}%
\psfrag{AAS}[cc][cc]{\matlabtextB $\ve{-25}$}%
\psfrag{AAT}[cc][cc]{\matlabtextB $\ve{-30}$}%
\psfrag{AAU}[cc][cc]{\matlabtextB $\ve{-35}$}%
\psfrag{AAV}[cc][cc]{\matlabtextB $\ve{-40}$}%
\psfrag{AAW}[cc][cc]{\matlabtextB $\ve{-45}$}%
\psfrag{AAX}[cc][cc]{\matlabtextB $\ve{-50}$}%
\psfrag{AAY}[cc][cc]{\matlabtextB $\ve{-55}$}%
\providecommand\matlabtextC{\color[rgb]{0.200,0.200,0.200}\fontsize{8}{8}\selectfont\strut}%
\psfrag{AAZ}[tc][tc]{\matlabtextC $10\log_{10}(\Pow_\probe/\mwatt)$ [dBm] $\longrightarrow$}%
\psfrag{ABA}[bc][bc]{\matlabtextC $\Rs$ [GBd] $\longrightarrow$}%
%
%
%
\def\matlabfragNegXTick{\mathord{\makebox[0pt][r]{$-$}}}
\providecommand\matlabtextD{\color[rgb]{0.200,0.200,0.200}\fontsize{8}{8}\selectfont\strut}%
\psfrag{AAA}[ct][ct]{\matlabtextD $0$}%
\psfrag{AAB}[ct][ct]{\matlabtextD $2$}%
\psfrag{AAC}[ct][ct]{\matlabtextD $4$}%
\psfrag{AAD}[ct][ct]{\matlabtextD $6$}%
\psfrag{AAE}[ct][ct]{\matlabtextD $8$}%
\psfrag{AAF}[ct][ct]{\matlabtextD $10$}%
\psfrag{AAG}[ct][ct]{\matlabtextD $12$}%
%
%
%
\psfrag{AAH}[rc][rc]{\matlabtextD $1$}%
\psfrag{AAI}[rc][rc]{\matlabtextD $20$}%
\psfrag{AAJ}[rc][rc]{\matlabtextD $40$}%
\psfrag{AAK}[rc][rc]{\matlabtextD $60$}%
\psfrag{AAL}[rc][rc]{\matlabtextD $80$}%
\psfrag{AAM}[rc][rc]{\matlabtextD $100$}%
%

%% file: figures/results_reglog-pert_fd_rs_vs_p/rs_vs_p.tex
%
%
\providecommand\matlabtextA{\color[rgb]{0.000,0.000,0.000}\fontsize{8}{8}\selectfont\strut}%
\psfrag{AAN}[bl][bl]{\matlabtextA $~10\log_{10}(\sigma_{\rm e}^2)$ [dB]}%
\providecommand\matlabtextB{\color[rgb]{0.000,0.000,0.000}\fontsize{8}{8}\selectfont\strut}%
\psfrag{AAO}[cc][cc]{\matlabtextB $\ve{-15}$}%
\psfrag{AAP}[cc][cc]{\matlabtextB $\ve{-20}$}%
\psfrag{AAQ}[cc][cc]{\matlabtextB $\ve{-25}$}%
\psfrag{AAR}[cc][cc]{\matlabtextB $\ve{-30}$}%
\psfrag{AAS}[cc][cc]{\matlabtextB $\ve{-35}$}%
\psfrag{AAT}[cc][cc]{\matlabtextB $\ve{-40}$}%
\psfrag{AAU}[cc][cc]{\matlabtextB $\ve{-45}$}%
\psfrag{AAV}[cc][cc]{\matlabtextB $\ve{-50}$}%
\providecommand\matlabtextC{\color[rgb]{0.200,0.200,0.200}\fontsize{8}{8}\selectfont\strut}%
\psfrag{AAW}[tc][tc]{\matlabtextC $10\log_{10}(\Pow_\probe/\mwatt)$ [dBm] $\longrightarrow$}%
\psfrag{AAX}[bc][bc]{\matlabtextC $\Rs$ [GBd] $\longrightarrow$}%
%
%
%
\def\matlabfragNegXTick{\mathord{\makebox[0pt][r]{$-$}}}
\providecommand\matlabtextD{\color[rgb]{0.200,0.200,0.200}\fontsize{8}{8}\selectfont\strut}%
\psfrag{AAA}[ct][ct]{\matlabtextD $0$}%
\psfrag{AAB}[ct][ct]{\matlabtextD $2$}%
\psfrag{AAC}[ct][ct]{\matlabtextD $4$}%
\psfrag{AAD}[ct][ct]{\matlabtextD $6$}%
\psfrag{AAE}[ct][ct]{\matlabtextD $8$}%
\psfrag{AAF}[ct][ct]{\matlabtextD $10$}%
\psfrag{AAG}[ct][ct]{\matlabtextD $12$}%
%
%
%
\psfrag{AAH}[rc][rc]{\matlabtextD $1$}%
\psfrag{AAI}[rc][rc]{\matlabtextD $20$}%
\psfrag{AAJ}[rc][rc]{\matlabtextD $40$}%
\psfrag{AAK}[rc][rc]{\matlabtextD $60$}%
\psfrag{AAL}[rc][rc]{\matlabtextD $80$}%
\psfrag{AAM}[rc][rc]{\matlabtextD $100$}%
%

%% file: figures/results_reglog-pert_td_rs_vs_p_lossy/rs_vs_p.tex
%
%
\providecommand\matlabtextA{\color[rgb]{0.000,0.000,0.000}\fontsize{8}{8}\selectfont\strut}%
\psfrag{AAN}[bl][bl]{\matlabtextA $~10\log_{10}(\sigma_{\rm e}^2)$ [dB]}%
\providecommand\matlabtextB{\color[rgb]{0.000,0.000,0.000}\fontsize{8}{8}\selectfont\strut}%
\psfrag{AAO}[cc][cc]{\matlabtextB $\ve{-20}$}%
\psfrag{AAP}[cc][cc]{\matlabtextB $\ve{-25}$}%
\psfrag{AAQ}[cc][cc]{\matlabtextB $\ve{-30}$}%
\psfrag{AAR}[cc][cc]{\matlabtextB $\ve{-35}$}%
\psfrag{AAS}[cc][cc]{\matlabtextB $\ve{-40}$}%
\psfrag{AAT}[cc][cc]{\matlabtextB $\ve{-45}$}%
\psfrag{AAU}[cc][cc]{\matlabtextB $\ve{-50}$}%
\psfrag{AAV}[cc][cc]{\matlabtextB $\ve{-55}$}%
\providecommand\matlabtextC{\color[rgb]{0.200,0.200,0.200}\fontsize{8}{8}\selectfont\strut}%
\psfrag{AAW}[tc][tc]{\matlabtextC $10\log_{10}(\Pow_\probe/\mwatt)$ [dBm] $\longrightarrow$}%
\psfrag{AAX}[bc][bc]{\matlabtextC $\Rs$ [GBd] $\longrightarrow$}%
%
%
%
\def\matlabfragNegXTick{\mathord{\makebox[0pt][r]{$-$}}}
\providecommand\matlabtextD{\color[rgb]{0.200,0.200,0.200}\fontsize{8}{8}\selectfont\strut}%
\psfrag{AAA}[ct][ct]{\matlabtextD $0$}%
\psfrag{AAB}[ct][ct]{\matlabtextD $2$}%
\psfrag{AAC}[ct][ct]{\matlabtextD $4$}%
\psfrag{AAD}[ct][ct]{\matlabtextD $6$}%
\psfrag{AAE}[ct][ct]{\matlabtextD $8$}%
\psfrag{AAF}[ct][ct]{\matlabtextD $10$}%
\psfrag{AAG}[ct][ct]{\matlabtextD $12$}%
%
%
%
\psfrag{AAH}[rc][rc]{\matlabtextD $1$}%
\psfrag{AAI}[rc][rc]{\matlabtextD $20$}%
\psfrag{AAJ}[rc][rc]{\matlabtextD $40$}%
\psfrag{AAK}[rc][rc]{\matlabtextD $60$}%
\psfrag{AAL}[rc][rc]{\matlabtextD $80$}%
\psfrag{AAM}[rc][rc]{\matlabtextD $100$}%
%

%% file: figures/results_reglog-pert_td_rho_vs_Nsp_lossy/rho_vs_nsp.tex
%
%
\providecommand\matlabtextA{\color[rgb]{0.000,0.000,0.000}\fontsize{8}{8}\selectfont\strut}%
\psfrag{AAQ}[bl][bl]{\matlabtextA $~10\log_{10}(\sigma_{\rm e}^2)$ [dB]}%
\providecommand\matlabtextB{\color[rgb]{0.000,0.000,0.000}\fontsize{8}{8}\selectfont\strut}%
\psfrag{AAR}[cc][cc]{\matlabtextB $\ve{-35}$}%
\psfrag{AAS}[cc][cc]{\matlabtextB $\ve{-40}$}%
\psfrag{AAT}[cc][cc]{\matlabtextB $\ve{-45}$}%
\psfrag{AAU}[cc][cc]{\matlabtextB $\ve{-50}$}%
\providecommand\matlabtextC{\color[rgb]{0.200,0.200,0.200}\fontsize{10}{10}\selectfont\strut}%
\psfrag{AAV}[tc][tc]{\matlabtextC $\Nsp\longrightarrow$}%
\psfrag{AAW}[bc][bc]{\matlabtextC $\rho\longrightarrow$}%
%
%
%
\def\matlabfragNegXTick{\mathord{\makebox[0pt][r]{$-$}}}
\providecommand\matlabtextD{\color[rgb]{0.200,0.200,0.200}\fontsize{8}{8}\selectfont\strut}%
\psfrag{AAA}[ct][ct]{\matlabtextD $1$}%
\psfrag{AAB}[ct][ct]{\matlabtextD $2$}%
\psfrag{AAC}[ct][ct]{\matlabtextD $3$}%
\psfrag{AAD}[ct][ct]{\matlabtextD $4$}%
\psfrag{AAE}[ct][ct]{\matlabtextD $5$}%
\psfrag{AAF}[ct][ct]{\matlabtextD $6$}%
\psfrag{AAG}[ct][ct]{\matlabtextD $7$}%
\psfrag{AAH}[ct][ct]{\matlabtextD $8$}%
\psfrag{AAI}[ct][ct]{\matlabtextD $9$}%
\psfrag{AAJ}[ct][ct]{\matlabtextD $10$}%
%
%
%
\psfrag{AAK}[rc][rc]{\matlabtextD $0$}%
\psfrag{AAL}[rc][rc]{\matlabtextD $0.2$}%
\psfrag{AAM}[rc][rc]{\matlabtextD $0.4$}%
\psfrag{AAN}[rc][rc]{\matlabtextD $0.6$}%
\psfrag{AAO}[rc][rc]{\matlabtextD $0.8$}%
\psfrag{AAP}[rc][rc]{\matlabtextD $1$}%
%

%% file: figures/E_td_sc_multspan_lossy.tex
%
%

\begin{tikzpicture}[font=\footnotesize]
\begin{axis}[%
width=\fwidth,
height=\fheight,
scale only axis,
xmin=1,
xmax=10,
major tick length=.1cm,
xtick={1,2,3,...,10},
xticklabels={},
xminorticks=true,
xlabel={\color[rgb]{0.2,0.2,0.2}$\Nsp$ $\longrightarrow$},
ylabel={$E_h$ $\longrightarrow$},
ytick={0,0.05,0.1,0.15,0.2,0.25,0.3},
yticklabels={{$0$},{},{$0.1$},{},{$0.2$},{},{$0.3$}},
xmajorgrids,
tick align=outside,
axis x line=none,
ymin=0,
ymax=0.3,
yminorticks=true,
xmajorgrids,
ymajorgrids,
yminorgrids,
grid style={solid,very thin},
legend style={legend cell align=left, align=left, draw=white!15!black}
]
\addplot [color=black,thick,mark=o,every mark/.append style={solid}]
  table[]{data/E_td_sc_multspan_lossy-1.tsv};
  \label{p1_tms}

\addplot [color=black,dashed,thick,mark=o,every mark/.append style={solid}]
  table[]{data/E_td_sc_multspan_lossy-2.tsv};
  \label{p2_tms}

\addplot [color=black,dotted,thick,mark=o,every mark/.append style={solid}]
  table[]{data/E_td_sc_multspan_lossy-3.tsv};
\label{p3_tms}
\node [draw,fill=white,anchor=north west] at (rel axis cs: 0.33,0.97) {\shortstack[l]{
\ref{p1_tms} $E_{h}^{\superSCI}$ \\
\ref{p2_tms} $E_{h,\phi}^{\superSCI}$\\
\ref{p3_tms} $E_{h,\Delta}^{\superSCI}$}};

\end{axis}
\begin{axis}[%
width=\fwidth,
height=\fheight,
scale only axis,
xmin=1,
xmax=10,
major tick length=.1cm,
xtick={1,2,3,...,10},
xminorticks=true,
ylabel={\color{black}{$E_h\times\Nsp^2$ $\longrightarrow$}},
xlabel={\color[rgb]{0.2,0.2,0.2}$\Nsp$ $\longrightarrow$},
xmajorgrids,
tick align=outside,
ymin=0,
ymax=3,
yminorticks=true,
xtick={1,2,3,...,10},
xmajorgrids,
ymajorgrids,
ylabel near ticks,
yticklabel pos=right,
grid style={solid,very thin},
legend style={legend cell align=left, align=left, draw=white!15!black}
]
\addplot [color=black,gray,thick,mark=x,every mark/.append style={solid}]
  table[]{data/ExNsp_td_sc_multspan_lossy-1.tsv};

\addplot [color=black,gray,dashed,thick,mark=x,every mark/.append style={solid}]
  table[]{data/ExNsp_td_sc_multspan_lossy-2.tsv};

\addplot [color=black,gray,dotted,thick,mark=x,every mark/.append style={solid}]
  table[]{data/ExNsp_td_sc_multspan_lossy-3.tsv};

\end{axis}
\end{tikzpicture}%

%% file: figures/E_td_2chan_ss_lossless.tex
%
%

\begin{tikzpicture}[font=\footnotesize]
\begin{axis}[%
width=\fwidth,
height=\fheight,
scale only axis,
xmin=76.8,
xmax=200,
major tick length=.1cm,
xtick={76.8,80,90,100,...,200},
xticklabels={{$76.8$},{},{},{$100$},{},{$120$},{},{$140$},{},{$160$},{},{$180$},{},{$200$}},
xminorticks=true,
xlabel={\color[rgb]{0.2,0.2,0.2}$\Delta \omega_1 / (2\pi)$ [GHz] $\longrightarrow$},
ylabel={$E_h$ $\longrightarrow$},
ytick={0,0.01,...,0.08},
yticklabels={{$0$},{},{$0.02$},{},{$0.04$},{},{$0.06$},{},{$0.08$}},
xmajorgrids,
tick align=outside,
ymin=0,
ymax=0.08,
yminorticks=true,
axis background/.style={fill=white},
xmajorgrids,
ymajorgrids,
yminorgrids,
scaled y ticks=false,
grid style={solid,very thin},
legend style={legend cell align=left, align=left, draw=white!15!black}
]
\addplot [color=black,thick]
  table[]{data/E_td_2chan_ss_lossless-1.tsv};
  \label{p1_2chan}

\addplot [color=black,dashed,thick]
  table[]{data/E_td_2chan_ss_lossless-2.tsv};
  \label{p2_2chan}

\addplot [color=black,dotted,thick]
  table[]{data/E_td_2chan_ss_lossless-3.tsv};
  \label{p3_2chan}
\node [draw,fill=white,anchor=north west] at (rel axis cs: 0.73,0.97) {\shortstack[l]{
\ref{p1_2chan} $E_{h}^{\superXCI}$ \\
\ref{p2_2chan} $E_{h,\phi}^{\superXCI}$\\
\ref{p3_2chan} $E_{h,\Delta}^{\superXCI}$}};

\end{axis}
\end{tikzpicture}%

%% file: figures/results_reglog-pert_td_rs_vs_p_2chan/chsp_vs_p1.tex
%
%
\providecommand\matlabtextA{\color[rgb]{0.000,0.000,0.000}\fontsize{8}{8}\selectfont\strut}%
\psfrag{AAO}[bl][bl]{\matlabtextA $~10\log_{10}(\sigma_{\rm e}^2)$ [dB]}%
\providecommand\matlabtextB{\color[rgb]{0.000,0.000,0.000}\fontsize{8}{8}\selectfont\strut}%
\psfrag{AAP}[cc][cc]{\matlabtextB $\ve{-30}$}%
\psfrag{AAQ}[cc][cc]{\matlabtextB $\ve{-35}$}%
\psfrag{AAR}[cc][cc]{\matlabtextB $\ve{-40}$}%
\psfrag{AAS}[cc][cc]{\matlabtextB $\ve{-45}$}%
\psfrag{AAT}[cc][cc]{\matlabtextB $\ve{-50}$}%
\providecommand\matlabtextC{\color[rgb]{0.200,0.200,0.200}\fontsize{8}{8}\selectfont\strut}%
\psfrag{AAU}[tc][tc]{\matlabtextC $10 \log_{10}(\Pow_1/\mwatt)$ [dBm] $\longrightarrow$}%
\psfrag{AAV}[bc][bc]{\matlabtextC $\Delta \omega_1 / (2\pi)$ [GHz] $\longrightarrow$}%
%
%
%
\def\matlabfragNegXTick{\mathord{\makebox[0pt][r]{$-$}}}
\providecommand\matlabtextD{\color[rgb]{0.200,0.200,0.200}\fontsize{8}{8}\selectfont\strut}%
\psfrag{AAA}[ct][ct]{\matlabtextD $0$}%
\psfrag{AAB}[ct][ct]{\matlabtextD $2$}%
\psfrag{AAC}[ct][ct]{\matlabtextD $4$}%
\psfrag{AAD}[ct][ct]{\matlabtextD $6$}%
\psfrag{AAE}[ct][ct]{\matlabtextD $8$}%
\psfrag{AAF}[ct][ct]{\matlabtextD $10$}%
\psfrag{AAG}[ct][ct]{\matlabtextD $12$}%
%
%
%
\psfrag{AAH}[rc][rc]{\matlabtextD $80$}%
\psfrag{AAI}[rc][rc]{\matlabtextD $100$}%
\psfrag{AAJ}[rc][rc]{\matlabtextD $120$}%
\psfrag{AAK}[rc][rc]{\matlabtextD $140$}%
\psfrag{AAL}[rc][rc]{\matlabtextD $160$}%
\psfrag{AAM}[rc][rc]{\matlabtextD $180$}%
\psfrag{AAN}[rc][rc]{\matlabtextD $200$}%
%

%% file: figures/results_reglog-pert_fd_rs_vs_p_2chan/chsp_vs_p1.tex
%
%
\providecommand\matlabtextA{\color[rgb]{0.000,0.000,0.000}\fontsize{8}{8}\selectfont\strut}%
\psfrag{AAO}[bl][bl]{\matlabtextA $~10\log_{10}(\sigma_{\rm e}^2)$ [dB]}%
\providecommand\matlabtextB{\color[rgb]{0.000,0.000,0.000}\fontsize{8}{8}\selectfont\strut}%
\psfrag{AAP}[cc][cc]{\matlabtextB $\ve{-25}$}%
\psfrag{AAQ}[cc][cc]{\matlabtextB $\ve{-30}$}%
\psfrag{AAR}[cc][cc]{\matlabtextB $\ve{-35}$}%
\psfrag{AAS}[cc][cc]{\matlabtextB $\ve{-40}$}%
\psfrag{AAT}[cc][cc]{\matlabtextB $\ve{-45}$}%
\psfrag{AAU}[cc][cc]{\matlabtextB $\ve{-50}$}%
\providecommand\matlabtextC{\color[rgb]{0.200,0.200,0.200}\fontsize{8}{8}\selectfont\strut}%
\psfrag{AAV}[tc][tc]{\matlabtextC $10 \log_{10}(\Pow_1/\mwatt)$ [dBm] $\longrightarrow$}%
\psfrag{AAW}[bc][bc]{\matlabtextC $\Delta \omega_1 / (2\pi)$ [GHz] $\longrightarrow$}%
%
%
%
\def\matlabfragNegXTick{\mathord{\makebox[0pt][r]{$-$}}}
\providecommand\matlabtextD{\color[rgb]{0.200,0.200,0.200}\fontsize{8}{8}\selectfont\strut}%
\psfrag{AAA}[ct][ct]{\matlabtextD $0$}%
\psfrag{AAB}[ct][ct]{\matlabtextD $2$}%
\psfrag{AAC}[ct][ct]{\matlabtextD $4$}%
\psfrag{AAD}[ct][ct]{\matlabtextD $6$}%
\psfrag{AAE}[ct][ct]{\matlabtextD $8$}%
\psfrag{AAF}[ct][ct]{\matlabtextD $10$}%
\psfrag{AAG}[ct][ct]{\matlabtextD $12$}%
%
%
%
\psfrag{AAH}[rc][rc]{\matlabtextD $80$}%
\psfrag{AAI}[rc][rc]{\matlabtextD $100$}%
\psfrag{AAJ}[rc][rc]{\matlabtextD $120$}%
\psfrag{AAK}[rc][rc]{\matlabtextD $140$}%
\psfrag{AAL}[rc][rc]{\matlabtextD $160$}%
\psfrag{AAM}[rc][rc]{\matlabtextD $180$}%
\psfrag{AAN}[rc][rc]{\matlabtextD $200$}%
%